\documentclass[a4paper,11pt]{article}
\usepackage{jheppub} 
\usepackage{lineno}

\usepackage[normalem]{ulem}
\usepackage{xcolor}
\usepackage{mathtools}
\usepackage{comment}
\usepackage{footnote}
\usepackage{mathrsfs}
\usepackage{physics}
\usepackage{amssymb,bm}
\usepackage{amsmath}
\usepackage{slashed}
\usepackage{graphics}
\usepackage{graphicx}
\usepackage{tikz}
\usetikzlibrary{decorations.pathmorphing,decorations.markings,arrows.meta}
\usepackage{dsfont}

\definecolor{lcolour}{rgb}{0,0.34,0.25}
\definecolor{citcolour}{rgb}{0.73,0.31,0.28}

\hypersetup{
  linkcolor=lcolour,
  citecolor=citcolour,
  urlcolor=blue
}
\usepackage[caption=false]{subfig}
\usepackage[capitalise]{cleveref} 

\newcounter{diag}
\newcounter{subdiag}[diag]
\newcommand{\namediag}[1]{\refstepcounter{diag} \thediag \label{#1}}

\renewcommand{\thediag}{(\alph{diag})}

\title{\boldmath Deep inelastic scattering as a photon shockwave}

\author[a,b]{Patricia Gimeno-Estivill,}
\author[a,b]{Tuomas Lappi,}
\author[a,b]{Xinyu Qin}
\author[b,c]{and Mikko Seppälä}
\affiliation[a]{Department of Physics,
P.O. Box 35, 40014 University of Jyväskylä, Finland}
\affiliation[b]{Helsinki Institute of Physics,
P.O. Box 64, 00014 University of Helsinki, Finland}
\affiliation[c]{Department of Physics, P.O. Box 64, 00014 University of Helsinki, Finland}

\emailAdd{patricia.p.gimenoestivill@jyu.fi}
\emailAdd{tuomas.v.v.lappi@jyu.fi}
\emailAdd{qinx@jyu.fi}
\emailAdd{mikko.seppala@helsinki.fi}

\abstract{
We use  light-cone quantization and an abelian shockwave scattering picture, akin to the standard CGC calculational framework, to calculate the next-to-leading order quark- and gluon-initiated contributions to the standard collinear factorization picture of semi-inclusive and inclusive deep inelastic scattering. In particular we show that treating the external partons as off-shell states removes collinear divergences from the hard factor and allows for a straightforward and physically transparent implementation of collinear factorization, matching the derivation of splitting functions from the UV-renormalization of parton densities. We also show that a change of variables from projectile to target momentum fractions is essential in order to obtain the correct form of the coefficient functions. 
We propose this calculational procedure as a method for higher-order calculations of high-energy dilute--dense processes with colored targets, such as forward particle production in high-energy proton--nucleus collisions.
}

\begin{document}
\maketitle
\flushbottom

\section{Introduction}
\label{sec:intro}
The Color Glass Condensate (CGC) effective theory \cite{Gelis:2010nm, Kovchegov:2012mbw} describes high-energy scattering of a dilute projectile off a dense target in terms of eikonal scattering of the projectile partons off the classical color field of the target, which acts as a ``shockwave''~\cite{Balitsky:2004rr,Balitsky:2005we} localized in light-cone time $x^+$. In the simplest processes, such as deep inelastic scattering (DIS) in the dipole picture, the projectile is a virtual photon, which at leading order (LO) fluctuates into a quark-antiquark state. At  next-to-leading order (NLO)   the dipole picture calculation involves  the corrections to the light-cone wave functions of the photon  generated by one real or virtual photon~\cite{Beuf:2017bpd, Hanninen:2017ddy}. A part of the phase space of these corrections (referred to as the leading log, LL, part) has to be subtracted from the cross section to be resummed by the small-$x$ evolution for the target, via e.g. the Balitsky-Kovchegov equation \cite{Balitsky:1995ub, Kovchegov:1999yj, Kovchegov:1999ua}.

Many processes of phenomenological interest, however, have colored particles in the initial or in the final states. The prime example is single inclusive hadron production in forward proton-nucleus collisions, whose NLO calculation \cite{Chirilli:2011km, Chirilli:2012jd} (see also \cite{Liu:2022ijp}) contains also collinear divergences that are absorbed into the DGLAP evolution of the parton distributions of the proton and of the fragmentation functions of the produced hadron. Since then, NLO calculations and phenomenological applications have been carried out for several such processes \cite{Mantysaari:2023vfh, Casuga:2026dif, Gimeno-Estivill:2024gbu, Gimeno-Estivill:2025rbw, Caucal:2026ymj}. Nevertheless, they remain few compared with processes initiated by a photon, and each of them is technically demanding. It is therefore desirable to develop methods that make the collinear factorization of the colored projectile partons in such dilute--dense calculations systematic and transparent. A better understanding of collinear factorization can also be useful outside the realm of small $x$, either as a calculational tool or as a conceptually and physically clearer exposition of what collinear factorization in fact means. 

A particular difficulty in higher order CGC calcualtions concerns projectile and target momentum fractions. In light-cone perturbation theory (LCPT) calculations in the CGC one quantizes the probe, and so one naturally has access to the fraction of its longitudinal momentum that participates in the scattering. In the shockwave picture one does not, on the other hand, control the momentum fraction coming in from the target: since the shockwave is instantaneous in light cone time $\sim \delta(x^+)$, it is taken to have an unlimited supply of light cone energy $k^-$,  as much as is needed to put the final state particles on shell. Thus, in order to access the target momentum fraction transferred in the collision, we have to reconstruct it from the final state particles~\cite{Bertilsson:2026lzx}. In higher-order calculations, understanding the process in terms of the target momentum fraction is nevertheless crucial for a stable perturbative expansion. This has turned out to be essential for obtaining reasonable phenomenological results both for NLO single inclusive hadron production in proton-nucleus collisions~\cite{Ducloue:2016shw, Iancu:2016vyg, Ducloue:2017mpb, Watanabe:2016gws, Shi:2021hwx, Liu:2020mpy, Liu:2022ijp} and for DIS~\cite{Ducloue:2017ftk} in the CGC formalism.

In order to develop calculational methods and clarify these two aspects of CGC  shockwave scattering, we consider in this paper a simple Abelian version of the CGC approach. That is, we will calculate, at NLO in the strong coupling, the scattering of   a colored projectile off a  photon shockwave. In terms of a physical scattering process this  corresponds to DIS off a proton or nucleus, viewed in the unusual picture  in which the partons of the proton are the projectile and the electron is the target. In the high-energy limit, $y\ll1$, the photon field of the electron acts as an Abelian shockwave, off which the partons of the proton scatter in the same way as off the gluon field of a nucleus in the CGC: we are doing small-$y$ physics to understand small-$x$ physics.  Because the target is a single photon, the calculation can be done analytically, and the results can be compared with the well-known NLO coefficient functions of inclusive DIS \cite{Ellis:1996mzs} and semi-inclusive DIS (SIDIS) \cite{deFlorian:1997zj}. In general, DIS cross sections get contributions from transverse and longitudinal polarization states of the virtual photon, which can experimentally be distinguished by the $y$-dependence of the cross section. However, in the limit taken here we only include the eikonal coupling of the partons in the proton to the photon, so that the  photon shockwave couples only to the plus component of the electromagnetic current of the partons. As a result we can only obtain the part of the structure function accessible at $y=0$. For inclusive DIS this means that we calculate the structure function $F_2$ and, in SIDIS, the combination $2F_1+F_L$, but not the longitudinal and transverse structure functions separately. 

The Abelian shockwave calculation serves as a testbed for two different potential methods for organizing higher order calculations for scattering off a CGC color field: taking external particles to be off-shell and reconstructing the target momentum fractions from the final state kinematics.

Taking external colored particles to be off-shell provides a different angle to understanding and implementing collinear factorization.  Even at low scales, the partons in a proton are spacelike off-shell, corresponding to the fact that they are not free, but parts of a bound state with a binding energy at some hadronic scale. 
In Ref.~\cite{Lappi:2026xaa}, the DGLAP splitting function for the quark parton distribution function (PDF) at two loops was calculated in LCPT from the ultraviolet renormalization of the parton distributions of off-shell partons. In this case the off-shellness enables the calculation by ensuring  that all the poles in dimensional regularization are related to  UV divergences.
Off-shell external partons are also used in state-of-the-art multiloop calculations of the splitting functions, which have recently been extended to four loops \cite{Gehrmann:2023cqm, Gehrmann:2026qbl}.  Partons taken from a renormalized parton distribution 
in fact physically not thought of as being on shell, but as carrying a virtuality of the order of the factorization scale.  This is  how the initial- and final-state parton showers of Monte Carlo event generators are de facto implemented: they follow from DGLAP evolution in a showering process where the virtuality increases towards the hard scattering. It is then natural to implement the idea of external partons as off-shell states explicitly in the calculation of the hard cross section.

 A spacelike virtuality of the incoming partons regulates the collinear behavior of the cross section.
The off-shell factorization scheme is defined by the renormalization condition that the parton distributions and fragmentation functions of  off-shell partons are trivial when the factorization scale equals the virtuality:
\begin{equation}
\left. f^R_{i/j(\mu_0)}(x,\mu_F)\right|_{\mu_F=\mu_0} = \delta(1-x)\delta_{ij}.
\end{equation}
Here $f^R$ is the UV-renormalized PDF of parton flavor $i$ in a dressed parton $j$ that is off-shell with a virtuality $\mu_0$, and $\mu_F$ is the factorization scale introduced by the UV renormalization of the bare PDF.
With this definition the partonic cross section for the scattering of a parton with an off-shellness equal to the factorization scale is directly the coefficient function. There are no infrared divergences in the calculation, and one directly obtains a factorized cross section without any subtractions of infrared poles.  
In the off-shell picture, the factorization scale appears explicitly in the calculation as a physical scale, rather than as a remnant of a regularization procedure and the subtraction of a collinear pole. In multiscale problems, this could bring out more naturally the large logarithms that are to be resummed by DGLAP evolution.  We also believe that this approach could provide a much more physically transparent understanding of e.g. the treatment of massive quarks in collinear factorization.

In this paper we calculate the quark- and gluon-initiated channels for DIS and SIDIS cross sections at $\mathcal O(\alpha_\mathrm{s})$ with off-shell incoming and outgoing partons. From these we obtain the coefficient functions in the off-shell scheme. We also relate the off-shell scheme coefficient functions to the conventional $\overline{\mathrm{MS}}$ ones by a finite renormalization of the parton distributions and fragmentation functions,  which we also reproduce by a direct calculation of the $\overline{\mathrm{MS}}$ parton distributions and fragmentation functions of off-shell partons.

The other aspect of our calculation, which is known to be essential for understanding high energy factorization, is the reconstruction of target momentum fractions from the final state. We start in the usual way in the shockwave picture, by formulating the calculation  in terms of the momentum fractions of the projectile. Indeed already at this stage the existence of a large collinear logarithm, and the associated splitting function, are manifest. But in order to get the correct expression for the measured DIS (or SIDIS) cross sections, one needs to calculate the cross section differentially in the experimentally measured Bjorken variable $x_\text{Bj}$ (and on the fragmentation momentum fraction $z_H$ in the case of SIDIS). In turns out that in the kinematics that we use here these are in fact related to \emph{target} (electron) momentum fractions. Thus, in the shockwave picture, they need to be reconstructed from the final state momenta in the scattering. More precisely, both  $x_\text{Bj}$ and  $z_H$ are inversely proportional to the total light cone energy transferred from the photon $q^-$, which we reconstruct from the invariant mass of the partonic state produced in the hard collision. 
We believe that our method for imposing these constraints can serve as a model for other high-energy collision systems. 
Curiously, while the splitting function and the collinear logarithm are already visible in terms of the projectile momentum fraction, the change to target momentum fractions is  essential to also obtain the correct, known, finite parts, the coefficient functions. In fact, this presents a test for different implementations of NLO calculations in the CGC framework: one should always ask if the method used reproduces the coefficient functions for DIS and SIDIS, when applied to an Abelian target.

The paper is organized as follows. In Sec.~\ref{sec:framework} we establish the preliminaries: we set up the kinematics and the collinear factorization of the partonic and proton structure functions, formulate LCPT with off-shell partons and define the off-shell factorization scheme, derive the photon shockwave of the electron, and compute the leading order. In Sec.~\ref{sec:NLO} we calculate the $\mathcal O(\alpha_\mathrm{s})$ partonic cross sections of the off-shell gluon and quark off the photon shockwave. These are integrated over the final state to obtain the coefficient functions of inclusive DIS in Sec.~\ref{sec:DIS} and of SIDIS in Sec.~\ref{sec:SIDIS}, where we also derive the transformation of the parton distributions and fragmentation functions between the off-shell and $\overline{\mathrm{MS}}$ schemes. We summarize in Sec.~\ref{sec:summary}. Details of the calculation are given in the appendices.

\section{Partons scattering off a photon shockwave}
\label{sec:framework}

In this section we set up the framework for scattering partons off a photon shockwave. We first   formulate in Sec.~\ref{sec:kinematics} the cross sections and their collinear factorization for a single parton, obtain from these the coefficient functions, and then apply them to a hadronic projectile described in terms of PDFS. We then describe the scattering in LCPT with off-shell partons, which defines the off-shell factorization scheme  in Sec.~\ref{sec:LCPT}, derive the photon shockwave of the electron in Sec.~\ref{sec:photon shockwave}, and compute the leading order coefficient functions for DIS and SIDIS in Sec.~\ref{sec:LO}.

\subsection{Collinear factorization of structure functions}
\label{sec:kinematics}

We describe DIS in a frame in which the electron moves with a large minus momentum $k_e^-$ and the proton, with momentum $P$, with a large plus momentum $P^+$. Rather than adopting the conventional picture in which a virtual photon is absorbed by a parton, we reinterpret the photon field of the electron as a shockwave localized at light-cone time $x^+=0$, off which the partons of the proton scatter. 

We are now working in the frame  where the incoming electron and parton have zero transverse momentum. The usual kinematics for high energy, i.e. small angle scattering then implies that  the virtuality of the photon is given by its transverse momentum.  The momentum transfer from the electron $q$ then has $Q^2\equiv-q^2=\boldsymbol{\mathrm{q}}^2$, and $y\equiv q^-/k_e^-$. In this frame we can take $q^+=0$ and the the scattering of the partons with the photon is eikonal. For a target parton with momentum $p$ and an observed final-state parton with momentum $k$, we define the partonic variables
\begin{align}
\label{eq:DIS:xhat}
    \hat{x}\equiv\frac{Q^2}{2p\cdot q}=\frac{Q^2}{2p^+q^-}\,,\qquad\hat{z}\equiv\frac{k^-}{q^-}\,,
\end{align}
where $q^-=\sum_{\rm out}k^--p^-$ is the light-cone energy transferred to the partonic final state, and $p\cdot q=p^+q^-$ since $q^+=0$ and $\boldsymbol{\mathrm{p}}=\boldsymbol 0$. In the one-photon exchange approximation and neglecting target-mass corrections, the inclusive cross section of the partonic target $i$  has the same form as for a hadronic target (see Ref.~\cite{Ellis:1996mzs}, Eqs.~(4.5) and (4.18)),
\begin{align}
\label{eq:DIS:xs}
    \frac{\mathrm{d}^2\hat\sigma^{i\gamma^*\to X}}{\mathrm{d}\hat{x}\,\mathrm{d}Q^2}=\frac{4\pi\alpha_{\mathrm{e.m.}}^2}{\hat{x}Q^4}\bigg[\bigg(1-y+\frac{y^2}{2}\bigg)F_2^i(\hat{x},Q^2)-\frac{y^2}{2}F_L^i(\hat{x},Q^2)\bigg],
\end{align}
and the semi-inclusive one, with the observed parton $j$, is (Ref.~\cite{deFlorian:1997zj}, Eq.~(31))
\begin{align}
\label{eq:SIDIS:xs}
    \frac{\mathrm{d}^3\hat\sigma^{i\gamma^*\to jX}}{\mathrm{d}\hat{x}\,\mathrm{d}y\,\mathrm{d}\hat{z}}=\frac{2\pi\alpha_{\mathrm{e.m.}}^2}{Q^2}\bigg[\frac{1+(1-y)^2}{y}\,2F_1^{i/j}(\hat{x},\hat{z},Q^2)+\frac{2(1-y)}{y}\,F_L^{i/j}(\hat{x},\hat{z},Q^2)\bigg].
\end{align}
Here $\hat\sigma^{i\gamma^*\to X}$ and $\hat\sigma^{i\gamma^*\to jX}$ are the cross sections of the parton off the photon shockwave, inclusive in the rest $X$ of the final state, and the parton $j$ has $\hat{z}=k_j^-/q^-$. The description of the electron as a recoilless classical source in Sec.~\ref{sec:photon shockwave} requires $q^-\ll k_e^-$, i.e.\ $y\ll1$. In this limit $1-y+y^2/2\to1$ and the $F_L^i$ term of Eq.\,\eqref{eq:DIS:xs} is suppressed by $y^2$, while in Eq.\,\eqref{eq:SIDIS:xs}, using $\mathrm{d}y=(y/Q^2)\,\mathrm{d}Q^2$ at fixed $\hat{x}$, both structure functions enter with the same weight, so that
\begin{align}
\label{eq:fact:F2}
    \frac{F_2^{i}(\hat{x},Q^2)}{\hat{x}}=\frac{Q^4}{4\pi\alpha_{\mathrm{e.m.}}^2}\,\frac{\mathrm{d}^2\hat\sigma^{i\gamma^*\to X}}{\mathrm{d}\hat{x}\,\mathrm{d}Q^2}\,,\qquad
    \Big[2F_1^{i/j}+F_L^{i/j}\Big](\hat{x},\hat{z},Q^2)=\frac{Q^4}{4\pi\alpha_{\mathrm{e.m.}}^2}\,\frac{\mathrm{d}^3\hat\sigma^{i\gamma^*\to jX}}{\mathrm{d}\hat{x}\,\mathrm{d}\hat{z}\,\mathrm{d}Q^2}\,.
\end{align}
This is consistent with the shockwave picture that we describe in more detail in  Sec.~\ref{sec:photon shockwave}: the potential $A^-$ couples to the current $J^+$, and for $q^+=0$ the component $W^{++}$ of the hadronic tensor is proportional to $F_2/\hat{x}=2F_1+F_L/\hat{x}$, where $F_L=F_2-2\hat{x}F_1$ as in Ref.~\cite{Ellis:1996mzs}. The SIDIS structure function $F_L^{i/j}$ of Ref.~\cite{deFlorian:1997zj} is normalized as $F_L/\hat{x}$, so that the inclusive and semi-inclusive cross sections probe the same combination.

Equation~\eqref{eq:fact:F2} is the starting point of the calculation. We compute its right-hand side, the partonic cross section, in LCPT. Its left-hand side is given by collinear factorization in terms of the PDFs and fragmentation functions (FFs) of the partonic target and the coefficient functions. Comparing the two sides order by order in $\alpha_\mathrm{s}$, the calculated partonic cross section determines the coefficient functions. Collinear factorization then states that the structure functions of the proton are obtained by replacing the  calculable  PDFs and FFs of the parton with the nonperturbative ones  of the proton and measured hadron. The  coefficient functions remain the same, since they describe the universal hard part of the scattering. 

The partonic cross sections on the r.h.s. of Eq.~\eqref{eq:fact:F2} are proportional to
\begin{align}
\label{eq:sigma0}
    \sigma_0\equiv\frac{4\pi\alpha_{\mathrm{e.m.}}^2e_f^2}{Q^4}
\end{align}
for a quark or antiquark of flavor $f$. At NLO we define the reduced partonic structure functions $\mathcal F_i$ for DIS and $\mathcal F_{ij}$ for SIDIS as the coefficients in the $\alpha_\mathrm{s}$ expansion of the cross sections,
\begin{align}
\label{eq:fact:C DIS}
    \frac{\mathrm{d}^2\hat\sigma^{q\gamma^*\to X}}{\mathrm{d}\hat{x}\,\mathrm{d}Q^2}&=\sigma_0\Big[\delta(1-\hat{x})+\frac{\alpha_\mathrm{s}}{2\pi}\,\mathcal F_q(\hat{x})\Big]\,,\qquad
    \frac{\mathrm{d}^2\hat\sigma^{g\gamma^*\to X}}{\mathrm{d}\hat{x}\,\mathrm{d}Q^2}=\sum_f\sigma_0\,\frac{\alpha_\mathrm{s}}{2\pi}\,\mathcal F_g(\hat{x})\,,\\
\label{eq:fact:C SIDIS}
    \frac{\mathrm{d}^3\hat\sigma^{q\gamma^*\to qX}}{\mathrm{d}\hat{x}\,\mathrm{d}\hat{z}\,\mathrm{d}Q^2}&=\sigma_0\Big[\delta(1-\hat{x})\,\delta(1-\hat{z})+\frac{\alpha_\mathrm{s}}{2\pi}\,\mathcal F_{qq}(\hat{x},\hat{z})\Big]\,,\nonumber\\
    \frac{\mathrm{d}^3\hat\sigma^{q\gamma^*\to gX}}{\mathrm{d}\hat{x}\,\mathrm{d}\hat{z}\,\mathrm{d}Q^2}&=\sigma_0\,\frac{\alpha_\mathrm{s}}{2\pi}\,\mathcal F_{gq}(\hat{x},\hat{z})\,,\nonumber\\
    \frac{\mathrm{d}^3\hat\sigma^{g\gamma^*\to qX}}{\mathrm{d}\hat{x}\,\mathrm{d}\hat{z}\,\mathrm{d}Q^2}&=\sigma_0\,\frac{\alpha_\mathrm{s}}{2\pi}\,\mathcal F_{qg}(\hat{x},\hat{z})\,,
\end{align}
and likewise for antiquarks.

\begin{figure}[tbp!]
\centering
\includegraphics[width=\linewidth]{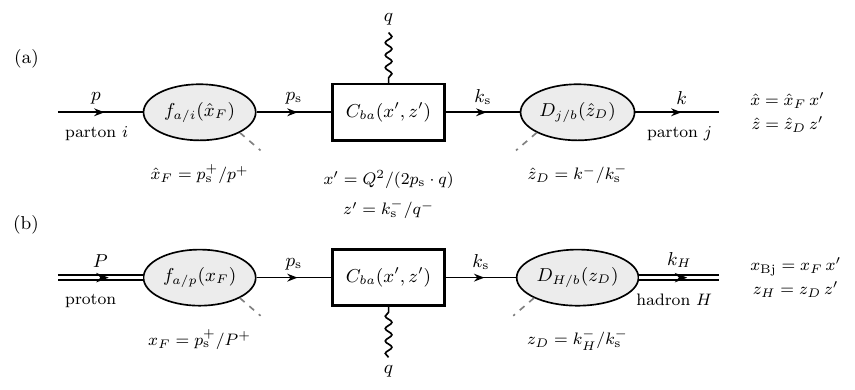}
\caption{Momenta and variables in collinear factorization for (a) a partonic target $i$ with an observed parton $j$, and (b) the proton with an observed hadron $H$. In both cases the photon scatters off a parton $a$ with momentum $p_{\rm s}$, which produces a parton $b$ with momentum $k_{\rm s}$, and the hard scattering $C_{ba}$ depends on the same variables $x'$ and $z'$. The distributions $f$ and the fragmentation functions $D$ connect it to the target and to the observed particle through the momentum fractions $\hat{x}_F,\hat{z}_D$ and $x_F,z_D$, and dashed lines indicate the unobserved rest of the final state. The variables of the target are the products of these fractions and the variables of the hard scattering, Eqs.~\eqref{eq:hard relations} and \eqref{eq:xBj relations}.}
\label{fig:factorization}
\end{figure}

Now let us  move to the l.h.s. of Eq.~\eqref{eq:fact:F2} and  see how these different photon-parton cross sections are related to PDFs, FFS and coefficient functions. In collinear factorization \cite{Collins:2011zzd, deFlorian:1997zj}, the hard scattering is separated from the collinear splittings of the incoming and outgoing partons. For the partonic target $i$ with momentum $p$ (see Fig.\,\ref{fig:factorization}a), the hard scattering of the photon happens off a parton of the target with momentum $p_{\rm s}$, and the hard scattering produces a parton with momentum $k_{\rm s}$, which gives the observed parton $j$ with momentum $k$. At leading order $p_{\rm s}=p$ and $k_{\rm s}=k$; at $\mathcal O(\alpha_\mathrm{s})$ they differ by the collinear splittings of the target and of the produced parton. The hard scattering has its own Bjorken variable $x'$  and momentum fraction of the produced parton $z'$, defined analogously as in Eq.\,\eqref{eq:DIS:xhat}. The scattering parton and the observed parton carry then the momentum fractions $\hat{x}_F$ of the target parton and $\hat{z}_D$ of the produced parton,
\begin{align}
\label{eq:hard variables}
    x'\equiv\frac{Q^2}{2p_{\rm s}\cdot q}\,,\qquad z'\equiv\frac{k_{\rm s}^-}{q^-}\,,\qquad\hat{x}_F\equiv\frac{p_{\rm s}^+}{p^+}\,,\qquad\hat{z}_D\equiv\frac{k^-}{k_{\rm s}^-}\,,
\end{align}
so that, by Eq.\,\eqref{eq:DIS:xhat},
\begin{align}
\label{eq:hard relations}
    \hat{x}=\hat{x}_F\,x'\,,\qquad\hat{z}=\hat{z}_D\,z'\,.
\end{align}
The variables for the partonic scattering are illustrated in Fig. \ref{fig:factorization}a). In terms of the LCPT terminology, the target and produced partons $i$ and $j$ are dressed partons, eigenstates of the interacting theory that are obtained by time evolution from single particle asymptotic states. The partons with momenta $p_s$ and $k_s$ that participate in the hard collision, on the other hand, are bare partons that are the eigenstates of the free theory. Perturbatively calculable light cone wave functions (LCWFs) relate the two sets of states to each other.

The structure functions of the partonic target are convolutions of the parton distributions $f_{a/i}(\hat{x}_F,\mu_F^2)$ of the partons $a$ in the target $i$ and the fragmentation functions $D_{j/b}(\hat{z}_D,\mu_D^2)$ of the partons $b$ into $j$  with coefficient functions of the hard scattering, which depend on $x'$ and $z'$. The relation between the bare and dressed parton number densities is UV divergent and must be renormalized, and as a result the PDFs and FFs depend on factorization scales $\mu_F$ and $\mu_D$, as discussed in more detail in Appendices~\ref{sec:app:PDF} and~\ref{sec:app:FF}.
At $\mathcal O(\alpha_\mathrm{s})$, the collinear factorization structure of the structure functions illustrated in Fig. \ref{fig:factorization} can be written into the form
\begin{align}
\label{eq:fact:general}
    \frac{F_2^{i}(\hat{x},Q^2)}{\hat{x}}&=\sum_{q,\bar q}e_q^2\bigg\{f_{q/i}(\hat{x},\mu_F^2)\nonumber\\
    &\quad+\frac{\alpha_\mathrm{s}}{2\pi}\Big[f_{q/i}\otimes C_q+\frac12\,f_{g/i}\otimes C_g\Big](\hat{x})\bigg\},\nonumber\\
    \Big[2F_1^{i/j}+F_L^{i/j}\Big](\hat{x},\hat{z},Q^2)&=\sum_{q,\bar q}e_q^2\bigg\{f_{q/i}(\hat{x},\mu_F^2)\,D_{j/q}(\hat{z},\mu_D^2)\nonumber\\
    &\quad+\frac{\alpha_\mathrm{s}}{2\pi}\Big[f_{q/i}\otimes C_{qq}\otimes D_{j/q}\nonumber\\
    &\quad+f_{q/i}\otimes C_{gq}\otimes D_{j/g}+f_{g/i}\otimes C_{qg}\otimes D_{j/q}\Big](\hat{x},\hat{z})\bigg\},
\end{align}
where the sums run over quarks and antiquarks, and the convolutions are
\begin{align}
\label{eq:DIS:convolution}
    \big[f\otimes C\big](\hat{x})&\equiv\int_{\hat{x}}^1\frac{\mathrm{d}x'}{x'}\,f\Big(\frac{\hat{x}}{x'},\mu_F^2\Big)\,C(x')\,,\\
\label{eq:SIDIS:convolution}
    \big[f\otimes C\otimes D\big](\hat{x},\hat{z})&\equiv\int_{\hat{x}}^1\frac{\mathrm{d}x'}{x'}\,f\Big(\frac{\hat{x}}{x'},\mu_F^2\Big)\int_{\hat{z}}^1\frac{\mathrm{d}z'}{z'}\,C(x',z')\,D\Big(\frac{\hat{z}}{z'},\mu_D^2\Big)\,,
\end{align}
with the arguments $\hat{x}/x'=\hat{x}_F$ and $\hat{z}/z'=\hat{z}_D$ of $f$ and $D$ by Eq.\,\eqref{eq:hard relations}. Here we have already used the explicit forms of the leading order coefficient functions 
$C_q^\text{LO}(x')=\delta(1-x')$, $C_{qq}^\text{LO}(x',z')=\delta(1-x')\delta(1-z')$ and $C_g^\text{LO}=C_{gq}^\text{LO}=C_{qg}^\text{LO}=0$.
Equation~\eqref{eq:fact:general} defines the NLO coefficient functions $C_{ba}(x',z';\mu_F,\mu_D)$ for producing the parton $b$ from an incoming parton $a$ (Fig.\,\ref{fig:factorization}), with $C_{ba}\equiv C^1_{ba}+C^L_{ba}$ in the notation of Ref.~\cite{deFlorian:1997zj}. The inclusive gluon coefficient function $C_g$  contains the contributions of a  quark and antiquark of one flavor. For a  more intuitive view it can be instructive to mentally transpose the indices in $f_{j/i}, C_{ji}, D_{j/i}$: the convolution 
$f_{j/i}\otimes C_{kj}\otimes D_{\ell/k}$ should be read as ``from incoming $i$ take a parton $j$, which transforms into $k$ in the  hard scattering and then fragments into a final state particle $\ell$.'' In this chain, there is no dependence on the target $i$ or the observed particle $\ell$ in the coefficient function. 
The coefficient functions depend on the scales only through $Q^2/\mu_F^2$ and $Q^2/\mu_D^2$; we suppress the scale arguments in the following. Associated with this dependence on the factorization scale, they also depend on the scheme.  The general factorization structure Eq.\,\eqref{eq:fact:general}, however, holds in any scheme, as long as  $f$, $D$ and $C$ are consistently evaluated in the same scheme. 

For the partonic target the distributions and fragmentation functions are perturbative, so at leading order we always have $f_{a/i}(x)=\delta_{ai}\delta(1-x)$ and $D_{j/b}(z)=\delta_{jb}\delta(1-z)$. At $\mathcal O(\alpha_\mathrm{s})$ the NLO coefficient functions therefore multiply only the leading order of the ``parton-in-parton'' PDFs and ``parton-to-parton'' FFs, with $\hat{x}_F=\hat{z}_D=1$. Thus, for a partonic target and at this order in perturbation theory, we can reduce the convolutions and write the structure function in terms of the coefficient function evaluated at the Bjorken variable and fragmentation momentum fraction  with $x'=\hat{x}$ and $z'=\hat{z}$:
\begin{align}
\label{eq:fact:parton}
    \frac{F_2^{q}(\hat{x},Q^2)}{\hat{x}}&=e_q^2\Big[f_{q/q}(\hat{x},\mu_F^2)+\frac{\alpha_\mathrm{s}}{2\pi}\,C_q(\hat{x})\Big]\,,\nonumber\\
    \frac{F_2^{g}(\hat{x},Q^2)}{\hat{x}}&=\sum_fe_f^2\Big[2f_{q/g}(\hat{x},\mu_F^2)+\frac{\alpha_\mathrm{s}}{2\pi}\,C_g(\hat{x})\Big]\,,\nonumber\\
    \Big[2F_1^{q/q}+F_L^{q/q}\Big](\hat{x},\hat{z},Q^2)&=e_q^2\Big[f_{q/q}(\hat{x},\mu_F^2)\,D_{q/q}(\hat{z},\mu_D^2)+\frac{\alpha_\mathrm{s}}{2\pi}\,C_{qq}(\hat{x},\hat{z})\Big]\,,\nonumber\\
    \Big[2F_1^{q/g}+F_L^{q/g}\Big](\hat{x},\hat{z},Q^2)&=e_q^2\Big[\delta(1-\hat{x})\,D_{g/q}(\hat{z},\mu_D^2)+\frac{\alpha_\mathrm{s}}{2\pi}\,C_{gq}(\hat{x},\hat{z})\Big]\,,\nonumber\\
    \Big[2F_1^{g/q}+F_L^{g/q}\Big](\hat{x},\hat{z},Q^2)&=e_q^2\Big[f_{q/g}(\hat{x},\mu_F^2)\,\delta(1-\hat{z})+\frac{\alpha_\mathrm{s}}{2\pi}\,C_{qg}(\hat{x},\hat{z})\Big]\,,
\end{align}
where in the gluon case the factor 2 counts the quark and the antiquark. Note that the SIDIS structure functions carry the labels of the target $i$ and the observed particle $j$ in this order: $F^{i/j}$. For the coefficient functions $C_{ba}$, on the other hand, the first index is the produced bare parton $b$ and the second one the incoming one $a$.  Thus,  e.g., $F^{g/q}$ contains $C_{qg}$. 

Now we are in a position to match the two sides of Eq.~\eqref{eq:fact:F2} to identify  the relation between the photon cross sections on the r.h.s. with the collinear factorization expressions for the l.h.s.
Comparing Eq.~\eqref{eq:fact:parton} with Eqs.~\eqref{eq:fact:F2}--\eqref{eq:fact:C SIDIS} we see that
\begin{align}
\label{eq:F from C}
    \mathcal F_q&=C_q+\frac{2\pi}{\alpha_\mathrm{s}}\Big[f_{q/q}(\hat{x})-\delta(1-\hat{x})\Big]\,,\nonumber\\
    \mathcal F_g&=C_g+\frac{2\pi}{\alpha_\mathrm{s}}\,2f_{q/g}(\hat{x})\,,\nonumber\\
    \mathcal F_{qq}&=C_{qq}+\frac{2\pi}{\alpha_\mathrm{s}}\Big\{\delta(1-\hat{z})\Big[f_{q/q}(\hat{x})-\delta(1-\hat{x})\Big]+\delta(1-\hat{x})\Big[D_{q/q}(\hat{z})-\delta(1-\hat{z})\Big]\Big\}\,,\nonumber\\
    \mathcal F_{gq}&=C_{gq}+\frac{2\pi}{\alpha_\mathrm{s}}\,\delta(1-\hat{x})\,D_{g/q}(\hat{z})\,,\nonumber\\
    \mathcal F_{qg}&=C_{qg}+\frac{2\pi}{\alpha_\mathrm{s}}\,\delta(1-\hat{z})\,f_{q/g}(\hat{x})\,,
\end{align}
up to $\mathcal O(\alpha_\mathrm{s})$, with the distributions at the factorization scale $\mu_F$ and the fragmentation functions at $\mu_D$. Like Eq.\,\eqref{eq:fact:general}, both sides of this equation are separately independent of the factorization scheme. 

For an on-shell parton the $\mathcal O(\alpha_\mathrm{s})$ terms of the cross sections $\mathcal F$  are collinear divergent, which is matched by a collinear divergence in $f_{a/i}$ and $D_{j/b}$  on the r.h.s. In Sec.~\ref{sec:LCPT} we take the partonic target and the observed parton off shell, which makes the cross section $\mathcal F$ finite. Then it is natural to use an off-shell scheme for renormalization, where, at the renormalization point where the factorization scale equals the off-shellness, $f_{a/i}(x)=\delta_{ai}\delta(1-x)$ and $D_{j/b}(z)=\delta_{jb}\delta(1-z)$  at all orders, rather than just at leading order. The scheme independence of the l.h.s. of Eq.~\eqref{eq:F from C} shows how a different UV renormalization scheme for $f_{a/i}$ and $D_{j/b}$, such as $\overline{\mathrm{MS}}$, must lead to different coefficient functions $C$.  We will in this paper compute the coefficient functions in the off-shell scheme of Sec.~\ref{sec:LCPT} and relate them to the $\overline{\mathrm{MS}}$ ones. We also explicitly check the resulting scheme transformation by a UV renormalization of the off-shell parton $f_{a/i}$ and $D_{j/b}$ in the $\overline{\mathrm{MS}}$ scheme.

With a general enough structure  it is now straightforward to move from a partonic target to  the proton (or any hadron). The key aspect is that the same coefficient functions that we calculate for a partonic target also enter the structure functions of the proton, $F_2^p$ and $2F_1^{p/H}+F_L^{p/H}$ for the observed hadron $H$.
The proton with momentum $P$ and the observed hadron $H$ with momentum $k_H$ (Fig.\,\ref{fig:factorization}b) take the place of the partonic target and the observed parton, while the hard scattering, with the momenta $p_{\rm s}$ and $k_{\rm s}$ and the variables $x'$ and $z'$ of Eq.\,\eqref{eq:hard variables}, is the same. The hadronic variables and the momentum fractions $x_F$ of the scattering parton in the proton and $z_D$ of the hadron in the produced parton are
\begin{align}
\label{eq:xBj}
    x_\text{Bj}\equiv\frac{Q^2}{2P\cdot q}\,,\qquad z_H\equiv\frac{k_H^-}{q^-}\,,\qquad x_F\equiv\frac{p_{\rm s}^+}{P^+}\,,\qquad z_D\equiv\frac{k_H^-}{k_{\rm s}^-}\,,
\end{align}
so that, by Eq.\,\eqref{eq:hard variables},
\begin{align}
\label{eq:xBj relations}
    x_\text{Bj}=x_F\,x'\,,\qquad z_H=z_D\,z'\,.
\end{align}
At $\mathcal O(\alpha_\mathrm{s})$ the structure functions are expressed in terms of the same coefficient functions,
\begin{align}
\label{eq:DIS:factorization}
    \frac{F_2^p(x_\text{Bj},Q^2)}{x_\text{Bj}}&=\sum_{q,\bar q}e_q^2\bigg\{f_{q/p}(x_\text{Bj},\mu_F^2)\nonumber\\
    &\quad+\frac{\alpha_\mathrm{s}}{2\pi}\Big[f_{q/p}\otimes C_q+\frac12\,f_{g/p}\otimes C_g\Big](x_\text{Bj})\bigg\},\\
    \Big[2F_1^{p/H}+F_L^{p/H}\Big](x_\text{Bj},z_H,Q^2)&=\sum_{q,\bar q}e_q^2\bigg\{f_{q/p}(x_\text{Bj},\mu_F^2)\,D_{H/q}(z_H,\mu_D^2)\nonumber\\
    &\quad+\frac{\alpha_\mathrm{s}}{2\pi}\Big[f_{q/p}\otimes C_{qq}\otimes D_{H/q}\nonumber\\
    &\quad+f_{q/p}\otimes C_{gq}\otimes D_{H/g}+f_{g/p}\otimes C_{qg}\otimes D_{H/q}\Big](x_\text{Bj},z_H)\bigg\},\label{eq:SIDIS:factorization}
\end{align}
where the convolutions are Eqs.~\eqref{eq:DIS:convolution} and \eqref{eq:SIDIS:convolution} with $\hat{x}\to x_\text{Bj}$ and $\hat{z}\to z_H$, again over the variables $x'$ and $z'$ of the hard scattering. They are related to the   arguments $x_\text{Bj}/x'=x_F$ and $z_H/z'=z_D$ of $f$ and $D$ by Eq.\,\eqref{eq:xBj relations}. Here the parton distributions $f_{a/p}(x_F,\mu_F^2)$ of the proton and the fragmentation functions $D_{H/b}(z_D,\mu_D^2)$ of the parton $b$ into the hadron $H$ are nonperturbative.

\subsection{Light-cone perturbation theory with off-shell partons}
\label{sec:LCPT}

We describe the scattering of a parton off the external field shockwave in LCPT, following the formulation originally developed in Ref.~\cite{Bjorken:1970ah}. The interaction with the field takes place at $x^+=0$, and the asymptotic states are dressed states evaluated  at this light-cone time, obtained from the free states with the interaction-picture time evolution operator,
\begin{align}
\label{eq: LCPT: dressed states}
    |\phi\rangle_\mathrm{D}^{\mathrm{in}}=U_\mathrm{I}(0,-\infty)|\phi\rangle\,,\qquad|\phi\rangle_\mathrm{D}^{\mathrm{out}}=U_\mathrm{I}^\dagger(\infty,0)|\phi\rangle\,.
\end{align}
Our single-particle states, both bosons and fermions,  are normalized as
\begin{align}
    \big\langle a(\vec{p},\lambda_p,i_p) \big| b(\vec{q},\lambda_q,i_q ) \big\rangle &= 2q^+(2\pi)^3 \delta^{(3)}(\vec{p}-\vec{q}\,)\delta_{i_p,i_q} \delta_{\lambda_p,\lambda_q} \,,\nonumber\\
    \delta^{(3)}(\vec{p}-\vec{q}\,)&=\delta({p}^+-{q}^+)\delta^{(2)}(\boldsymbol{\mathrm{p}}-\boldsymbol{\mathrm{q}})\,,\label{eq: LCPT: orthonormalization states}
\end{align}
where $\lambda$ is the light-front helicity, $i$ the color index, and boldface symbols denote transverse vectors. For a light-cone Hamiltonian $\hat H=\hat H_0+\hat V$, the dressed states are given by the perturbative series \cite{Brodsky:1997de, Li:2022tmi}
\begin{align}
    \label{eq: LCPT: Definition Fock state}
    |a\rangle_\mathrm{D}^{\mathrm{in,out}} =\Bigg[\prod_{i=1}^{N_a}\sqrt{Z_{i}}\Bigg]\Bigg[  |a\rangle + {\sum_{n}}'|n\rangle \frac{\langle n | \hat{V}| a\rangle}{E_a-E_n \pm i\delta}+\dots\Bigg] ,
\end{align}
where $E$ are the eigenvalues of $\hat H_0$. The prime excludes intermediate states in which every particle has the same momentum as in $|a\rangle$; these contributions are included in the single-particle normalization factors $\sqrt{Z_i}$, which are fixed by the normalization of the dressed states and can be chosen to be real. The sign of $i\delta$ is different for the in and out states, but is irrelevant at the order considered in this paper. The phase-space sum over an intermediate state $n$ with $N_n$ particles is
\begin{align}
    {\sum_{n}}'&\coloneqq\sum_{\substack{\mathrm{helicity}\\\mathrm{colour}}}\prod_{i=1}^{N_n} \int \frac{\mathrm{d}k_i^+}{2\pi}\frac{\theta(k_i^+)}{2k_i^+}\,2p^+(2\pi)\delta\Big(p^+-\textstyle\sum_{i}k_i^+\Big)\nonumber\\
    &\quad\times\displaystyle\prod_{l=1}^{N_n}\int \frac{\mathrm{d}^{D-2}\boldsymbol{\mathrm{k}}_l}{(2\pi)^{D-2}}\,(2\pi)^{D-2}\delta^{(D-2)}\Big(\boldsymbol{\mathrm{p}}-\textstyle\sum_l\boldsymbol{\mathrm{k}}_l\Big) \,,\label{eq: LCPT: Definition phase space}
\end{align}
where $(p^+,\boldsymbol{\mathrm{p}})$ is the total momentum of the parent state. The coefficients of the expansion are known as the light-cone wave functions; at first order
\begin{align}
    \label{eq: LCPT: Definition LCWF}
 2k^+_a (2\pi)^3 \delta^{(3)}(\vec{k}_a-\vec{k}_n\,)   \Psi^{a\to n}=\frac{\langle n | \hat{V}| a\rangle}{\mathrm{ED}_{a\to n}}\,,\qquad\mathrm{ED}_{a\to n}=E_a-E_n\,.
\end{align}
Here it is convenient to factor out the overall momentum delta function, which was already included in the phase space measure~\eqref{eq: LCPT: Definition phase space}.\footnote{Since $\hat{V}$ is a light cone energy operator, i.e. a minus component of a Lorentz 4-vector, its matrix elements in the states normalized as in Eq.~\eqref{eq: LCPT: orthonormalization states} turn out to be proportional to 
$(2\pi)^3 \delta^{(3)}(\vec{k}_a-\vec{k}_n\,)$. Thus factoring out the boost invariant delta function that includes $2k_a^+$ leaves out the factor $2k_a^+$, which naturally combines with the energy denominator leading to the combination $2k_a^+\mathrm{ED}_{a\to n}$ that has the dimensions of transverse momentum squared.
}
For a multiparton state $|n\rangle$, the same expansion contains the components with fewer partons, i.e.\ the merging wave functions $2k^+_a (2\pi)^3 \delta^{(3)}(\vec{k}_a-\vec{k}_n\,) \Psi^{n\to a}=\langle a|\hat V|n\rangle/(\mathrm{ED}_{n\to a})$. Since $\hat V$ is Hermitian and $\mathrm{ED}_{n\to a}=E_n-E_a=-\mathrm{ED}_{a\to n}$,
\begin{align}
\label{eq:LCPT:merging}
    \Psi^{n\to a}=-\big(\Psi^{a\to n}\big)^\dagger\,,
\end{align}
which is also the condition for the dressed states to remain orthogonal, ${}_\mathrm{D}\langle a|n\rangle_\mathrm{D}=\Psi^{n\to a}+(\Psi^{a\to n})^\dagger=0$ at first order.

\subsection{Off-shell regularization}
\label{sec:offshell}

The off-shell regularization of collinear divergences is formulated in terms of these energy denominators. Consider a parent parton with momentum $p=(p^+,p^-,\boldsymbol{0})$ splitting into on-shell daughters $b$ and $c$ with plus-momentum fractions $z$ and $1-z$ and relative transverse momentum $\boldsymbol{\ell}$, i.e.\ $\boldsymbol{\mathrm{k}}_b=-\boldsymbol{\mathrm{k}}_c=\boldsymbol{\ell}$. Then
\begin{align}
\label{eq:ED general}
    \mathrm{ED}_{a\to bc}=p^--k_b^--k_c^-=p^--\frac{\boldsymbol{\ell}^2}{2z(1-z)p^+}\,.
\end{align}
For an on-shell parent, $p^-=0$, the energy denominator vanishes in the collinear limit $\boldsymbol{\ell}\to0$, and this is the origin of the collinear divergences of the cross section. We regulate it by taking the incoming parton off shell with a spacelike virtuality $\mu_{\mathrm{in}}^2$,
\begin{align} \label{eq: off-shell regulator}
    p=\bigg(p^+,\,p^-=-\frac{\mu_{\mathrm{in}}^2}{2p^+},\,\boldsymbol{0}\bigg),\qquad p^2=-\mu_{\mathrm{in}}^2\,,
\end{align}
with $p=p_q$ for an incoming quark and $p=p_g$ for an incoming gluon. The energy denominator of the initial-state splitting then becomes
\begin{align}
\label{eq:ED regulated}
    \mathrm{ED}_{a\to bc}=-\frac{1}{2p^+}\,\frac{\boldsymbol{\ell}^2+z(1-z)\mu_{\mathrm{in}}^2}{z(1-z)}\,,
\end{align}
which is bounded away from zero. In terms of the invariant mass $M^2=\boldsymbol{\ell}^2/[z(1-z)]\geq0$ of the pair, Eq.\,\eqref{eq:ED general} reads $\mathrm{ED}_{a\to bc}=(p^2-M^2)/(2p^+)$. The spacelike parent lies below the two-parton threshold $M^2=0$, so that it cannot decay into its Fock components, and the energy denominator has a definite sign. A timelike virtuality, $p^2=+\mu_{\mathrm{in}}^2$, would instead put the parent above the threshold, and the energy denominator would vanish at $M^2=\mu_{\mathrm{in}}^2$. This pole corresponds to the energy conservation delta function in the decay of a massive parent state into  an on-shell pair (see e.g. the discussion in Ref.~\cite{Beuf:2021srj}. In the light-cone wave function the virtuality is only introduced in the calculation through the light cone energy, and thus only affects the energy denominator $\boldsymbol{\ell}^2\to\boldsymbol{\ell}^2+z(1-z)\mu_{\mathrm{in}}^2$. The vertices are all expressed in terms of three-momentum and are not affected by the virtuality. This is one of the attractive features of the LCPT approach of Hamiltonian perturbation theory in a fixed gauge, explicitly in terms of physical polarization states.

The spacelike virtuality of the single particle state regulates initial state collinear divergences, corresponding to the DGLAP evolution of the PDF. Final state collinear divergences are regulated by putting the outgoing one or two parton state off shell. Here we must note that in LCPT the off-shellness is really a property of the whole multiparticle state, not something that can be attributed to a single parton. It is natural to take a single particle final state off shell with a spacelike virtuality $\mu_{\mathrm{out}}^2$. The natural choice for a two-particle final state is to take it as the  dressed two-particle state that is orthogonal to the spacelike off-shell single particle one, i.e. keeping the orthogonality condition~\eqref{eq:LCPT:merging} satisfied. In terms of the light cone energy this means that we add a positive off-shellness to the LC energy of the ougoing two-particle state (e.g. in diagram~\ref{diag:real_external} in Fig.~\ref{fig:cross_section}). Since for the outgoing two-particle state the wavefunction is in fact a merging, rather than splitting, one, the LC energy denominator (the sign is $\mathrm{ED}_{a\to n}=E_a-E_n$) is now positive, and the positive LC energy added to the dressed state makes it positive definite. To conclude, the natural way to remove final state divergence and only have energy denominators with a definite sign is to have outgoing two-particle states being time-like off shell with $p_{a+b}^2 = \mu^2_{\mathrm{out}}>0$ and outgoing one-particle states spacelike off-shell $p^2 = - \mu^2_{\mathrm{out}} < 0$; these are actually the dressed states that are orthogonal to each other. 

Since a regulator only shifts $\boldsymbol{\ell}^2$ in one energy denominator, it matters only in the collinear region $\boldsymbol{\ell}^2\lesssim z(1-z)\mu_{\mathrm{in}}^2$ or $z(1-z)\mu_{\mathrm{out}}^2$ of the corresponding wave function. Elsewhere it is a power correction in $\mu_{\mathrm{in}}^2/Q^2$ or $\mu_{\mathrm{out}}^2/Q^2$ and we drop it (see Appendix~\ref{sec:app:DIS}). The collinear regions are identified by the transverse momenta of the final state.  Let us first fix the kinematics of the final state particles, so that at $\mathcal O(\alpha_\mathrm{s})$ the pair consists of a parton with plus-momentum fraction $z$ and transverse momentum $\boldsymbol{\mathrm{k}}$ and a parton with $1-z$ and $\boldsymbol{\mathrm{q}}-\boldsymbol{\mathrm{k}}$. We also choose so that if the splitting happens before the absorption of the photon, it is the parton with final momentum $\boldsymbol{\mathrm{k}}$ that absorbed the photon. 
\begin{itemize}
\item 
In an initial state splitting the parent has transverse momentum $\boldsymbol 0$ and the pair $\pm\boldsymbol{\ell}$. The spectator keeps $-\boldsymbol{\ell}=\boldsymbol{\mathrm{q}}-\boldsymbol{\mathrm{k}}$ and the absorption of the photon turns the momentum of the absorber from $ \boldsymbol{\ell}=\boldsymbol{\mathrm{k}}-\boldsymbol{\mathrm{q}}$ to $\boldsymbol{\mathrm{k}}$. The limit $\boldsymbol{\ell}\to \boldsymbol 0$ is the initial state collinear region.
\item In a final state splitting one first absorbs the total transverse momentum $\boldsymbol{\mathrm{q}}$. The splitting then happens with a  relative transverse momentum $\boldsymbol{\mathrm{m}}=\boldsymbol{\mathrm{k}}-z\boldsymbol{\mathrm{q}}$. The momentum $\boldsymbol{\mathrm{m}}$ is the one in the energy denominator of the final state merging wavefunction, and $\boldsymbol{\mathrm{m}}\to \boldsymbol 0$ is the final state collinear region.
\end{itemize}
To summarize, the initial and final state splitting wavefunctions depend on the momenta
\begin{align}
\label{eq:l m}
    \boldsymbol{\ell}=\boldsymbol{\mathrm{k}}-\boldsymbol{\mathrm{q}}\,,\qquad\boldsymbol{\mathrm{m}}=(1-z)\,\boldsymbol{\mathrm{k}}-z\,(\boldsymbol{\mathrm{q}}-\boldsymbol{\mathrm{k}})=\boldsymbol{\mathrm{k}}-z\boldsymbol{\mathrm{q}}\,.
\end{align}

Now let us see what these regions imply in terms of the the partonic Bjorken variable $\hat{x}$ and momentum fraction $\hat{z}$. With $q^-=\sum_{\rm out}\boldsymbol{\mathrm{k}}^2/(2k^+)$, up to the power correction from $p^-$, the light-cone energy transferred to the final state is
\begin{align}
\label{eq:collinear qminus}
    q^-=\frac{1}{2p^+}\bigg[\frac{\boldsymbol{\mathrm{k}}^2}{z}+\frac{(\boldsymbol{\mathrm{q}}-\boldsymbol{\mathrm{k}})^2}{1-z}\bigg]=\frac{Q^2+M^2}{2p^+}\,,\qquad\hat{x}=\frac{Q^2}{Q^2+M^2}\,,\qquad M^2=\frac{\boldsymbol{\mathrm{m}}^2}{z(1-z)}\,.
\end{align}
We see that the Bjorken variable $\hat{x}<1$ measures the invariant mass of the final pair. 
\begin{itemize}
\item The initial splitting collinear limit  $\boldsymbol{\ell}\to0$ has  $\boldsymbol{\mathrm{k}}=\boldsymbol{\mathrm{q}}$ and  $M^2=(1-z)Q^2/z,$ i.e. $z=Q^2/(Q^2+M^2)$.
 The parton with fraction $z$ carries $k^-=\boldsymbol{\mathrm{k}}^2/(2zp^+)$, which in this limit is $k^-=q^-$.
\item  The final state collinear region $\boldsymbol{\mathrm{m}}\to0$, on the other hand, has $\boldsymbol{\mathrm{k}}=z\boldsymbol{\mathrm{q}}$ and  $M^2=0$.
 Now struck parton energy   $k^-=\boldsymbol{\mathrm{k}}^2/(2zp^+)$ is
 $k^-=z\,q^-$. 
\end{itemize}
Thus in these two collinear regions expressed in terms of  the variables defined in  Eq.\,\eqref{eq:DIS:xhat} are
\begin{align}
\label{eq:collinear}
    \text{initial-state splitting, }\boldsymbol{\ell}\to0\ (\mu_{\mathrm{in}}):&\quad\hat{x}=z\,,\quad\hat{z}=1\,,\nonumber\\
    \text{final-state splitting, }\boldsymbol{\mathrm{m}}\to0\ (\mu_{\mathrm{out}}):&\quad\hat{x}=1\,,\quad\hat{z}=z\,,
\end{align}
with $\hat{z}$ that of the parton with fraction $z$. At $\mathcal O(\alpha_\mathrm{s})$ the gluon channel has only the initial-state splitting $g\to q\bar q$, whereas the quark channel has both.

Now that we know how to deal with off-shell states and collinear regions, let us connect this discussion to the building blocks of collinear factorization discussed in Sec.~\ref{sec:kinematics}. We recall that what is needed is a well-defined partonic target, for which we can calculate the cross section for scattering off a virtual photon, the l.h.s.  of
Eqs.~\eqref{eq:fact:parton}--\eqref{eq:F from C}. We take this target to be the off-shell parton $i^*(\mu_{\mathrm{in}})$. We denote by $f_{a/i^*(\mu_{\mathrm{in}})}(x,\mu_F^2)$ the distribution of the parton $a$ at the factorization scale $\mu_F$ in the target $i^*$ of virtuality $\mu_{\mathrm{in}}^2$, and by $D_{j^*(\mu_{\mathrm{out}})/b}(z,\mu_D^2)$ the fragmentation function of the parton $b$ into the observed parton $j^*$, whose outgoing dressed state has the virtuality $\mu_{\mathrm{out}}^2$. Here the dressed off-shell parton $i^*$ takes the place of the proton, $j^*$ takes the place of the observed hadron, while the partons $a$ and $b$ of the hard scattering are bare. We also recall that the quantities on the r.h.s. of Eqs.~\eqref{eq:fact:parton}--\eqref{eq:F from C}, the coefficient functions, PDFs and FFs are scheme dependent. We now fix this scheme dependence  by defining the off-shell factorization scheme with the  renormalization condition that the PDFs and FFs are trivial when the scale equals the virtuality \cite{Lappi:2026xaa},
\begin{align}
\label{eq:rencond}
    f^{\rm off}_{a/i^*(\mu_{\mathrm{in}})}(x,\mu_F^2)\Big|_{\mu_F=\mu_{\mathrm{in}}}=\delta_{ai}\,\delta(1-x)\,,\qquad
    D^{\rm off}_{j^*(\mu_{\mathrm{out}})/b}(z,\mu_D^2)\Big|_{\mu_D=\mu_{\mathrm{out}}}=\delta_{jb}\,\delta(1-z)\,.
\end{align}
At other scales they follow from DGLAP evolution with Eq.\,\eqref{eq:rencond} as the initial condition. At $\mathcal O(\alpha_\mathrm{s})$ the evolution is driven by the leading-order splitting functions acting on the trivial distributions, so that
\begin{align}
\label{eq:offPDF}
    f^{\rm off}_{a/i^*(\mu_{\mathrm{in}})}(x,\mu_F^2)&=\delta_{ai}\,\delta(1-x)+\frac{\alpha_\mathrm{s}}{2\pi}\,\mathcal P_{a\leftarrow i}(x)\ln\frac{\mu_F^2}{\mu_{\mathrm{in}}^2}\,,\nonumber\\
    D^{\rm off}_{j^*(\mu_{\mathrm{out}})/b}(z,\mu_D^2)&=\delta_{jb}\,\delta(1-z)+\frac{\alpha_\mathrm{s}}{2\pi}\,\mathcal P_{j\leftarrow b}(z)\ln\frac{\mu_D^2}{\mu_{\mathrm{out}}^2}\,,
\end{align}
up to $\mathcal O(\alpha_\mathrm{s}^2)$, where $\mathcal P_{a\leftarrow i}$ are the leading-order DGLAP splitting functions, e.g.\ $\mathcal P_{q\leftarrow q}(x)=C_\mathrm{F}\big[(1+x^2)/(1-x)_++\frac32\delta(1-x)\big]$. We denote by $P_{a\leftarrow i}$ the real-emission splitting functions, without plus prescriptions and $\delta$ terms; they coincide with $\mathcal P_{a\leftarrow i}$ for $a\neq i$, and $\mathcal P_{q\leftarrow q}=[P_{q\leftarrow q}]_+$. Inserted into Eq.\,\eqref{eq:F from C}, the distributions \eqref{eq:offPDF} only add collinear logarithms between the virtualities and the factorization scales, which vanish when the scales equal, so that
\begin{align}
\label{eq:C from F}
    C^{\rm off}(\hat{x},\hat{z};\mu_F,\mu_D)=\mathcal F(\hat{x},\hat{z};\mu_{\mathrm{in}},\mu_{\mathrm{out}})\Big|_{\mu_{\mathrm{in}}\to\mu_F,\ \mu_{\mathrm{out}}\to\mu_D}\,,
\end{align}
up to power corrections in $\mu_{\mathrm{in}}^2/Q^2$ and $\mu_{\mathrm{out}}^2/Q^2$. the same happens in inclusive DIS, where only $\mu_{\mathrm{in}}$ and $\mu_F$ appear. In particular, at $\mu_F=\mu_{\mathrm{in}}$ and $\mu_D=\mu_{\mathrm{out}}$ the partonic cross sections of the off-shell partons are directly the coefficient functions, and collinear factorization follows without additional subtractions. 

The off-shell distributions \eqref{eq:offPDF} and the $\overline{\mathrm{MS}}$ distributions of the same off-shell partons differ by target-independent finite terms (Secs.~\ref{sec:DIS} and \ref{sec:SIDIS}), so that Eq.\,\eqref{eq:rencond} defines a factorization scheme, related to $\overline{\mathrm{MS}}$ by a finite renormalization. The regulator is not \textit{ad hoc}: it can be interpreted as the binding energy of a parton bound inside a hadron. In general dealing with off-shell states in higher orders in perturbation theory can be calculationally more complicated than on-shell massless ones. However, it is routine in the context of small-$x$ DIS  in the dipole picture, where the physical situation requires dealing with the incoming photon that is spacelike off shell. Since off-shell matrix elements are gauge dependent, the light-cone gauge $A^+=0$, the restriction to transverse gluon polarizations and the average over the $D-2$ transverse polarization states are regarded as part of the definition of the scheme.

Finally, to actually calculate  cross sections in the shockwave formalism, we  start from the definition of a scattering amplitude as  the matrix element of the scattering operator $\hat S$ between the dressed in- and out-states,
\begin{align}
    \label{eq: LCPT: Definition scattering amplitude}
    \prescript{\mathrm{out}}{\mathrm{D}}{\big\langle} b \,\big| \hat S - 1 \big| \, a \big\rangle_\mathrm{D}^{\mathrm{in}} =2p^+(2\pi)\delta\Big(p^+-\textstyle\sum\limits_{k\in b}k^+\Big)\,i \mathcal{M}^{a\rightarrow b} ,
\end{align}
where $p^+$ is the plus momentum of the incoming parton, and the cross section is $\mathrm{d}\hat\sigma^{a\to b}=|\mathcal M^{a\to b}|^2\,\mathrm{d}(\mathrm{P.S.})_b$ with
\begin{align}
\label{eq: LCPT: PS}
    \mathrm{d}(\textrm{P.S.})_b= 2p^+ (2\pi)\delta\Big(p^+-\textstyle\sum\limits_{k\in b}k^+\Big)\displaystyle\prod_{l\in b} \frac{\mathrm{d}k_l^+ \mathrm{d}^{D-2}\boldsymbol{\mathrm{k}}_l}{2k_l^+(2\pi)^{D-1}} \,.
\end{align}
For a shockwave localized at $x^+=0$, $\hat S$ is diagonal in transverse coordinates. The amplitude is then obtained by Fourier transforming the dressed states to mixed space $(k^+,\boldsymbol{\mathrm{x}})$ and multiplying each parton by the eikonal factor of the field at its transverse position, which for the photon shockwave of the electron is derived in the next subsection. Spinor matrix elements for light-front helicity states in light-cone gauge can be found, e.g., in Refs.~\cite{Lappi:2016oup, Beuf:2022ndu}; we use the decomposition of Ref.~\cite{Beuf:2021qqa}, Appendix~A, which holds in arbitrary $D$.

\subsection{Photonic shockwave of the electron}
\label{sec:photon shockwave}

We are discussing high energy scattering off an electron target, which we take to be leftmoving. Analoguously to the standard CGC~\cite{Gelis:2010nm} picture of dilute-dense scattering, we will describe the electromagnetic field of the electron as a shockwave. This  gives the correct result in the limit $y \ll 1$, which is the high energy limit in this case.
The ultrarelativistic electron is the source of a current $J^\mu=(J^+,J^-,\boldsymbol{J})=(0,j^-,\boldsymbol{0})$, which in the light-cone gauge $A^+=0$ creates the potential $A^\mu=(0,A^-,\boldsymbol{0})$. For a source independent of $x^-$ the only nonvanishing field strength is $F^{i-}=\partial^iA^-=-\partial_iA^-$, and Maxwell's equations reduce to $-\nabla_\perp^2A^-=j^-$. We treat the electron as a classical point-like source, whose charge is localized at $x^+\sim1/k_e^-\to0$ and at the transverse origin:
\begin{align}
    j^-(x^+,0,\boldsymbol{\mathrm{x}})=e\,\delta(x^+)\,\delta^{(2)}(\boldsymbol{\mathrm{x}})
    \quad\Longrightarrow\quad
    A^-(x^+,0,\boldsymbol{\mathrm{x}})=-e\,\delta(x^+)\,\frac{\delta^{(2)}(\boldsymbol{\mathrm{x}})}{\nabla_\perp^2}\,.
\end{align}
In the light-cone gauge the shockwave acts diagonally in transverse coordinate space and does not transfer plus momentum. A quark of charge $e_fe$ crossing it at transverse position $\boldsymbol{\mathrm{x}}$ picks up the eikonal phase $V_q(\boldsymbol{\mathrm{x}})=\exp\big(ie_fe\int\mathrm{d}x^+A^-(x^+,0,\boldsymbol{\mathrm{x}})\big)$ \cite{Bjorken:1970ah}, and an antiquark the conjugate phase. With the potential above these Abelian versions of the Wilson lines are
\begin{align}
\label{eq: LCPT: eikonal amplitude quark}
    V_q(\boldsymbol{\mathrm{x}})=\exp\bigg(-ie_fe^2\,\frac{\delta^{(2)}(\boldsymbol{\mathrm{x}})}{\nabla_\perp^2}\bigg),\qquad
    V_{\bar q}^\dagger(\boldsymbol{\mathrm{y}})=\exp\bigg(+ie_fe^2\,\frac{\delta^{(2)}(\boldsymbol{\mathrm{y}})}{\nabla_\perp^2}\bigg).
\end{align}
The gluon is electrically neutral and is not affected by the shockwave. We keep only single-photon exchange, i.e.\ the lowest order in $e^2$, which corresponds to standard one-photon-exchange in DIS. With $ie^2\delta^{(2)}(\boldsymbol{\mathrm{x}})/\nabla_\perp^2=\int\frac{\mathrm{d}^2\boldsymbol{\mathrm{q}}}{(2\pi)^2}e^{i\boldsymbol{\mathrm{q}}\cdot\boldsymbol{\mathrm{x}}}(-ie^2/\boldsymbol{\mathrm{q}}^2)$ one finds for a single quark and for a quark--antiquark pair
\begin{align}
\label{eq: LCPT: amplitude quark}
    1-V_q(\boldsymbol{\mathrm{x}})&\simeq\int\frac{\mathrm{d}^2\boldsymbol{\mathrm{q}}}{(2\pi)^2}\bigg(-i\frac{e^2}{\boldsymbol{\mathrm{q}}^2}\bigg)e_f\,e^{i\boldsymbol{\mathrm{q}}\cdot\boldsymbol{\mathrm{x}}}\,,\\
\label{eq: LCPT: amplitude dipole}
    1-V_q(\boldsymbol{\mathrm{x}})V_{\bar q}^\dagger(\boldsymbol{\mathrm{y}})&\simeq\int\frac{\mathrm{d}^2\boldsymbol{\mathrm{q}}}{(2\pi)^2}\bigg(-i\frac{e^2}{\boldsymbol{\mathrm{q}}^2}\bigg)e_f\Big(e^{i\boldsymbol{\mathrm{q}}\cdot\boldsymbol{\mathrm{x}}}-e^{i\boldsymbol{\mathrm{q}}\cdot\boldsymbol{\mathrm{y}}}\Big),
\end{align}
where in the second line the term $(1-V_q)(1-V_{\bar q}^\dagger)$, of order $e_f^2e^4$, has been dropped. The factor $-ie^2/\boldsymbol{\mathrm{q}}^2$ combines the photon propagator and the electron vertex. The relative sign in Eq.\,\eqref{eq: LCPT: amplitude dipole} reflects the opposite charges of the quark and the antiquark, and the two terms describe photon absorption by the quark or by the antiquark.  Since the photon field of a point-like charge is a superposition of different transverse momentum $\boldsymbol{\mathrm{q}}$ modes, the  transferred momentum and the photon virtuality $Q^2=\boldsymbol{\mathrm{q}}^2$ are fixed by the final state. Since the shockwave carries $q^+=0$, the plus momenta and the light-cone helicities of the partons are conserved in the interaction.

\subsection{Leading order}
\label{sec:LO}

At leading order the incoming quark absorbs the photon, $q(p_q)+\gamma^*(q)\to q(k_q)$. At this order both dressed states in Eq.\,\eqref{eq: LCPT: Definition scattering amplitude} reduce to the bare one-quark states. The shockwave conserves the plus momentum and acts on the quark at transverse position $\boldsymbol{\mathrm{x}}$ through the eikonal phase $V_q(\boldsymbol{\mathrm{x}})$, so that, with $\boldsymbol{\mathrm{p}}_q=\boldsymbol 0$,
\begin{align}
    {}^{\rm out}_{\ \ \mathrm D}\big\langle q(k_q)\big|\hat S-1\big|q(p_q)\big\rangle_{\mathrm D}^{\rm in}=2p_q^+(2\pi)\delta(k_q^+-p_q^+)\int\mathrm{d}^2\boldsymbol{\mathrm{x}}\,e^{-i\boldsymbol{\mathrm{k}}_q\cdot\boldsymbol{\mathrm{x}}}\Big[V_q(\boldsymbol{\mathrm{x}})-1\Big]\,.
\end{align}
Comparing with Eq.\,\eqref{eq: LCPT: Definition scattering amplitude} and inserting Eq.\,\eqref{eq: LCPT: amplitude quark},
\begin{align}
    i\mathcal M^{q\gamma^*\to q}_{\rm LO}=-\int\mathrm{d}^2\boldsymbol{\mathrm{x}}\,e^{-i\boldsymbol{\mathrm{k}}_q\cdot\boldsymbol{\mathrm{x}}}\Big[1-V_q(\boldsymbol{\mathrm{x}})\Big]
    =i\int\mathrm{d}^2\boldsymbol{\mathrm{q}}\,\frac{e_fe^2}{\boldsymbol{\mathrm{q}}^2}\,\delta^{(2)}\big(\boldsymbol{\mathrm{q}}-\boldsymbol{\mathrm{k}}_q\big)\,.
\end{align}
so that
\begin{align}
\label{eq:LO:amplitude}
    \mathcal M^{q\gamma^*\to q}_{\rm LO}=\frac{e_fe^2}{\boldsymbol{\mathrm{q}}^2}\,,\qquad\boldsymbol{\mathrm{q}}=\boldsymbol{\mathrm{k}}_q\,.
\end{align}
In general $\boldsymbol{\mathrm{q}}=\sum_{\rm out}\boldsymbol{\mathrm{k}}$ is the total transverse momentum of the final state, and the photon virtuality $Q^2=\boldsymbol{\mathrm{q}}^2$ is not an independent variable but a function of the final-state momenta. To obtain a cross sections differential in $Q^2$ we therefore  insert $\delta(Q^2-\boldsymbol{\mathrm{q}}^2)$ into the phase-space integral. For a multi-particle final state one trades one of the final-state transverse momenta for $\boldsymbol{\mathrm{q}}$, and with $\mathrm{d}^2\boldsymbol{\mathrm{q}}=\frac12\mathrm{d}Q^2\,\mathrm{d}\phi_q$ the integral over the overall azimuth $\phi_q$ gives $\mathrm{d}^2\boldsymbol{\mathrm{q}}\to\pi\,\mathrm{d}Q^2$; we will see how this works in practice  for the real emissions in Sec.~\ref{sec:NLO}.

At leading order the one-particle phase space \eqref{eq: LCPT: PS} reduces, after the $k_q^+$ integral, to
\begin{align}
    \mathrm{d}(\mathrm{P.S.})_q=2p_q^+(2\pi)\delta(p_q^+-k_q^+)\,\frac{\mathrm{d}k_q^+\,\mathrm{d}^2\boldsymbol{\mathrm{k}}_q}{2k_q^+(2\pi)^3}=\frac{\mathrm{d}^2\boldsymbol{\mathrm{k}}_q}{(2\pi)^2}\,,
\end{align}
and with $\boldsymbol{\mathrm{q}}=\boldsymbol{\mathrm{k}}_q$,
\begin{align}
    \frac{\mathrm{d}\hat\sigma^{q\gamma^*\to q}_{\rm LO}}{\mathrm{d}Q^2}=\int\frac{\mathrm{d}^2\boldsymbol{\mathrm{k}}_q}{(2\pi)^2}\,\big|\mathcal M^{q\gamma^*\to q}_{\rm LO}\big|^2\,\delta\big(Q^2-\boldsymbol{\mathrm{k}}_q^2\big)=\frac{1}{4\pi}\,\frac{e_f^2e^4}{Q^4}\,,
\end{align}
which with $e^2=4\pi\alpha_{\mathrm{e.m.}}$ is $\sigma_0$ of Eq.\,\eqref{eq:sigma0}.

To obtain the cross sections differential in the partonic variables \eqref{eq:DIS:xhat}, we insert the corresponding delta functions. Light-cone energy conservation across the shockwave gives
\begin{align}
    q^-=k_q^--p_q^-=\frac{\boldsymbol{\mathrm{k}}_q^2+\mu_{\mathrm{in}}^2}{2p_q^+}\,,
\end{align}
so that, at $\boldsymbol{\mathrm{k}}_q^2=Q^2$,
\begin{align}
    \hat{x}=\hat{z}=\frac{Q^2}{Q^2+\mu_{\mathrm{in}}^2}\,.
\end{align}
Both are fixed by $Q^2$, and
\begin{align}
\label{eq:LO:sigma delta}
    \frac{\mathrm{d}^3\hat\sigma^{q\gamma^*\to q}_{\rm LO}}{\mathrm{d}\hat{x}\,\mathrm{d}\hat{z}\,\mathrm{d}Q^2}&=\int\frac{\mathrm{d}^2\boldsymbol{\mathrm{k}}_q}{(2\pi)^2}\,\big|\mathcal M^{q\gamma^*\to q}_{\rm LO}\big|^2\,\delta\big(Q^2-\boldsymbol{\mathrm{k}}_q^2\big)\,\delta\bigg(\hat{x}-\frac{Q^2}{2p_q^+q^-}\bigg)\,\delta\bigg(\hat{z}-\frac{k_q^-}{q^-}\bigg)\nonumber\\
    &=\sigma_0\,\delta\bigg(\hat{x}-\frac{Q^2}{Q^2+\mu_{\mathrm{in}}^2}\bigg)\,\delta\bigg(\hat{z}-\frac{Q^2}{Q^2+\mu_{\mathrm{in}}^2}\bigg)\,.
\end{align}
Up to power corrections of order $\mu_{\mathrm{in}}^2/Q^2$, the incoming quark carries the full partonic Bjorken variable and the outgoing quark the full light-cone energy transferred by the shockwave, $\hat{x}=\hat{z}=1$, so that the SIDIS and DIS cross sections are
\begin{align}
\label{eq:LO:partonic}
    \frac{\mathrm{d}^3\hat\sigma^{q\gamma^*\to q}_{\rm LO}}{\mathrm{d}\hat{x}\,\mathrm{d}\hat{z}\,\mathrm{d}Q^2}=\sigma_0\,\delta(1-\hat{x})\,\delta(1-\hat{z})\,,\qquad
    \frac{\mathrm{d}^2\hat\sigma^{q\gamma^*\to q}_{\rm LO}}{\mathrm{d}\hat{x}\,\mathrm{d}Q^2}=\sigma_0\,\delta(1-\hat{x})\,,
\end{align}
where the latter one follows by integrating over $\hat{z}$. These are the leading terms of the partonic cross sections Eqs.~\eqref{eq:fact:C DIS} and \eqref{eq:fact:C SIDIS}, with the leading-order coefficient functions $\delta(1-\hat{x})$ and $\delta(1-\hat{x})\,\delta(1-\hat{z})$. At the hadron level, Eqs.~\eqref{eq:DIS:factorization} and \eqref{eq:SIDIS:factorization} they reduce to the parton model expressions
\begin{align}
\label{eq:LO:F}
    F_2^p\Big|_{\rm LO}&=x_\text{Bj}\sum_{q,\bar q}e_q^2\,f_{q/p}(x_\text{Bj},\mu_F^2)\,,\nonumber\\
    \Big[2F_1^{p/H}+F_L^{p/H}\Big]_{\rm LO}&=\sum_{q,\bar q}e_q^2\,f_{q/p}(x_\text{Bj},\mu_F^2)\,D_{H/q}(z_H,\mu_{D}^2)\,.
\end{align}
At this order the off-shell and $\overline{\mathrm{MS}}$ schemes coincide; the scheme dependence first appears at $\mathcal O(\alpha_\mathrm{s})$.

\section{Partonic cross sections at next-to-leading order}
\label{sec:NLO}

In this section we compute the $\mathcal O(\alpha_\mathrm{s})$ cross sections of the off-shell gluon (Sec.~\ref{sec:gluon channel partonic}) and quark (Sec.~\ref{sec:quark channel partonic}) scattering off the photon shockwave, differentially in the final-state momenta. The inclusive DIS and SIDIS in Secs.~\ref{sec:DIS} and \ref{sec:SIDIS} are then obtained from these cross sections by imposing the appropriate target momentum fraction kinematics. Throughout this section, the final-state partons carry momenta $k_q$, $k_{\bar q}$ and $k_g$. Intermediate partons in the fluctuation of the incoming state, before the shockwave, are denoted by a prime ($k_q'$, $k_g'$, \dots), and intermediate partons after the shockwave but before a final-state splitting by a double prime ($k_q''$, \dots). Amplitudes and cross sections carry the partonic process as a superscript, e.g.\ $\mathcal M^{g\gamma^*\to q\bar q}$, while light-cone wave functions carry the splitting, e.g.\ $\Psi^{g\to q\bar q}$. Details on the evaluation of the light cone wave functions are given in Appendix \ref{sec:app:partonic}. 

\subsection{Gluon-initiated channel}
\label{sec:gluon channel partonic}

The incoming off-shell gluon, with momentum $p_g$ as in Eq.\,\eqref{eq: off-shell regulator}, color $a$ and transverse polarization $\lambda$, does not couple to the shockwave. The interaction proceeds through its $q\bar q$ fluctuation formed before $x^+=0$,
\begin{align}
\label{eq:g:Fock}
    \big|g(p_g,\lambda,a)\big\rangle_\mathrm{D}^{\mathrm{in}}=\sqrt{Z_g^{\mathrm{in}}}\bigg[\big|g(p_g,\lambda,a)\big\rangle+{\sum_{q\bar q}}'\,\Psi^{g\to q\bar q}_{h_q,h_{\bar q};\lambda}(\boldsymbol{\ell},z;\mu_{\mathrm{in}}^2)\,\big|q(k_q',h_q,i)\,\bar q(k_{\bar q}',h_{\bar q},j)\big\rangle+\dots\bigg]\,,
\end{align}
where the quark carries the plus-momentum fraction $z$ and the relative transverse momentum $\boldsymbol{\ell}$, i.e.\ $\boldsymbol{\mathrm{k}}_q'=-\boldsymbol{\mathrm{k}}_{\bar q}'=\boldsymbol{\ell}$. The light-cone wave function, derived in Appendix~\ref{sec:app:g LCWF}, is
\begin{align}
\label{eq:g:LCWF}
    \Psi^{g\to q\bar q}_{h_q,h_{\bar q};\lambda}(\boldsymbol{\rho},z;\mu_0^2)&=-\frac{g\,t^a_{ij}}{4p_g^+}\,\varepsilon^n_\lambda\,\bar u_{h_q}\,\gamma^+\Big[-2(1-2z)\,\delta^{mn}+[\gamma^m,\gamma^n]\Big]v_{h_{\bar q}}\,\frac{\boldsymbol{\rho}^m}{\boldsymbol{\rho}^2+z(1-z)\mu_0^2}\,,
\end{align}
where $\boldsymbol{\rho}=(1-z)\boldsymbol{\mathrm{k}}_1-z\boldsymbol{\mathrm{k}}_2$ is the relative transverse momentum of a pair whose quark and antiquark have the transverse momenta $\boldsymbol{\mathrm{k}}_1$ and $\boldsymbol{\mathrm{k}}_2$ and the  fractions $z$ and $1-z$ of the plus-momentum of the gluon, $\mu_0$ is the virtuality of the gluon, and $m,n$ are transverse indices in $D-2$ dimensions. The Dirac structures $\bar u\gamma^+\Gamma v$, with $\Gamma$ a product of transverse $\gamma$ matrices, depend only on the plus momenta and the helicities \cite{Beuf:2021qqa}, and the spinors carry the momenta of the quark and the antiquark of the pair.   In the fluctuation \eqref{eq:g:Fock}, the momentum argument is $\boldsymbol{\rho}=\boldsymbol{\ell}$ and $\mu_0=\mu_{\mathrm{in}}$.

\begin{figure*}[tbp!]
\centering
\includegraphics[width=0.45\textwidth]{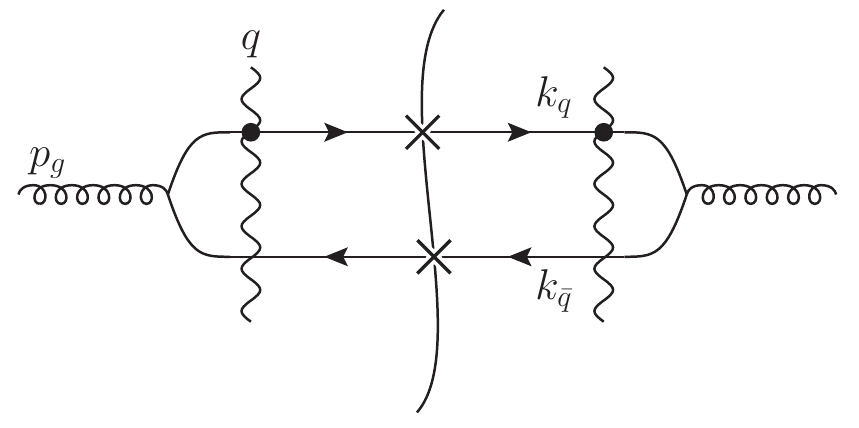}
\begin{tikzpicture}[overlay]
\node[font=\footnotesize,anchor=south west] at (-1cm,-0.5cm) {\namediag{diag:g qq}};
\end{tikzpicture}
\rule{2em}{0pt}
\includegraphics[width=0.45\textwidth]{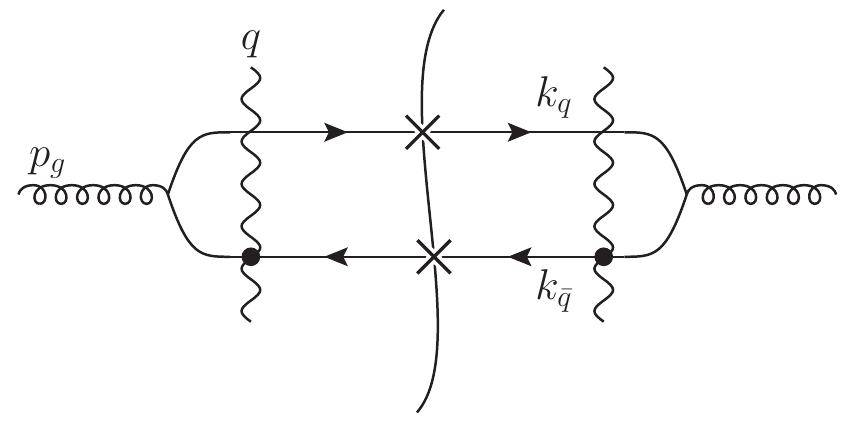}
\begin{tikzpicture}[overlay]
\node[font=\footnotesize,anchor=south west] at (-1cm,-0.5cm) {\namediag{diag:g qbqb}};
\end{tikzpicture}
\\
\vspace{1cm}
\includegraphics[width=0.45\textwidth]{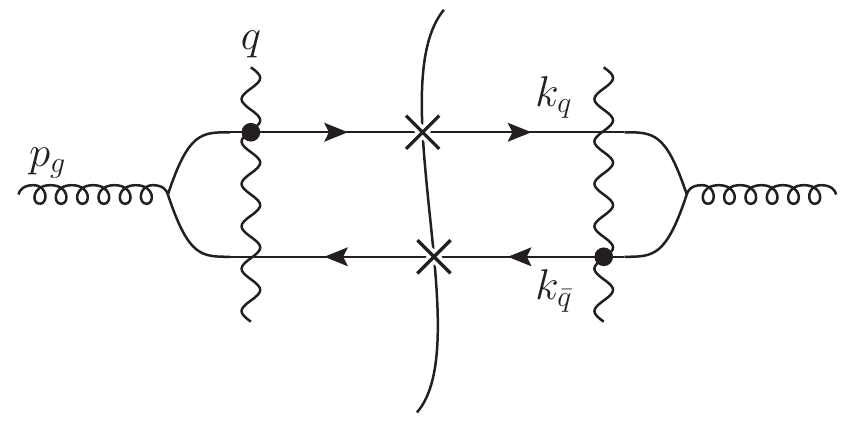}
\begin{tikzpicture}[overlay]
\node[font=\footnotesize,anchor=south west] at (-1cm,-0.5cm) {\namediag{diag:g qqb}};
\end{tikzpicture}
\caption{Contributions to the gluon-channel cross section at $\mathcal O(\alpha_\mathrm{s})$. The off-shell gluon fluctuates into a $q\bar q$ pair before the shockwave, and the photon is absorbed (dot) by the quark or by the antiquark in the amplitude (left of the cut) and in the complex-conjugate amplitude (right of the cut): \ref{diag:g qq} by the quark in both, \ref{diag:g qbqb} by the antiquark in both, \ref{diag:g qqb} interference (plus its complex conjugate). The wavy lines denote the photon shockwave at $x^+=0$, the curved line the final-state cut, and the crosses the final-state partons, whose momenta are fixed in the differential cross section \eqref{eq:g:partonic xs}; the DIS and SIDIS cross sections follow from it by integrating with the delta functions of $\hat{x}$ and $\hat{z}$.}
\label{fig:gluon channel}
\end{figure*}

The amplitude for $g(p_g)+\gamma^*(q)\to q(k_q)+\bar q(k_{\bar q})$ receives two contributions, in which the photon is absorbed by the quark or by the antiquark (Fig.\,\ref{fig:gluon amplitude} in Appendix~\ref{sec:app:gluon}), corresponding to the two terms of Eq.\,\eqref{eq: LCPT: amplitude dipole}. There is no contribution with a splitting after the shockwave, beucause the gluon has no electric charge, and consequently no off-shellness is needed for the final state particles.   Since the shockwave conserves plus momentum, $z=k_q^+/p_g^+$. When the quark absorbs the photon, the spectator antiquark fixes $\boldsymbol{\ell}=\boldsymbol{\ell}_1\equiv-\boldsymbol{\mathrm{k}}_{\bar q}$; when the antiquark absorbs it, $\boldsymbol{\ell}=\boldsymbol{\ell}_2\equiv\boldsymbol{\mathrm{k}}_q$.  These two contributions have opposite signs because of the opposite electric charges. With $\boldsymbol{\mathrm{q}}=\boldsymbol{\mathrm{k}}_q+\boldsymbol{\mathrm{k}}_{\bar q}$ one obtains (Appendix~\ref{sec:app:gluon})
\begin{align}
    \mathcal M^{g\gamma^*\to q\bar q}_{h_q,h_{\bar q};\lambda}&=\frac{e_fe^2}{\boldsymbol{\mathrm{q}}^2}\Big[\Psi^{g\to q\bar q}_{h_q,h_{\bar q};\lambda}(\boldsymbol{\ell}_1,z;\mu_{\mathrm{in}}^2)-\Psi^{g\to q\bar q}_{h_q,h_{\bar q};\lambda}(\boldsymbol{\ell}_2,z;\mu_{\mathrm{in}}^2)\Big]\nonumber\\
    &=-ge_f\,\frac{e^2}{\boldsymbol{\mathrm{q}}^2}\,\frac{t^a_{ij}}{4p_g^+}\,\varepsilon^n_\lambda\,\bar u_{h_q}(k_q)\,\gamma^+\Big[-2(1-2z)\,\delta^{mn}+[\gamma^m,\gamma^n]\Big]v_{h_{\bar q}}(k_{\bar q})\,\boldsymbol{\mathrm{V}}^m\,,\label{eq:g:amplitude}
\end{align}
with
\begin{align}
\label{eq:g:V}
    \boldsymbol{\mathrm{V}}\equiv\frac{\boldsymbol{\ell}_1}{\boldsymbol{\ell}_1^2+z(1-z)\mu_{\mathrm{in}}^2}-\frac{\boldsymbol{\ell}_2}{\boldsymbol{\ell}_2^2+z(1-z)\mu_{\mathrm{in}}^2}\,,\qquad\boldsymbol{\ell}_1=\boldsymbol{\mathrm{k}}_q-\boldsymbol{\mathrm{q}}\,,\qquad\boldsymbol{\ell}_2=\boldsymbol{\mathrm{k}}_q\,.
\end{align}
In the squared amplitude, the two terms give the contributions shown in Fig.\,\ref{fig:gluon channel}. Averaging over the polarization and color of the gluon and summing over the final-state helicities and colors, with $T_\mathrm{R}=1/2$,
\begin{align}
\label{eq:g:Msq}
    \overline{|\mathcal M^{g\gamma^*\to q\bar q}|^2}=2g^2e^4e_f^2\,T_\mathrm{R}\,\frac{z(1-z)}{\boldsymbol{\mathrm{q}}^4}\,\frac{(1-2z)^2+D-3}{D-2}\,\boldsymbol{\mathrm{V}}^2\,.
\end{align}
Since all collinear singularities are regulated by $\mu_{\mathrm{in}}^2$, we set $D=4$ in the following; the $D$-dimensional polarization average is needed only for the comparison with the $\overline{\mathrm{MS}}$ scheme. With the two-body phase space $\mathrm{d}(\mathrm{P.S.})_{q\bar q}=\frac{\mathrm{d}z}{4\pi z(1-z)}\frac{\mathrm{d}^{D-2}\boldsymbol{\mathrm{k}}_q}{(2\pi)^{D-2}}\frac{\mathrm{d}^{D-2}\boldsymbol{\mathrm{k}}_{\bar q}}{(2\pi)^{D-2}}$, the factor $z(1-z)$ cancels, and after changing variables from $\boldsymbol{\mathrm{k}}_{\bar q}$ to $\boldsymbol{\mathrm{q}}$ and integrating over the overall azimuth, the partonic cross section in $D=4$ becomes
\begin{align}
\label{eq:g:partonic xs}
    \frac{\mathrm{d}\hat\sigma^{g\gamma^*\to q\bar q}}{\mathrm{d}z\,\mathrm{d}^2\boldsymbol{\mathrm{k}}_q\,\mathrm{d}Q^2}=\frac{\alpha_{\mathrm{e.m.}}^2}{Q^4}\,\frac{\alpha_\mathrm{s}}{2\pi}\,4e_f^2\,P_{q\leftarrow g}(z)\,\boldsymbol{\mathrm{V}}^2\,,\qquad P_{q\leftarrow g}(z)=T_\mathrm{R}\big[z^2+(1-z)^2\big]\,.
\end{align}

Now we start to see the emergence of DGLAP evolution for the PDF. The whole cross section Eq.~\eqref{eq:g:partonic xs} expressed in terms of the projectile momentum fraction $z$ is proportional to the splitting function $P_{q\leftarrow g}(z)$, and the coefficient $\boldsymbol{\mathrm{V}}^2$ from Eq.~\eqref{eq:g:V} has  logarithmic singularities at $\boldsymbol{\mathrm{k}}_q = \boldsymbol{0}$ and  $\boldsymbol{\mathrm{k}}_q = \boldsymbol{\mathrm{q}}$, regulated by the off-shellness of the incoming gluon. It is clear that these singularities correspond to the part of DGLAP evolution where the collinear logarithm is absorbed into the quark (or antiquark)  distribution. However, we know that in the usual collinear factorization picture the gluon-initiated channel also has a finite coefficient function, without a logarithm and not proportional to the splitting function. To discover this we will need to change to target momentum fraction variables, which we will do in Secs.~\ref{sec:DIS} and~\ref{sec:SIDIS}. In addition to the splitting function $P_{q\leftarrow g}(\hat{x})$ (recall from Sec.~\ref{sec:LCPT} that the initial state collinear region is $\hat{x}=z$), this variable change will bring explicitly into view also the finite parts. 

\subsection{Quark-initiated channel}
\label{sec:quark channel partonic}

\begin{figure*}[tbp!]
\centering
\includegraphics[width=0.45\textwidth]{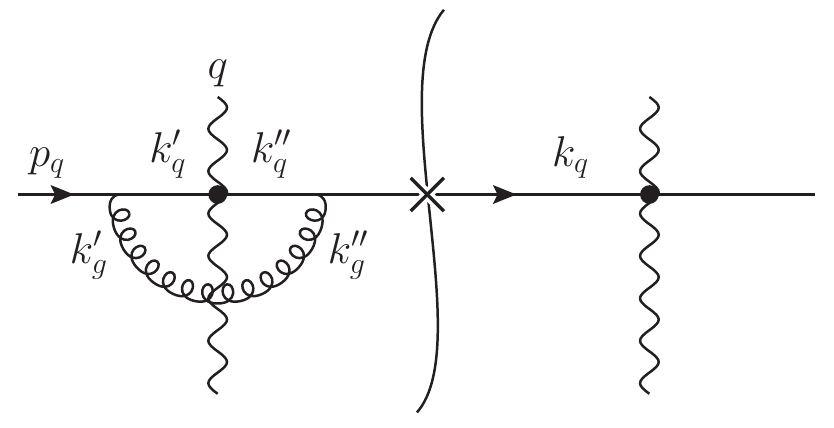}
\begin{tikzpicture}[overlay]
\node[font=\footnotesize,anchor=south west] at (-1cm,-0.5cm) {\namediag{diag:virtual}};
\end{tikzpicture}
\rule{2em}{0pt}
\includegraphics[width=0.45\textwidth]{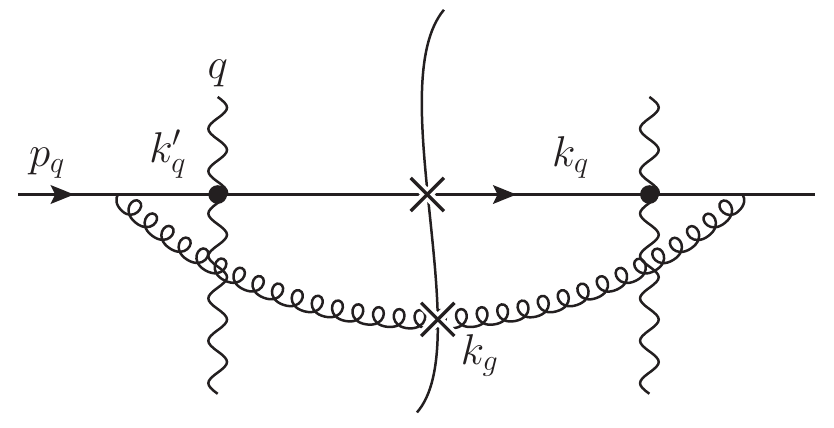}
\begin{tikzpicture}[overlay]
\node[font=\footnotesize,anchor=south west] at (-1cm,-0.5cm) {\namediag{diag:real_internal}};
\end{tikzpicture}
\\
\vspace{1cm}
\includegraphics[width=0.45\textwidth]{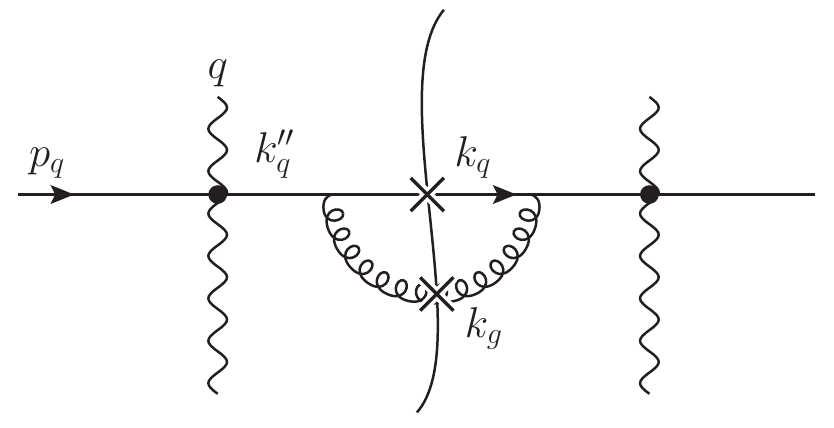}
\begin{tikzpicture}[overlay]
\node[font=\footnotesize,anchor=south west] at (-1cm,-0.5cm) {\namediag{diag:real_external}};
\end{tikzpicture}
\rule{2em}{0pt}
\includegraphics[width=0.45\textwidth]{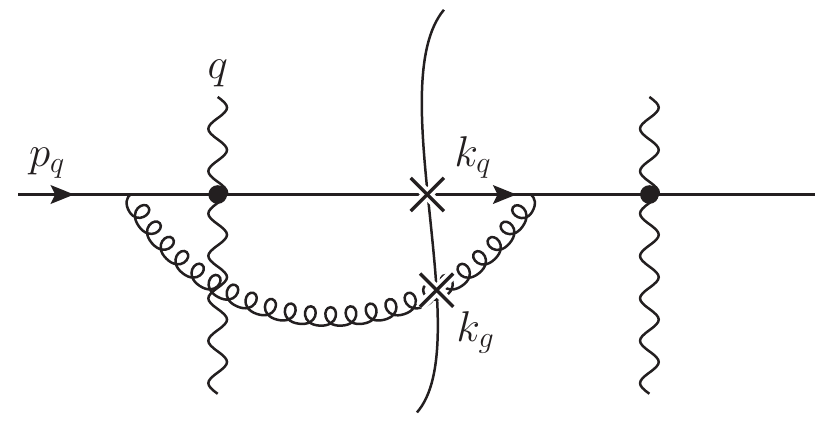}
\begin{tikzpicture}[overlay]
\node[font=\footnotesize,anchor=south west] at (-1cm,-0.5cm) {\namediag{diag:real_crossed}};
\end{tikzpicture}
\caption{
Contributions to the quark-channel cross section at $\mathcal O(\alpha_\mathrm{s})$: \ref{diag:virtual} virtual correction; \ref{diag:real_internal} gluon emission before the shockwave, squared; \ref{diag:real_external} gluon emission after the shockwave, squared; \ref{diag:real_crossed} interference of the two emissions (plus its complex conjugate). The wavy lines denote the photon shockwave at $x^+=0$ in the amplitude (left of the cut) and in the complex-conjugate amplitude (right of the cut), the dots the absorption of the photon by the quark, the curved line the final-state cut, and the crosses the final-state partons, whose momenta are fixed in the differential cross sections \eqref{eq:q:virtual result} and \eqref{eq:q:real result}; the DIS and SIDIS cross sections follow from them by integrating with the delta functions of $\hat{x}$ and $\hat{z}$.
}\label{fig:cross_section}
\end{figure*}

We now turn to the quark channel, $q+\gamma^*\to q+X$, with the incoming quark of momentum $p_q$ as in Eq.\,\eqref{eq: off-shell regulator} and helicity $h$. The contributions to the cross section at $\mathcal O(\alpha_\mathrm{s})$ are shown in Fig.\,\ref{fig:cross_section} where the gluon does not couple to the shockwave but affects the kinematics of the quark that does. They are all generated by the $qg$ components of the incoming and outgoing dressed quark states. For an incoming quark with zero transverse momentum the Fock state decomposition reads
\begin{align}
\label{eq:q:Fock}
    \big|q(p_q,h,j)\big\rangle^{\mathrm{in}}_\mathrm{D} =\sqrt{Z_q^{\text{in}}}\bigg[\big|q(p_q,h,j)\big\rangle + {\sum_{qg}}'\,\Psi^{q\to qg}_{h_q,\sigma;h}(\boldsymbol{\ell},\zeta;\mu_{\mathrm{in}}^2)\,\big|q(k_q',h_q,i)\,g(k_g',\sigma,a)\big\rangle + \dots \bigg] \,,
\end{align}
where the intermediate quark carries the plus-momentum fraction $\zeta={k_q'}^+/p_q^+$ and $\boldsymbol{\mathrm{k}}_q'=-\boldsymbol{\mathrm{k}}_g'=\boldsymbol{\ell}$. The outgoing dressed quark has the same structure, with $Z_q^{\rm in}\to Z_q^{\rm out}$ and $\mu_{\mathrm{in}}^2\to\mu_{\mathrm{out}}^2$. The light-cone wave function for gluon emission, derived in Appendix~\ref{sec:app:q LCWF}, is
\begin{align}
\label{eq:q:LCWF}
    \Psi^{q\to qg}_{h_q,\sigma;h}(\boldsymbol{\rho},\zeta;\mu_0^2)&=\frac{g\,t^a_{ij}}{4p_q^+}\,\varepsilon^{*n}_\sigma\,\bar u_{h_q}\,\gamma^+\Big[2(1+\zeta)\,\delta^{mn}-(1-\zeta)\,[\gamma^m,\gamma^n]\Big]u_h\,\frac{\boldsymbol{\rho}^m}{\boldsymbol{\rho}^2+\zeta(1-\zeta)\mu_0^2}\,,
\end{align}
where, as in Eq.\,\eqref{eq:g:LCWF}, $\boldsymbol{\rho}=(1-\zeta)\boldsymbol{\mathrm{k}}_1-\zeta\boldsymbol{\mathrm{k}}_2$ is the relative transverse momentum of the $qg$ pair, with $\boldsymbol{\mathrm{k}}_1$ and $\boldsymbol{\mathrm{k}}_2$ the transverse momenta of its quark and gluon, $\zeta$ is the plus-momentum fraction of its quark with respect to the single quark, $\mu_0$ is the virtuality of the single quark, and the spinors carry the momenta of the quark of the pair and of the single quark. 

With the momenta labeled as in Fig.\,\ref{fig:cross_section}, calculating the cross section requires  three wave functions, which can all be expressed in terms of the gluon emission wavefunction~\eqref{eq:q:LCWF} as
\begin{align}
\label{eq:q:LCWF cases}
    \text{incoming splitting:}&\quad\Psi^{q\to qg}_{h_q,\sigma;h}(\boldsymbol{\mathrm{k}}_q',\zeta;\mu_{\mathrm{in}}^2)\,,&&\bar u_{h_q}(k_q')\,,\ u_h(p_q)\,,\nonumber\\
    \text{outgoing splitting:}&\quad\Psi^{q\to qg}_{h_q,\sigma;h}(\boldsymbol{\mathrm{k}}_q''-\zeta\boldsymbol{\mathrm{k}}_q,\zeta;\mu_{\mathrm{out}}^2)\,,&&\bar u_{h_q}(k_q'')\,,\ u_h(k_q)\,,\nonumber\\
    \text{outgoing merging:}&\quad-\Big[\Psi^{q\to qg}_{h_q,\sigma;h}(\boldsymbol{\mathrm{m}},z;\mu_{\mathrm{out}}^2)\Big]^*\,,&&\bar u_{h_q}(k_q)\,,\ u_h(k_q'')\,,
\end{align}
where we also indicate the spinors corresponding to these wavefunctions. The first one, with  $\boldsymbol{\rho}=\boldsymbol{\mathrm{k}}_q'=\boldsymbol{\ell}$ in Eq.\,\eqref{eq:q:LCWF}, enters the in-state in diagrams \ref{diag:virtual}, \ref{diag:real_internal}
and \ref{diag:real_crossed} in Fig.\,\ref{fig:cross_section}. The second, with, $\boldsymbol{\rho}=\boldsymbol{\mathrm{k}}_q''-\zeta\boldsymbol{\mathrm{k}}_q$ is the $qg$ component of the outgoing quark in diagram 
 \ref{diag:virtual}. Here one must remember that the wavefunction should be read from the asymptotic state towards the shockwave, and thus a gluon reabsorbed in the final state in the virtual correction corresponds to a splitting wavefunction. The third one, with $\boldsymbol{\rho}=\boldsymbol{\mathrm{m}}=\boldsymbol{\mathrm{k}}_q-z\boldsymbol{\mathrm{q}}$ is the single-quark component of the outgoing $qg$ pair, so reading the figure backwards from the cut it is a merging wavefunction in the final state,  appearing in the final state emission diagram \ref{diag:real_external}, and the interference diagram  \ref{diag:real_crossed} in the conjugate amplitude.
The momentum fraction of the virtual gluon in diagram \ref{diag:virtual} is $\zeta$, and in the real emission $\zeta=z$.
See Appendix~\ref{sec:app:q LCWF} for a more detailed discussion.

The virtual contribution, diagram~\ref{diag:virtual}, consists of the $qg$ fluctuation of the incoming quark crossing the shockwave and projected onto the $qg$ component of the outgoing quark. The quark absorbs the photon, the gluon is a spectator, and the outgoing pair has relative momentum $\boldsymbol{\mathrm{k}}_q''-\zeta\boldsymbol{\mathrm{k}}_q=\boldsymbol{\ell}+(1-\zeta)\boldsymbol{\mathrm{q}}$, Eq.\,\eqref{eq:q:LCWF cases} with $\boldsymbol{\mathrm{k}}_q''=\boldsymbol{\ell}+\boldsymbol{\mathrm{q}}$ and $\boldsymbol{\mathrm{k}}_q=\boldsymbol{\mathrm{q}}$. The final state is a single quark, $\boldsymbol{\mathrm{q}}=\boldsymbol{\mathrm{k}}_q$, and the amplitude is proportional to the leading-order one \eqref{eq:LO:amplitude} (Appendix~\ref{sec:app:q virtual}),
\begin{align}
\label{eq:q:MV}
    \mathcal M^{q\gamma^*\to q}_{\rm V}&=\mathcal M^{q\gamma^*\to q}_{\rm LO}\int_0^1\frac{\mathrm{d}\zeta}{4\pi\zeta(1-\zeta)}\int\frac{\mathrm{d}^{D-2}\boldsymbol{\ell}}{(2\pi)^{D-2}}\nonumber\\
    &\quad\times\sum_{i,a,h_q,\sigma}\Big[\Psi^{q\to qg}_{h_q,\sigma;h}\big(\boldsymbol{\ell}+(1-\zeta)\boldsymbol{\mathrm{q}},\zeta;\mu_{\mathrm{out}}^2\big)\Big]^\dagger\Psi^{q\to qg}_{h_q,\sigma;h}(\boldsymbol{\ell},\zeta;\mu_{\mathrm{in}}^2)\,.
\end{align}
The normalization factors of the dressed states are solved from the condition ${_\mathrm{D}^\mathrm{in}}\langle q | q\rangle_\mathrm{D}^\mathrm{in}=\langle q|q\rangle$, so at this order they are fixed by the same wave functions,
\begin{align}
\label{eq:q:Z}
    Z_q^{\rm in}&=1-\int_0^1\frac{\mathrm{d}\zeta}{4\pi\zeta(1-\zeta)}\int\frac{\mathrm{d}^{D-2}\boldsymbol{\ell}}{(2\pi)^{D-2}}\sum_{i,a,h_q,\sigma}\Big|\Psi^{q\to qg}_{h_q,\sigma;h}(\boldsymbol{\ell},\zeta;\mu_{\mathrm{in}}^2)\Big|^2\,,\nonumber\\
    Z_q^{\rm out}&=1-\int_0^1\frac{\mathrm{d}\zeta}{4\pi\zeta(1-\zeta)}\int\frac{\mathrm{d}^{D-2}\boldsymbol{\mathrm{k}}_q''}{(2\pi)^{D-2}}\sum_{i,a,h_q,\sigma}\Big|\Psi^{q\to qg}_{h_q,\sigma;h}(\boldsymbol{\mathrm{k}}_q''-\zeta\boldsymbol{\mathrm{k}}_q,\zeta;\mu_{\mathrm{out}}^2)\Big|^2\,.
\end{align}
After a shift of the integration variable, $Z_q^{\rm out}$ differs from $Z_q^{\rm in}$ only by $\mu_{\mathrm{in}}^2\to\mu_{\mathrm{out}}^2$. At $\mathcal O(\alpha_\mathrm{s})$ the virtual cross section is $|\mathcal M^{q\gamma^*\to q}_{\rm LO}|^2(Z_q^{\rm in}+Z_q^{\rm out}-2)+2\,{\rm Re}\big[\mathcal M^{q\gamma^*\to q}_{\rm V}\big(\mathcal M^{q\gamma^*\to q}_{\rm LO}\big)^*\big]$ times the one-particle phase space. The three terms combine into a perfect square. With the shorthand notation
\begin{align}
\label{eq:q:tilde}
    \tilde{\mu}_{\mathrm{in}}^2\equiv\zeta(1-\zeta)\mu_{\mathrm{in}}^2\,,\qquad\tilde{\mu}_{\mathrm{out}}^2\equiv\zeta(1-\zeta)\mu_{\mathrm{out}}^2\,,\qquad\tilde{\boldsymbol{\mathrm{q}}}\equiv(1-\zeta)\boldsymbol{\mathrm{q}}\,,
\end{align}
Eq.\,\eqref{eq:q:LCWF} and\footnote{By azimuthal symmetry we know that the cross section will in the end not depend on the azimuthal angle of the total transverse momentum  $\boldsymbol{\mathrm{q}}$ but only on its absolute value. Thus we can make this replacement already here even if we keep $\boldsymbol{\mathrm{q}}$ as a vector in the integrand.} $\mathrm{d}^2\boldsymbol{\mathrm{q}}=\pi\,\mathrm{d}Q^2$, one finds, in terms of $\sigma_0$ of Eq.\,\eqref{eq:sigma0},
\begin{align}
\label{eq:q:virtual square}
    \frac{\mathrm{d}\hat\sigma^{q\gamma^*\to q}_{\rm V}}{\mathrm{d}z\,\mathrm{d}Q^2}&=-\sigma_0\,\alpha_\mathrm{s}C_\mathrm{F}\,\delta(1-z)\int_0^1\mathrm{d}\zeta\,\frac{(1+\zeta)^2+(1-\zeta)^2(D-3)}{1-\zeta}\nonumber\\
    &\quad\times\int\frac{\mathrm{d}^{D-2}\boldsymbol{\ell}}{(2\pi)^{D-2}}\bigg[\frac{\boldsymbol{\ell}}{\boldsymbol{\ell}^2+\tilde{\mu}_{\mathrm{in}}^2}-\frac{\boldsymbol{\ell}+\tilde{\boldsymbol{\mathrm{q}}}}{(\boldsymbol{\ell}+\tilde{\boldsymbol{\mathrm{q}}})^2+\tilde{\mu}_{\mathrm{out}}^2}\bigg]^2\,,
\end{align}
where $z=k_q^+/p_q^+$. The first and second terms in the square bracket come from the incoming and outgoing fluctuations; their squares originate from $Z_q^{\rm in}$ and $Z_q^{\rm out}$, and the cross term from the loop diagram. Each of these contributions is separately ultraviolet divergent, but expression inside the square bracket falls off as $1/\boldsymbol{\ell}^2$ and the $\boldsymbol{\ell}$ integral is finite in a typical cancellation between ``emission before'' and ``emission after'' diagrams. The virtuality $\mu_{\mathrm{out}}^2$ of the outgoing quark enters only through the energy denominator of the outgoing fluctuation. Since the expression is UV (and IR) finite, we can evaluate it in $D=4$, with  the result (see Appendix~\ref{sec:app:q virtual})
\begin{align}
\label{eq:q:virtual result}
    \frac{\mathrm{d}\hat\sigma^{q\gamma^*\to q}_{\rm V}}{\mathrm{d}z\,\mathrm{d}Q^2}=\sigma_0\,\frac{\alpha_\mathrm{s}}{2\pi}\,C_\mathrm{F}\,\delta(1-z)\int_0^1\mathrm{d}\zeta\,\frac{1+\zeta^2}{1-\zeta}\bigg[2-\frac{2\operatorname{artanh}\beta}{\beta}\bigg],
\end{align}
where
\begin{align}
\label{eq:q:beta}
    \beta\equiv\sqrt{1-\frac{4\tilde{\mu}_{\mathrm{in}}^2\tilde{\mu}_{\mathrm{out}}^2}{(\tilde{\mu}_{\mathrm{in}}^2+\tilde{\mu}_{\mathrm{out}}^2+\tilde{\boldsymbol{\mathrm{q}}}^2)^2}}\,.
\end{align}
The $\zeta$ integral in Eq.\,\eqref{eq:q:virtual result} is not convergent by itself at $\zeta\to1$, where the gluon becomes soft. For $\mu_{\mathrm{in}}\neq\mu_{\mathrm{out}}$ the virtualities $\tilde{\mu}_{\mathrm{in}}^2,\tilde{\mu}_{\mathrm{out}}^2\propto1-\zeta$ dominate over $\tilde{\boldsymbol{\mathrm{q}}}^2\propto(1-\zeta)^2$, so that $\beta\to|\mu_{\mathrm{in}}^2-\mu_{\mathrm{out}}^2|/(\mu_{\mathrm{in}}^2+\mu_{\mathrm{out}}^2)$ and the square bracket tends to the nonzero constant $2-\frac{\mu_{\mathrm{in}}^2+\mu_{\mathrm{out}}^2}{\mu_{\mathrm{in}}^2-\mu_{\mathrm{out}}^2}\ln\frac{\mu_{\mathrm{in}}^2}{\mu_{\mathrm{out}}^2}$. This soft divergence will  cancel against the real emission at $z\to1$, see Eq.\,\eqref{eq:q:R+V} below.

The real contributions, diagrams~\ref{diag:real_internal}--\ref{diag:real_crossed}, consist of the gluon emission before the shockwave, where the gluon collinear to the incoming quark is regulated by $\mu_{\mathrm{in}}^2$, and the emission after the shockwave, where the gluon collinear to the outgoing quark is regulated by $\mu_{\mathrm{out}}^2$. In the first case the quark of the incoming fluctuation absorbs the photon and the gluon is a spectator, so that the relative momentum is $\boldsymbol{\ell}=-\boldsymbol{\mathrm{k}}_g=\boldsymbol{\mathrm{k}}_q-\boldsymbol{\mathrm{q}}$. In the second case the bare quark absorbs the photon and is projected on the single-quark component of the outgoing dressed $qg$ state,
\begin{align}
\label{eq:q:Fock qg}
    \big|q\,g\big\rangle^{\mathrm{out}}_\mathrm{D}=\sqrt{Z_q^{\rm out}Z_g^{\rm out}}\bigg[\big|q\,g\big\rangle+\sum_{h,j}\Psi^{qg\to q}_{h;h_q,\sigma}(\boldsymbol{\mathrm{m}},z;\mu_{\mathrm{out}}^2)\,\big|q(k_q'',h,j)\big\rangle+\dots\bigg]\,,
\end{align}
for the final-state quark $q(k_q,h_q,i)$ and gluon $g(k_g,\sigma,a)$, where the quark $k_q''$ carries the total momentum of the pair, $\boldsymbol{\mathrm{k}}_q''=\boldsymbol{\mathrm{q}}$, and the virtuality $\mu_{\mathrm{out}}^2$. By Eq.\,\eqref{eq:LCPT:merging} the merging wave function is $\Psi^{qg\to q}=-(\Psi^{q\to qg})^\dagger$ at the relative momentum $\boldsymbol{\mathrm{m}}=\boldsymbol{\mathrm{k}}_q-z\boldsymbol{\mathrm{q}}$ of the pair, Eq.\,\eqref{eq:l m}. In the bra state ${}^{\rm out}_{\ \,\rm D}\langle qg|$ this component therefore enters with $-\Psi^{q\to qg}(\boldsymbol{\mathrm{m}},z;\mu_{\mathrm{out}}^2)$. One finds (Appendix~\ref{sec:app:q real})
\begin{align}
\label{eq:q:real amplitude}
    \mathcal M^{q\gamma^*\to qg}_{h_q,\sigma;h}=\frac{e_fe^2}{\boldsymbol{\mathrm{q}}^2}\Big[\Psi^{q\to qg}_{h_q,\sigma;h}(\boldsymbol{\ell},z;\mu_{\mathrm{in}}^2)-\Psi^{q\to qg}_{h_q,\sigma;h}(\boldsymbol{\mathrm{m}},z;\mu_{\mathrm{out}}^2)\Big]\,,\qquad\boldsymbol{\mathrm{q}}=\boldsymbol{\mathrm{k}}_q+\boldsymbol{\mathrm{k}}_g\,,
\end{align}
with $z=k_q^+/p_q^+$. This amplitude again features the typical negative relative sign between the ``emission before'' and ``emission after'' contributions, which leads to a cancellation of the leading UV behavior between the two. In our case we will impose a constraint on the total ligh cone energy of the final state and the contributions would also be UV finite separately. It is a feature of the shockwave formalism that, due to the infinite light cone energy available from the target, in the absence of such a constraint on $q^-$ the terms would separately be UV divergent but combine into a UV finite cross section. The two terms have the same spinor structure and, with Eq.\,\eqref{eq:q:LCWF}, combine into
\begin{align}
\label{eq:q:real amplitude W}
    \mathcal M^{q\gamma^*\to qg}_{h_q,\sigma;h}=-g\,e_f\,\frac{e^2}{\boldsymbol{\mathrm{q}}^2}\,\frac{t^a_{ij}}{4p_q^+}\,\varepsilon^{*n}_\sigma\,\bar u_{h_q}(k_q)\,\gamma^+\Big[2(1+z)\,\delta^{mn}-(1-z)\,[\gamma^m,\gamma^n]\Big]u_h(p_q)\,\boldsymbol{\mathrm{W}}^m\,,
\end{align}
with
\begin{align}
\label{eq:q:W}
    \boldsymbol{\mathrm{W}}\equiv-\frac{\boldsymbol{\ell}}{\boldsymbol{\ell}^2+z(1-z)\mu_{\mathrm{in}}^2}+\frac{\boldsymbol{\mathrm{m}}}{\boldsymbol{\mathrm{m}}^2+z(1-z)\mu_{\mathrm{out}}^2}\,.
\end{align}
The first term of $\boldsymbol{\mathrm{W}}$ corresponds to initial-state and the second to final-state emission. 
The relative minus sign  between them  makes the amplitude vanish without momentum transfer: for $\boldsymbol{\mathrm{q}}=0$ one has $\boldsymbol{\mathrm{m}}=\boldsymbol{\ell}$, so that $\boldsymbol{\mathrm{W}}=0$ for $\mu_{\mathrm{out}}=\mu_{\mathrm{in}}$, where the in- and out-states coincide.
Averaging over the helicity and color of the incoming quark and summing over the final state,
\begin{align}
\label{eq:q:Msq}
    \overline{|\mathcal M^{q\gamma^*\to qg}|^2}=g^2e^4e_f^2\,C_\mathrm{F}\,\frac{z}{\boldsymbol{\mathrm{q}}^4}\Big[(1+z)^2+(1-z)^2(D-3)\Big]\boldsymbol{\mathrm{W}}^2\,.
\end{align}
With the two-body phase space $\mathrm{d}(\mathrm{P.S.})_{qg}=\frac{\mathrm{d}z}{4\pi z(1-z)}\frac{\mathrm{d}^{D-2}\boldsymbol{\mathrm{k}}_q}{(2\pi)^{D-2}}\frac{\mathrm{d}^{D-2}\boldsymbol{\mathrm{k}}_g}{(2\pi)^{D-2}}$, the factor $z$ cancels, and after changing variables from $\boldsymbol{\mathrm{k}}_g$ to $\boldsymbol{\mathrm{q}}$ and integrating over the azimuthal angle of the total momentum $\boldsymbol{\mathrm{q}}$, the partonic cross section in $D=4$ becomes
\begin{align}
\label{eq:q:real result}
    \frac{\mathrm{d}\hat\sigma^{q\gamma^*\to qg}}{\mathrm{d}z\,\mathrm{d}^2\boldsymbol{\mathrm{k}}_q\,\mathrm{d}Q^2}=\frac{\alpha_{\mathrm{e.m.}}^2}{Q^4}\,\frac{\alpha_\mathrm{s}}{2\pi}\,4e_f^2\,P_{q\leftarrow q}(z)\,\boldsymbol{\mathrm{W}}^2\,,\qquad P_{q\leftarrow q}(z)=C_\mathrm{F}\,\frac{1+z^2}{1-z}\,,
\end{align}
where $P_{q\leftarrow q}(z)$ denotes the real-emission splitting function for $z<1$. 

The real and virtual contributions are closely related. Since $\boldsymbol{\mathrm{m}}=\boldsymbol{\ell}+(1-z)\boldsymbol{\mathrm{q}}$ by Eq.\,\eqref{eq:l m}, $\boldsymbol{\mathrm{W}}$ is minus the square bracket of Eq.\,\eqref{eq:q:virtual square} with $\zeta\to z$. Integrating the real cross section Eq.\,\eqref{eq:q:real result} over the transverse momentum of the final-state quark at fixed $\boldsymbol{\mathrm{q}}$ therefore gives the same transverse integral as in the virtual correction, which is finite and was evaluated there. The result is (Appendix~\ref{sec:app:q real})
\begin{align}
\label{eq:q:real integrated}
    \frac{\mathrm{d}\hat\sigma^{q\gamma^*\to qg}}{\mathrm{d}z\,\mathrm{d}Q^2}=\sigma_0\,\frac{\alpha_\mathrm{s}}{2\pi}\,C_\mathrm{F}\,\frac{1+z^2}{1-z}\bigg[\frac{2\operatorname{artanh}\beta}{\beta}-2\bigg]\,,
\end{align}
with $\beta$ of Eq.\,\eqref{eq:q:beta} evaluated at $\zeta=z$. This is, for $z<1$, minus the integrand of the virtual correction \eqref{eq:q:virtual result}, which in turn sits at $z=1$. The sum of the two is
\begin{align}
\label{eq:q:R+V}
    \frac{\mathrm{d}\hat\sigma^{q\gamma^*\to qg}}{\mathrm{d}z\,\mathrm{d}Q^2}+\frac{\mathrm{d}\hat\sigma^{q\gamma^*\to q}_{\rm V}}{\mathrm{d}z\,\mathrm{d}Q^2}=\sigma_0\,\frac{\alpha_\mathrm{s}}{2\pi}\,C_\mathrm{F}\,\bigg[\frac{1+z^2}{1-z}\bigg(\frac{2\operatorname{artanh}\beta}{\beta}-2\bigg)\bigg]_+\,,
\end{align}
with $[f]_+\equiv f(z)-\delta(1-z)\int_0^1\mathrm{d}z'f(z')$. 

As we discussed in the case of the gluon initiated channel, expanding Eq. \eqref{eq:q:R+V} to leading logarithm in $\mu_{\mathrm{in}}/\mathbf{q}$ and $\mu_{\mathrm{out}}/\mathbf{q}$, we obtain the quark-quark splitting function $\mathcal P_{q\leftarrow q}(z)$  multiplying large logarithms related to initial and final state radiation. The off-shell regularization thus naturally leads to  collinear factorization already at the partonic level. However, as we have light-cone quantized the probe, the emergent splitting function is given naturally in terms of the momentum fraction of the probe $z$. To obtain the correct DIS and SIDIS structure functions, we need to change variables to the target momentum fraction. Indeed, the integral of Eq.\,\eqref{eq:q:R+V} over $z$ vanishes: at fixed photon momentum the $\mathcal O(\alpha_\mathrm{s})$ corrections only redistribute the final state in $z$, as required by unitarity. Instead, since in DIS and SIDIS the partonic variables $\hat{x}$ and $\hat{z}$ also depend on the gluon momentum, the integration over the parton's phase space has to be implemented at fixed $\hat{x}$ and $\hat{z}$, which leads to the cross section having also finite parts, the coefficient functions. This is done in the following sections.

\section{Inclusive DIS}
\label{sec:DIS}

Let us now see how integrating the partonic cross sections of Sec.~\ref{sec:NLO} over the final state at fixed $\hat{x}$ gives the reduced partonic structure functions of inclusive DIS, which using Eq.\,\eqref{eq:C from F} are  the coefficient functions in the off-shell scheme. In the gluon channel (Sec.~\ref{sec:gluon channel DIS}) there is only a collinear region related to initial-state splitting. In the quark channel (Sec.~\ref{sec:quark channel DIS}) there are also collinear regions from  final-state emissions  and the virtual correction contribute in addition. However, in inclusive DIS these  cancel at the endpoint $\hat{x}=1$ associated with final state splitting, since there are no fragmentation functions to absorb corresponding collinear divergences. We will also compare the coefficient functions with the $\overline{\mathrm{MS}}$ scheme, which gives the transformation of the parton distributions between the two schemes. Further details on the computations of this section are given in Appendix \ref{sec:app:DIS all}.

\subsection{Gluon channel}
\label{sec:gluon channel DIS}

The gluon contributes to inclusive DIS through the partonic cross section $\mathrm{d}^2\hat\sigma^{g\gamma^*\to X}/\mathrm{d}\hat{x}\,\mathrm{d}Q^2$ of Eq.\,\eqref{eq:fact:C DIS}, which we obtain from the differential cross section \eqref{eq:g:partonic xs} by integrating over the final state at fixed $\hat{x}$. To this end, we express the partonic Bjorken variable in terms of the final-state momenta. By Eqs.~\eqref{eq:l m} and \eqref{eq:collinear qminus}, with $z$ the plus-momentum fraction of the quark and $\boldsymbol{\mathrm{k}}=\boldsymbol{\mathrm{k}}_q$,
\begin{align}
\label{eq:g:xhat}
    \hat{x}(\boldsymbol{\mathrm{m}},z)=\frac{z(1-z)Q^2}{\boldsymbol{\mathrm{m}}^2+z(1-z)Q^2}\,,\qquad\boldsymbol{\mathrm{m}}=\boldsymbol{\mathrm{k}}_q-z\boldsymbol{\mathrm{q}}\,,\qquad\boldsymbol{\mathrm{k}}_{\bar q}=(1-z)\boldsymbol{\mathrm{q}}-\boldsymbol{\mathrm{m}}\,.
\end{align}
Note that $\hat{x}$ differs from the plus-momentum fraction $z$, which is integrated over. To obtain the cross section differentially in $\hat{x}$, we insert the definition of $\hat{x}$ with a delta function,
\begin{align}
\label{eq:g:DIS partonic}
    \frac{\mathrm{d}^2\hat\sigma^{g\gamma^*\to q\bar q}}{\mathrm{d}\hat{x}\,\mathrm{d}Q^2}=\int_0^1\mathrm{d}z\int\mathrm{d}^2\boldsymbol{\mathrm{k}}_q\,\delta\big(\hat{x}-\hat{x}(\boldsymbol{\mathrm{m}},z)\big)\,\frac{\mathrm{d}\hat\sigma^{g\gamma^*\to q\bar q}}{\mathrm{d}z\,\mathrm{d}^2\boldsymbol{\mathrm{k}}_q\,\mathrm{d}Q^2}\,.
\end{align}
At fixed $z$ we write $\mathrm{d}^2\boldsymbol{\mathrm{k}}_q=\mathrm{d}^2\boldsymbol{\mathrm{m}}=\frac12\mathrm{d}m_\perp^2\,\mathrm{d}\phi$, where $\phi$ is the angle between $\boldsymbol{\mathrm{m}}$ and $\boldsymbol{\mathrm{q}}$. The delta function fixes the magnitude $m_\perp$ and gives the Jacobian
\begin{align}
\label{eq:g:Jacobian}
    \int\mathrm{d}m_\perp^2\,\delta\big(\hat{x}-\hat{x}(\boldsymbol{\mathrm{m}},z)\big)=\frac{z(1-z)Q^2}{\hat{x}^2}\,,\qquad m_\perp^2=z(1-z)Q^2\,\frac{1-\hat{x}}{\hat{x}}\,,
\end{align}
so that only the integral over $\phi$ remains. With Eq.\,\eqref{eq:g:partonic xs} and $\sigma_0$ of Eq.\,\eqref{eq:sigma0},
\begin{align}
\label{eq:g:DIS sigma}
    \frac{\mathrm{d}^2\hat\sigma^{g\gamma^*\to q\bar q}}{\mathrm{d}\hat{x}\,\mathrm{d}Q^2}=\sigma_0\,\frac{\alpha_\mathrm{s}}{2\pi}\,\mathcal F_g(\hat{x};r_\mathrm{in})\,,\qquad r_\mathrm{in}\equiv\frac{\mu_{\mathrm{in}}^2}{Q^2}\,,
\end{align}
the reduced partonic cross section is
\begin{align}
\label{eq:g:Cg def}
    \mathcal F_g(\hat{x};r_\mathrm{in})\equiv\int_0^1\mathrm{d}z\,P_{q\leftarrow g}(z)\,\frac{z(1-z)Q^2}{\hat{x}^2}\int_0^{2\pi}\frac{\mathrm{d}\phi}{2\pi}\,\boldsymbol{\mathrm{V}}^2\Big|_{m_\perp^2=z(1-z)Q^2(1-\hat{x})/\hat{x}}\,.
\end{align}
The final state contains both the quark and the antiquark of one flavor $f$, so that Eq.\,\eqref{eq:g:DIS sigma} is the inclusive cross section of the off-shell gluon appearing in Eq.\,\eqref{eq:fact:C DIS}, and $\mathcal F_g$ is its reduced partonic structure function.

Denoting the relative momentum  in the splitting $g\to q\bar q$ before the shockwave by $\boldsymbol{\ell}$, the first term in $\boldsymbol{\mathrm{V}}$ (defined in Eq.~\eqref{eq:g:V} in  Sec.~\ref{sec:gluon channel partonic}) has $\boldsymbol{\ell}=-\boldsymbol{\mathrm{k}}_{\bar q}$ and the second one $\boldsymbol{\ell}=\boldsymbol{\mathrm{k}}_q$. The collinear singularities in the limit $\boldsymbol{\ell}\to0$ are both regulated by $\mu_{\mathrm{in}}$. 
Using Eq.\,\eqref{eq:collinear} we see that the collinear region $\boldsymbol{\ell}\to0$ lies at $\hat{x}=z$  for the first term and at $\hat{x}=1-z$ for the second one.\footnote{In Sec.~\ref{sec:offshell} $z$ was defined to be the momentum fraction of the parton that absorbs the photon, which is the quark for the first term and the antiquark for the second term. Now we define $z$ to be the momentum fraction of the quark, so the collinear region for the second term has $z$ replaced with $1-z$ in Eq.\,\eqref{eq:collinear}.} There is no final-state splitting, so that no wave function depends on $\boldsymbol{\mathrm{m}}$, and $\boldsymbol{\mathrm{m}}\to0$, i.e.\ $\hat{x}\to1$, is not singular. With no virtual correction at this order, $\mathcal F_g$ is therefore regular at $\hat{x}=1$, and no endpoint treatment is needed. 

The angular integral in Eq.\,\eqref{eq:g:Cg def} can be done exactly, and the remaining $z$ integral is evaluated in the limit $r_\mathrm{in}\to0$ in Appendix~\ref{sec:app:DIS}. The result is
\begin{align}
\label{eq:g:Fg}
    \mathcal F_g(\hat{x};r_\mathrm{in})=2P_{q\leftarrow g}(\hat{x})\ln\frac{Q^2}{\mu_{\mathrm{in}}^2}-4P_{q\leftarrow g}(\hat{x})\ln\hat{x}+8\hat{x}(1-\hat{x})-2+\mathcal O\big(r_\mathrm{in}\ln(1/r_\mathrm{in})\big)\,.
\end{align}
The coefficient of the collinear logarithm is the splitting function, as required by DGLAP evolution of the quark distribution. After the change of variables from $z$ to $\hat{x}$ there is now also a finite term that is not proportional to the collinear logarithm, which includes a part that is not proportional to the splitting function. By setting the off-shellness equal to the factorization scale (see Eq.\,\eqref{eq:C from F}), the reduced partonic structure function becomes the coefficient function
\begin{align}
\label{eq:g:Cgoff}
    C_g^{\rm off}(\hat{x})=2P_{q\leftarrow g}(\hat{x})\ln\frac{Q^2}{\mu_F^2}-4P_{q\leftarrow g}(\hat{x})\ln\hat{x}+8\hat{x}(1-\hat{x})-2\,.
\end{align}
This is the main result of this section, the gluon channel DIS coefficient function in the off-shell scheme.

The corresponding $\overline{\mathrm{MS}}$ coefficient function, at the same factorization scale and normalized in the same way (i.e.\ multiplying $\sum_fe_f^2$ and including both the $q$ and $\bar q$ contributions), is \cite{Ellis:1996mzs}
\begin{align}
\label{eq:g:CgMS}
    C_g^{\overline{\mathrm{MS}}}(\hat{x})=2P_{q\leftarrow g}(\hat{x})\ln\bigg(\frac{Q^2}{\mu_F^2}\,\frac{1-\hat{x}}{\hat{x}}\bigg)+8\hat{x}(1-\hat{x})-1\,.
\end{align}
The collinear logarithms of Eqs.~\eqref{eq:g:Cgoff} and \eqref{eq:g:CgMS} coincide, and the difference between the two schemes is
\begin{align}
\label{eq:g:DeltaCg}
    \Delta C_g(\hat{x})\equiv C_g^{\rm off}(\hat{x})-C_g^{\overline{\mathrm{MS}}}(\hat{x})=-2P_{q\leftarrow g}(\hat{x})\ln\big[\hat{x}(1-\hat{x})\big]-1\,,
\end{align}
which is symmetric under $\hat{x}\to1-\hat{x}$.

Since the structure functions are scheme independent, the two schemes are related by a redefinition of the parton distributions (Appendix~\ref{sec:app:PDF}). We denote by $f^{\rm S}_{j/T}$ the distributions in the scheme ${\rm S}={\rm off},\overline{\mathrm{MS}}$, with the target $T$ being either the proton, $T=p$, or the off-shell gluon, $T=g^*(\mu_{\mathrm{in}})$. As shown in Eq.\,\eqref{eq:F from C} the scheme-independent reduced partonic structure function is $\mathcal F_g=C_g^{\rm S}+(2\pi/\alpha_\mathrm{s})\,2f^{\rm S}_{q/g^*(\mu_{\mathrm{in}})}(\hat{x},\mu_F^2)$ in either scheme. The form  of the off-shell scheme quark PDF in a gluon $f^{\rm off}_{q/g^*(\mu_{\mathrm{in}})}$ obtained in Eq.\,\eqref{eq:offPDF} is simple, consisting only  of the renormalization condition and the splitting function. The corresponding distribution in the
$\overline{\mathrm{MS}}$ scheme, obtained by performing the UV renormalization of the operator definition  as discussed in Appendix~\ref{sec:app:PDF} (see Eq.~\eqref{eq:msbarbareqing}), is
\begin{align}
\label{eq:g:fqg}
    f^{\overline{\mathrm{MS}}}_{q/g^*(\mu_{\mathrm{in}})}(x,\mu_F^2)=\frac{\alpha_\mathrm{s}}{2\pi}\bigg\{P_{q\leftarrow g}(x)\ln\frac{\mu_F^2}{\mu_{\mathrm{in}}^2x(1-x)}-\frac12\bigg\}\,,
\end{align}
The constant $-1/2$ comes from the $\mathcal O(\varepsilon)$ part of the $D$-dimensional polarization average in Eq.\,\eqref{eq:g:Msq}.
Since $\mathcal F_g$ is the same in both schemes, Eq.\,\eqref{eq:F from C} leads to a consistency relation
\begin{align}
\label{eq:g:Delta fqg}
    \Delta f_{q/g^*}(x)\equiv f^{\rm off}_{q/g^*(\mu_{\mathrm{in}})}(x,\mu_F^2)-f^{\overline{\mathrm{MS}}}_{q/g^*(\mu_{\mathrm{in}})}(x,\mu_F^2)=-\frac{\alpha_\mathrm{s}}{2\pi}\,\frac12\Delta C_g(x)\,,
\end{align}
independent of $\mu_{\mathrm{in}}$ and $\mu_F$. This consistency condition is indeed satisfied, as is clear from a comparison of Eqs.~\eqref{eq:g:DeltaCg} and \eqref{eq:g:fqg} which originate in separate calculations of, on the one hand, $\Delta C_g(x)$  in this section and, on the other hand, $\Delta f_{q/g^*}(x)$ in Appendix~\ref{sec:app:PDF}. The factor $\frac12$ in front of  $\Delta C_g(x)$ can be interpreted as the gluonic contribution being shifted in equal parts to the quark and antiquark distributions  in the scheme transformation,  preserving the total cross section.  For the quark PDF of the proton, every gluon in the proton contributes the same difference, weighted with its distribution $f_{g/p}$. This gives the gluon contribution to the scheme transformation~\eqref{eq:q:PDF shift} for the quark distribution in the proton.

\subsection{Quark channel}
\label{sec:quark channel DIS}

The quark contributes to inclusive DIS through the partonic cross section $\mathrm{d}^2\hat\sigma^{q\gamma^*\to X}/\mathrm{d}\hat{x}\,\mathrm{d}Q^2$ of Eq.\,\eqref{eq:fact:C DIS}. At $\mathcal O(\alpha_\mathrm{s})$ it receives NLO corrections from real emission \eqref{eq:q:real result}, with the $qg$ final state, and the virtual correction \eqref{eq:q:virtual result}, with a single quark in the final state. We calculate these keeping the initial and final virtualities independent, with $r_\mathrm{out}\equiv\mu_{\mathrm{out}}^2/Q^2$.

As in the gluon channel, $\hat{x}(\boldsymbol{\mathrm{m}},z)$ is given by Eq.\,\eqref{eq:g:xhat}, with $z$ the plus-momentum fraction of the final-state quark and $\boldsymbol{\mathrm{k}}_g$ in place of $\boldsymbol{\mathrm{k}}_{\bar q}$. The two wave functions in the amplitude Eq.\,\eqref{eq:q:real amplitude} depend on the relative momentum of the initial-state emission, $\boldsymbol{\ell}=-\boldsymbol{\mathrm{k}}_g$, measured from the incoming quark, and of the final-state emission, $\boldsymbol{\mathrm{m}}$, measured from the outgoing quark. We now know from Eq.\,\eqref{eq:collinear} that the initial-state collinear region $\boldsymbol{\mathrm{k}}_g\to0$ lies at $\hat{x}=z$, at every $\hat{x}$, as in the gluon channel, whereas $\boldsymbol{\mathrm{m}}\to0$ is now the final-state collinear region, at the endpoint $\hat{x}=1$. Since the one-quark final state of the virtual correction has $\hat{x}=1$ up to power corrections, as at leading order, we write
\begin{align}
\label{eq:q:DIS sigma}
    \frac{\mathrm{d}^2\hat\sigma^{q\gamma^*\to X}}{\mathrm{d}\hat{x}\,\mathrm{d}Q^2}&=\sigma_0\Big[\delta(1-\hat{x})+\frac{\alpha_\mathrm{s}}{2\pi}\,\mathcal F_q(\hat{x};r_\mathrm{in},r_\mathrm{out})\Big]\,,\nonumber\\
    \mathcal F_q(\hat{x};r_\mathrm{in},r_\mathrm{out})&=H_q(\hat{x};r_\mathrm{in},r_\mathrm{out})+V_q(r_\mathrm{in},r_\mathrm{out})\,\delta(1-\hat{x})\,.
\end{align}
The parton level real cross section was computed in  Eq.\,\eqref{eq:q:real result}, and with the Jacobian \eqref{eq:g:Jacobian} it becomes
\begin{align}
\label{eq:q:H def}
    H_q(\hat{x};r_\mathrm{in},r_\mathrm{out})\equiv\int_0^1\mathrm{d}z\,P_{q\leftarrow q}(z)\,\frac{z(1-z)Q^2}{\hat{x}^2}\int_0^{2\pi}\frac{\mathrm{d}\phi}{2\pi}\,\boldsymbol{\mathrm{W}}^2\Big|_{m_\perp^2=z(1-z)Q^2(1-\hat{x})/\hat{x}}\,,
\end{align}
with $\boldsymbol{\mathrm{W}}$ from Eq.\eqref{eq:q:W}. 
The virtual correction from  \eqref{eq:q:virtual result} is given by
\begin{align}
\label{eq:q:V}
    V_q(r_\mathrm{in},r_\mathrm{out})=C_\mathrm{F}\int_0^1\mathrm{d}\zeta\,\frac{1+\zeta^2}{1-\zeta}\bigg[2-\frac{2\operatorname{artanh}\beta}{\beta}\bigg]\,,
\end{align}
with $\beta$ of Eq.\,\eqref{eq:q:beta} and $\tilde{\boldsymbol{\mathrm{q}}}^2=(1-\zeta)^2Q^2$.

For $\hat{x}<1$ only the real emission contributes, and in the limit $r_\mathrm{in}\to0$ we find (Appendix~\ref{sec:app:q DIS})
\begin{align}
\label{eq:q:Hbulk}
    H_q^{\rm bulk}(\hat{x})=P_{q\leftarrow q}(\hat{x})\bigg[\ln\frac{Q^2}{\mu_{\mathrm{in}}^2}-2\ln\hat{x}-1\bigg]+C_\mathrm{F}\,\frac{\frac12+\hat{x}-3\hat{x}^2}{1-\hat{x}}\,,\qquad\hat{x}<1\,.
\end{align}
As for  the gluon channel, the coefficient of the collinear logarithm is the splitting function, here originating from the initial-state singularity at $z=\hat{x}$. The final-state collinear region, the soft-gluon region and the virtual correction all sit at $1-\hat{x}=\mathcal O(r_\mathrm{in})$. However, they do  not need to be computed separately but can be reconstructed from a unitarity argument.  We recall that integrating Eq.\,\eqref{eq:q:R+V} over $z$ gives zero, i.e.\ the total $\mathcal O(\alpha_\mathrm{s})$ correction at fixed $Q^2$ vanishes. Thus we know that
\begin{align}
\label{eq:q:DIS unitarity}
    \int_0^1\mathrm{d}\hat{x}\,\mathcal F_q(\hat{x};r_\mathrm{in},r_\mathrm{out})=0\,.
\end{align}
Acting on a smooth test function, the contributions at $1-\hat{x}=\mathcal O(r_\mathrm{in})$ enter only through its value at $\hat{x}=1$. Thus the whole reduced partonic structure function, which according to Eq.\,\eqref{eq:C from F} is the same as the quark coefficient function in the off-shell scheme, can be obtained from Eq.~\eqref{eq:q:Hbulk} by subtracting a contribution proportional to $\delta(1-\hat{x})$, with a coefficient determined by the constraint~\eqref{eq:q:DIS unitarity}. The reduced partonic structure function is therefore the plus distribution of the bulk result,
\begin{align}
\label{eq:q:Fq}
    \mathcal F_q(\hat{x};r_\mathrm{in},r_\mathrm{out})=\big[H_q^{\rm bulk}(\hat{x})\big]_++\mathcal O\big(r_\mathrm{in}\ln(1/r_\mathrm{in})\big)\,,
\end{align}
independent of $\mu_{\mathrm{out}}$. The resulting coefficient function is
\begin{align}
\label{eq:q:Cqoff}
    C_q^{\rm off}(\hat{x})&=\mathcal P_{q\leftarrow q}(\hat{x})\ln\frac{Q^2}{\mu_F^2}+C_\mathrm{F}\bigg[-2\,\frac{1+\hat{x}^2}{1-\hat{x}}\ln\hat{x}-\frac{7}{2}\,\frac{1}{(1-\hat{x})_+}\nonumber\\
    &\quad+3+4\hat{x}-\bigg(\frac52+\frac{2\pi^2}{3}\bigg)\delta(1-\hat{x})\bigg]\,,
\end{align}
noting that  $\int_0^1\mathrm{d}\hat{x}\,\mathcal P_{q\leftarrow q}(\hat{x})=0$.  The constraint $\int_0^1\mathrm{d}\hat{x}\,C_q^{\rm off}(\hat{x})=0$ satisfied by Eq.~\eqref{eq:q:Cqoff} can also be interpreted as quark-number conservation.

The corresponding $\overline{\mathrm{MS}}$ coefficient function at the same factorization scale, normalized in the same way, is \cite{Ellis:1996mzs}
\begin{align}
\label{eq:q:CqMS}
    C_q^{\overline{\mathrm{MS}}}(\hat{x})&=\mathcal P_{q\leftarrow q}(\hat{x})\ln\frac{Q^2}{\mu_F^2}+C_\mathrm{F}\bigg[2\bigg(\frac{\ln(1-\hat{x})}{1-\hat{x}}\bigg)_+-\frac32\,\frac{1}{(1-\hat{x})_+}-(1+\hat{x})\ln(1-\hat{x})\nonumber\\
    &\quad-\frac{1+\hat{x}^2}{1-\hat{x}}\ln\hat{x}+3+2\hat{x}-\bigg(\frac92+\frac{\pi^2}{3}\bigg)\delta(1-\hat{x})\bigg]\,.
\end{align}
The collinear logarithms of Eqs.~\eqref{eq:q:Cqoff} and \eqref{eq:q:CqMS} coincide. For $\hat{x}<1$ the terms with $\ln(1-\hat{x})$ in Eq.\,\eqref{eq:q:CqMS} combine into $P_{q\leftarrow q}(\hat{x})\ln(1-\hat{x})$, and since both coefficient functions integrate to zero, the difference between the two schemes is
\begin{align}
\label{eq:q:DeltaCq}
    \Delta C_q(\hat{x})&\equiv C_q^{\rm off}(\hat{x})-C_q^{\overline{\mathrm{MS}}}(\hat{x})\nonumber\\
    &=\Big[-P_{q\leftarrow q}(\hat{x})\big(\ln\big[\hat{x}(1-\hat{x})\big]+1\big)-C_\mathrm{F}(1-\hat{x})\Big]_+\nonumber\\
    &=C_\mathrm{F}\bigg[-\frac{1+\hat{x}^2}{1-\hat{x}}\ln\hat{x}-(1+\hat{x}^2)\bigg(\frac{\ln(1-\hat{x})}{1-\hat{x}}\bigg)_+-\frac{1+\hat{x}^2}{(1-\hat{x})_+}\nonumber\\
    &\quad-(1-\hat{x})+\bigg(2-\frac{\pi^2}{3}\bigg)\delta(1-\hat{x})\bigg]\,.
\end{align}
As in the gluon channel, the difference is related to the scheme dependence of the parton distribution of the off-shell target (Appendix~\ref{sec:app:q PDF}). Using Eq.\,\eqref{eq:F from C}, $\mathcal F_q=C_q^{\rm S}+(2\pi/\alpha_\mathrm{s})\big[f^{\rm S}_{q/q^*(\mu_{\mathrm{in}})}(\hat{x},\mu_F^2)-\delta(1-\hat{x})\big]$ in either scheme, with $f^{\rm off}_{q/q^*(\mu_{\mathrm{in}})}$ given by Eq.\,\eqref{eq:offPDF}. In the $\overline{\mathrm{MS}}$ scheme (see Appendix~\ref{sec:app:q PDF}),
\begin{align}
\label{eq:q:fqq}
    f^{\overline{\mathrm{MS}}}_{q/q^*(\mu_{\mathrm{in}})}(x,\mu_F^2)=\delta(1-x)+\frac{\alpha_\mathrm{s}}{2\pi}\bigg\{P_{q\leftarrow q}(x)\bigg[\ln\frac{\mu_F^2}{\mu_{\mathrm{in}}^2x(1-x)}-1\bigg]-C_\mathrm{F}(1-x)\bigg\}_+\,,
\end{align}
where $-P_{q\leftarrow q}(x)$ comes from the factor $(1-\varepsilon)$ of the transverse momentum integral, $-C_\mathrm{F}(1-x)$ from the $\mathcal O(\varepsilon)$ part of the $D$-dimensional polarization sum in Eq.\,\eqref{eq:q:Msq}, and the plus prescription from the normalization factor $Z_q^{\rm in}$, which ensures quark number conservation. Since $\mathcal F_q$ is the same in both schemes the scheme dependent parts of the coefficient functions and PDFs are related by
\begin{align}
\label{eq:q:Delta fqq}
    \Delta f_{q/q^*}(x)\equiv f^{\rm off}_{q/q^*(\mu_{\mathrm{in}})}(x,\mu_F^2)-f^{\overline{\mathrm{MS}}}_{q/q^*(\mu_{\mathrm{in}})}(x,\mu_F^2)=-\frac{\alpha_\mathrm{s}}{2\pi}\,\Delta C_q(x)\,,
\end{align}
which Eqs.~\eqref{eq:offPDF}, \eqref{eq:q:DeltaCq} and \eqref{eq:q:fqq} indeed satisfy, again independently of $\mu_{\mathrm{in}}$ and $\mu_F$. For the quark PDF in a proton, every quark contributes the same factor, weighted with its distribution $f_{q/p}$. Together with the gluon term of Sec.~\ref{sec:gluon channel DIS} the quark distributions of the proton in the two schemes are related by
\begin{align}
\label{eq:q:PDF shift}
    f^{\rm off}_{q/p}(x,\mu_F^2)=f^{\overline{\mathrm{MS}}}_{q/p}(x,\mu_F^2)-\frac{\alpha_\mathrm{s}}{2\pi}\int_x^1\frac{\mathrm{d}\xi}{\xi}\bigg[f_{q/p}\Big(\frac x\xi,\mu_F^2\Big)\,\Delta C_q(\xi)+f_{g/p}\Big(\frac x\xi,\mu_F^2\Big)\,\frac12\Delta C_g(\xi)\bigg]\,,
\end{align}
and likewise for the antiquark.

We have now computed the coefficient functions for DIS at NLO for both the quark and gluon channels in the off-shell scheme. We have also verified that, when the definitions of the PDFs are transformed to the  $\overline{\mathrm{MS}}$ scheme, we recover the standard results in the literature. Let us now move to performing a similar analysis for SIDIS.

\section{SIDIS}
\label{sec:SIDIS}

In SIDIS the momentum fraction $\hat{z}$ of the observed parton is fixed, in addition to $\hat{x}$. We recall from  Eq.\,\eqref{eq:collinear} that the initial-state collinear regions lie at $\hat{z}=1$ and the final-state ones at $\hat{x}=1$. Thus  the parton distribution logarithms appear at $\hat{z}=1$ and the fragmentation function logarithms at $\hat{x}=1$. The gluon channel (Sec.~\ref{sec:gluon channel SIDIS}) only has the former; the quark channel (Sec.~\ref{sec:quark channel SIDIS}) has both. The comparison with the $\overline{\mathrm{MS}}$ scheme will also involve the transformation of the fragmentation functions, in addition to that of the parton distributions. Additional calculational details for the calculations in this section are given in Appendix \ref{sec:app:SIDIS all}.

\subsection{Gluon channel}
\label{sec:gluon channel SIDIS}

The gluon channel contributes through the partonic cross section $\mathrm{d}^3\hat\sigma^{g\gamma^*\to qX}/\mathrm{d}\hat{x}\,\mathrm{d}\hat{z}\,\mathrm{d}Q^2$ of Eq.\,\eqref{eq:fact:C SIDIS}, with the quark of $g\to q\bar q$ observed (the case of the tagged antiquark is analogous). Only the phase-space integration differs from inclusive DIS. With Eq.\,\eqref{eq:DIS:xhat} and the expressions for $k_q^-$ and $q^-$ given in Sec.~\ref{sec:gluon channel DIS},
\begin{align}
\label{eq:g:zhat}
    \hat{z}(\boldsymbol{\mathrm{m}},z)=\frac{k_q^-}{q^-}=\frac{\hat{x}\,\boldsymbol{\mathrm{k}}_q^2}{zQ^2}\,,\qquad\boldsymbol{\mathrm{k}}_q=\boldsymbol{\mathrm{m}}+z\boldsymbol{\mathrm{q}}\,.
\end{align}
Inserting the definitions of both $\hat{x}$ and $\hat{z}$ with delta functions we start from
\begin{align}
\label{eq:g:Coff def}
    \frac{\mathrm{d}^3\hat\sigma^{g\gamma^*\to q\bar q}}{\mathrm{d}\hat{x}\,\mathrm{d}\hat{z}\,\mathrm{d}Q^2}&=\int_0^1\mathrm{d}z\int\mathrm{d}^2\boldsymbol{\mathrm{k}}_q\,\delta\big(\hat{x}-\hat{x}(\boldsymbol{\mathrm{m}},z)\big)\,\delta\big(\hat{z}-\hat{z}(\boldsymbol{\mathrm{m}},z)\big)\,\frac{\mathrm{d}\hat\sigma^{g\gamma^*\to q\bar q}}{\mathrm{d}z\,\mathrm{d}^2\boldsymbol{\mathrm{k}}_q\,\mathrm{d}Q^2}\nonumber\\
    &\equiv\sigma_0\,\frac{\alpha_\mathrm{s}}{2\pi}\,\mathcal F_{qg}(\hat{x},\hat{z};r_\mathrm{in})\,,
\end{align}
which, according to Eq.\,\eqref{eq:fact:C SIDIS}, defines the reduced partonic structure function $\mathcal F_{qg}$ of the off-shell gluon. With Eq.\,\eqref{eq:g:partonic xs},
\begin{align}
\label{eq:g:Coff m}
    \mathcal F_{qg}(\hat{x},\hat{z};r_\mathrm{in})=\frac1\pi\int_0^1\mathrm{d}z\,P_{q\leftarrow g}(z)\int\mathrm{d}^2\boldsymbol{\mathrm{m}}\,\delta\big(\hat{x}-\hat{x}(\boldsymbol{\mathrm{m}},z)\big)\,\delta\big(\hat{z}-\hat{z}(\boldsymbol{\mathrm{m}},z)\big)\,\boldsymbol{\mathrm{V}}^2\,,
\end{align}
with $\boldsymbol{\mathrm{V}}$ given by Eq.~\eqref{eq:g:V}.

As in inclusive DIS, we do the the $m_\perp^2$ integral  with the first delta function, Eq.\,\eqref{eq:g:Jacobian} that related the total invariant mass of the final state to the partonic Bjorken variable $\hat{x}$. Now the constraint on $\hat{z}$ depends on the angle $\phi$ between the integration variable $\boldsymbol{\mathrm{m}}$ and the fixed momentum $\boldsymbol{\mathrm{q}}$, 
\begin{align}
\label{eq:g:zhat phi}
    \hat{z}=\hat{x}z+(1-\hat{x})(1-z)+2\sqrt{\hat{x}(1-\hat{x})z(1-z)}\,\cos\phi\,.
\end{align}
Thus the second delta function fixes $\phi$ (which for the inclusive DIS case was integrated over). Owing to these two constraints the transverse momenta are determined by $(\hat{x},\hat{z},z)$, $\boldsymbol{\mathrm{k}}_q^2=z\hat{z}Q^2/\hat{x}$ and $\boldsymbol{\mathrm{k}}_{\bar q}^2=(1-z)(1-\hat{z})Q^2/\hat{x}$. Changing the remaining integration variable from $z$ to an angle $\psi$ (Appendix~\ref{sec:app:SIDIS}), we obtain the compact representation
\begin{align}
\label{eq:g:Coff psi}
    \mathcal F_{qg}(\hat{x},\hat{z};r_\mathrm{in})=\frac{1}{\pi\hat{x}^2}\int_0^\pi\mathrm{d}\psi\,\Big[P_{q\leftarrow g}(z)\,z(1-z)Q^2\,\boldsymbol{\mathrm{V}}^2\Big]_{z=z_0+\delta\cos\psi}\,,
\end{align}
with $z_0=\hat{x}\hat{z}+(1-\hat{x})(1-\hat{z})$ and $\delta=2\sqrt{\hat{x}(1-\hat{x})\hat{z}(1-\hat{z})}$. By Eq.\,\eqref{eq:collinear}, the initial state collinear region $\boldsymbol{\mathrm{k}}_{\bar q}\to0$, where the tagged quark absorbs the photon, lies at $\hat{z}=1$, while the region $\boldsymbol{\mathrm{k}}_q\to0$, where the tagged quark itself is collinear to the target, has $\hat{z}=0$ and lies outside the current fragmentation region $\hat{z}\geq z_H>0$. We therefore drop the incoming state regulator for $\hat{z}<1$ and keep it only near $\hat{z}=1$, where the endpoint region of width $\hat{x}^2r_\mathrm{in}$ regularizes the pole in $1/(1-\hat{z})$. Combining the two regions  gives $\mathcal F_{qg}$, in which the collinear logarithm $\ln(Q^2/\mu_{\mathrm{in}}^2)$ multiplies $\delta(1-\hat{z})\,P_{q\leftarrow g}(\hat{x})$, up to corrections of $\mathcal O(r_\mathrm{in}\ln(1/r_\mathrm{in}))$. After evaluating the integral~\eqref{eq:g:Coff psi}  (see Appendix~\ref{sec:app:SIDIS} for the calculation) we get the partonic structure function, which by Eq.\,\eqref{eq:C from F} is the coefficient function in the off-shell scheme, as 
\begin{align}
\label{eq:g:Coff result}
    C^{\rm off}_{qg}(\hat{x},\hat{z})&=\delta(1-\hat{z})\,P_{q\leftarrow g}(\hat{x})\bigg[\ln\frac{Q^2}{\mu_F^2}-2\ln\hat{x}-1\bigg]\nonumber\\
    &\quad+P_{q\leftarrow g}(\hat{x})\bigg[\frac{1}{(1-\hat{z})_+}+\frac1{\hat{z}}-2\bigg]+4\hat{x}(1-\hat{x})\,.
\end{align}

The corresponding $\overline{\mathrm{MS}}$ coefficient function at the same factorization scale is given by Eqs.~(C.4) and (C.7) of Ref.~\cite{deFlorian:1997zj} (see also Eqs.~(33), (34) there),
\begin{align}
    C^{\overline{\mathrm{MS}}}_{qg}(\hat{x},\hat{z})&=\delta(1-\hat{z})\bigg[P_{q\leftarrow g}(\hat{x})\ln\bigg(\frac{Q^2}{\mu_F^2}\,\frac{1-\hat{x}}{\hat{x}}\bigg)+\hat{x}(1-\hat{x})\bigg]\nonumber\\
    &\quad+P_{q\leftarrow g}(\hat{x})\bigg[\frac{1}{(1-\hat{z})_+}+\frac1{\hat{z}}-2\bigg]+4\hat{x}(1-\hat{x})\,.\label{eq:g:CqgMS}
\end{align}
Integrating over $\hat{z}$ with weight $\hat{z}$ and summing over the tagged $q$ and $\bar q$ reproduces the inclusive coefficient function $C_g^{\overline{\mathrm{MS}}}(\hat{x})$ of Eq.\,\eqref{eq:g:CgMS}, as required by the momentum sum rule of the fragmentation functions.

The terms of Eqs.~\eqref{eq:g:Coff result} and \eqref{eq:g:CqgMS} that are not proportional to $\delta(1-\hat{z})$ coincide. The two schemes differ only in the initial state collinear region $\hat{z}=1$,
\begin{align}
\label{eq:g:DeltaCqg}
    C^{\rm off}_{qg}(\hat{x},\hat{z})-C^{\overline{\mathrm{MS}}}_{qg}(\hat{x},\hat{z})=\Big\{-P_{q\leftarrow g}(\hat{x})\ln\big[\hat{x}(1-\hat{x})\big]-\frac12\Big\}\,\delta(1-\hat{z})=\frac12\Delta C_g(\hat{x})\,\delta(1-\hat{z})\,,
\end{align}
where $P_{q\leftarrow g}(\hat{x})+\hat{x}(1-\hat{x})=1/2$ was used. This is the gluon term of the shift \eqref{eq:q:PDF shift} of the quark distribution, which enters SIDIS through the leading-order quark channel coefficient function $\delta(1-\hat{x})\delta(1-\hat{z})$; the factor $1/2$ reflects that $C_g$ contains both the $q$ and $\bar q$ contributions of one flavor, whereas $C_{qg}$ refers to a single tagged quark. The scheme difference is localized at $\boldsymbol{\mathrm{k}}_{\bar q}\to0$, the only region in which the two regularizations differ. Since the gluon channel does not have a final state collinear logarithm at this order, it does not contribute to a scheme transformation of the fragmentation functions.

\subsection{Quark channel}
\label{sec:quark channel SIDIS}

The quark channel contributes through the partonic cross sections $\mathrm{d}^3\hat\sigma^{q\gamma^*\to qX}$ and $\mathrm{d}^3\hat\sigma^{q\gamma^*\to gX}$ of Eq.\,\eqref{eq:fact:C SIDIS}, in which the quark or the gluon fragments into the observed hadron. The real emission \eqref{eq:q:real result} is unchanged, and as in the gluon channel only the phase-space integration differs. The momentum fraction $\hat{z}(\boldsymbol{\mathrm{m}},z)$ of the tagged quark is given by Eq.\,\eqref{eq:g:zhat}, with $z$ now the plus-momentum fraction of the final-state quark, and the gluon carries $1-\hat{z}(\boldsymbol{\mathrm{m}},z)$, since $k_q^-+k_g^-=q^-$. Inserting both delta functions, the reduced partonic structure functions \eqref{eq:fact:C SIDIS} of the off-shell quark are
\begin{align}
\label{eq:q:Cqq def}
    \mathcal F_{qq}(\hat{x},\hat{z};r_\mathrm{in},r_\mathrm{out})&=\frac1\pi\int_0^1\mathrm{d}z\,P_{q\leftarrow q}(z)\int\mathrm{d}^2\boldsymbol{\mathrm{m}}\,\delta\big(\hat{x}-\hat{x}(\boldsymbol{\mathrm{m}},z)\big)\,\delta\big(\hat{z}-\hat{z}(\boldsymbol{\mathrm{m}},z)\big)\,\boldsymbol{\mathrm{W}}^2\nonumber\\
    &\quad+V_q(r_\mathrm{in},r_\mathrm{out})\,\delta(1-\hat{x})\,\delta(1-\hat{z})\,,\nonumber\\
    \mathcal F_{gq}(\hat{x},\hat{z};r_\mathrm{in},r_\mathrm{out})&=\frac1\pi\int_0^1\mathrm{d}z\,P_{q\leftarrow q}(z)\int\mathrm{d}^2\boldsymbol{\mathrm{m}}\,\delta\big(\hat{x}-\hat{x}(\boldsymbol{\mathrm{m}},z)\big)\,\delta\big(1-\hat{z}-\hat{z}(\boldsymbol{\mathrm{m}},z)\big)\,\boldsymbol{\mathrm{W}}^2\,,
\end{align}
with $\boldsymbol{\mathrm{W}}$ given by Eq.~\eqref{eq:q:W}. 
The virtual correction, with a single quark in the final state, contributes only to $\mathcal F_{qq}$ at $\hat{x}=\hat{z}=1$. We recall from Eq.\,\eqref{eq:collinear} that for the tagged quark the initial-state collinear region $\boldsymbol{\mathrm{k}}_g\to0$ lies at $\hat{z}=1$ and the final-state one $\boldsymbol{\mathrm{m}}\to0$ at $\hat{x}=1$. For a tagged gluon the region $\boldsymbol{\mathrm{k}}_g\to0$ has $\hat{z}=0$ and lies outside the current fragmentation region while $\boldsymbol{\mathrm{m}}\to0$ again lies at $\hat{x}\to1$: this channel only has an outgoing state collinear logarithm. The regulators are dropped outside the collinear regions; in particular, the soft-gluon limit $z\to1$ at fixed $\hat{x}<1$, which lies at $\hat{z}=\hat{x}$, is not collinear. Similarly to the quark channel for inclusive DIS, the coefficient of $\delta(1-\hat{x})\delta(1-\hat{z})$, which contains the virtual correction, follows from the conservation of the quark number, $\int_0^1\mathrm{d}\hat{z}\,\mathcal F_{qq}(\hat{x},\hat{z};r_\mathrm{in},r_\mathrm{out})=\mathcal F_q(\hat{x};r_\mathrm{in},r_\mathrm{out})$. As a result of this structure the initial-state collinear logarithm $\ln(Q^2/\mu_{\mathrm{in}}^2)$ multiplies $\delta(1-\hat{z})$ and the final-state one $\ln(Q^2/\mu_{\mathrm{out}}^2)$ multiplies $\delta(1-\hat{x})$, up to corrections of $\mathcal O(r_{\rm in,out}\ln(1/r_{\rm in,out}))$. 
The detailed calculation of  $\mathcal F_{qq}$ and $\mathcal F_{gq}$ is shown in Appendix~\ref{sec:app:q SIDIS}, and again directly gives the coefficient functions by Eq.\,\eqref{eq:C from F}. The resulting  coefficient functions in the off-shell scheme are
\begin{align}
\label{eq:q:Cqq result}
    C^{\rm off}_{qq}(\hat{x},\hat{z})&=\delta(1-\hat{z})\bigg\{\mathcal P_{q\leftarrow q}(\hat{x})\ln\frac{Q^2}{\mu_F^2}-C_\mathrm{F}\bigg[2\,\frac{1+\hat{x}^2}{1-\hat{x}}\ln\hat{x}+\frac{1+\hat{x}^2}{(1-\hat{x})_+}\bigg]\bigg\}\nonumber\\
    &\quad+\delta(1-\hat{x})\bigg\{\mathcal P_{q\leftarrow q}(\hat{z})\ln\frac{Q^2}{\mu_{D}^2}-C_\mathrm{F}\,\frac{1+\hat{z}^2}{(1-\hat{z})_+}\bigg\}\nonumber\\
    &\quad+C_\mathrm{F}\bigg[\frac{2}{(1-\hat{x})_+(1-\hat{z})_+}-\frac{1+\hat{z}}{(1-\hat{x})_+}-\frac{1+\hat{x}}{(1-\hat{z})_+}+2(1+\hat{x}\hat{z})+4\hat{x}\hat{z}\nonumber\\
    &\quad-\bigg(4+\frac{2\pi^2}{3}\bigg)\delta(1-\hat{x})\,\delta(1-\hat{z})\bigg]\,,\\
\label{eq:q:Cgq result}
    C^{\rm off}_{gq}(\hat{x},\hat{z})&=P_{g\leftarrow q}(\hat{z})\bigg\{\delta(1-\hat{x})\bigg[\ln\frac{Q^2}{\mu_{D}^2}-1\bigg]+\frac{1}{(1-\hat{x})_+}\bigg\}\nonumber\\
    &\quad+C_\mathrm{F}\bigg[2(1+\hat{x}-\hat{x}\hat{z})-\frac{1+\hat{x}}{\hat{z}}+4\hat{x}(1-\hat{z})\bigg]\,,
\end{align}
with $P_{g\leftarrow q}(\hat{z})=C_\mathrm{F}\,[1+(1-\hat{z})^2]/\hat{z}$. As a check, the momentum sum rule $\int_0^1\mathrm{d}\hat{z}\,\hat{z}\,\big[C^{\rm off}_{qq}+C^{\rm off}_{gq}\big](\hat{x},\hat{z})=C_q^{\rm off}(\hat{x})$ holds, with $\mu_{D}$ dropping out since $\int_0^1\mathrm{d}\hat{z}\,\hat{z}\,\big[\mathcal P_{q\leftarrow q}(\hat{z})+\mathcal P_{g\leftarrow q}(\hat{z})\big]=0$.

The corresponding $\overline{\mathrm{MS}}$ coefficient functions at the same factorization scales $\mu_F$ and $\mu_{D}$ for the initial and final state are given by Eqs.~(C.2), (C.3), (C.5) and (C.6) of Ref.~\cite{deFlorian:1997zj},
\begin{align}
    C^{\overline{\mathrm{MS}}}_{qq}(\hat{x},\hat{z})&=\delta(1-\hat{z})\bigg\{\mathcal P_{q\leftarrow q}(\hat{x})\ln\frac{Q^2}{\mu_F^2}\nonumber\\
    &\quad+C_\mathrm{F}\bigg[(1+\hat{x}^2)\bigg(\frac{\ln(1-\hat{x})}{1-\hat{x}}\bigg)_+-\frac{1+\hat{x}^2}{1-\hat{x}}\ln\hat{x}+1-\hat{x}\bigg]\bigg\}\nonumber\\
    &\quad+\delta(1-\hat{x})\bigg\{\mathcal P_{q\leftarrow q}(\hat{z})\ln\frac{Q^2}{\mu_{D}^2}\nonumber\\
    &\quad+C_\mathrm{F}\bigg[(1+\hat{z}^2)\bigg(\frac{\ln(1-\hat{z})}{1-\hat{z}}\bigg)_++\frac{1+\hat{z}^2}{1-\hat{z}}\ln\hat{z}+1-\hat{z}\bigg]\bigg\}\nonumber\\
    &\quad+C_\mathrm{F}\bigg[\frac{2}{(1-\hat{x})_+(1-\hat{z})_+}-\frac{1+\hat{z}}{(1-\hat{x})_+}-\frac{1+\hat{x}}{(1-\hat{z})_+}+2(1+\hat{x}\hat{z})+4\hat{x}\hat{z}\nonumber\\
    &\quad-8\,\delta(1-\hat{x})\,\delta(1-\hat{z})\bigg]\,,\label{eq:q:CqqMS}\\
    C^{\overline{\mathrm{MS}}}_{gq}(\hat{x},\hat{z})&=P_{g\leftarrow q}(\hat{z})\bigg\{\delta(1-\hat{x})\ln\frac{Q^2\hat{z}(1-\hat{z})}{\mu_{D}^2}+\frac{1}{(1-\hat{x})_+}\bigg\}\nonumber\\
    &\quad+C_\mathrm{F}\bigg[\hat{z}\,\delta(1-\hat{x})+2(1+\hat{x}-\hat{x}\hat{z})-\frac{1+\hat{x}}{\hat{z}}+4\hat{x}(1-\hat{z})\bigg]\,.\label{eq:q:CgqMS}
\end{align}
The terms of Eqs.~\eqref{eq:q:Cqq result}--\eqref{eq:q:CgqMS} that are proportional to neither $\delta(1-\hat{x})$ nor $\delta(1-\hat{z})$ coincide. The two schemes differ only at the collinear endpoints,
\begin{align}
\label{eq:q:DeltaCqq}
    C^{\rm off}_{qq}-C^{\overline{\mathrm{MS}}}_{qq}&=\delta(1-\hat{z})\,\Delta C_q(\hat{x})+\delta(1-\hat{x})\,\Delta C_q(\hat{z})\,,\nonumber\\
    C^{\rm off}_{gq}-C^{\overline{\mathrm{MS}}}_{gq}&=\delta(1-\hat{x})\,\Delta C_{gq}(\hat{z})\,,\nonumber\\
    \Delta C_{gq}(\hat{z})&\equiv-P_{g\leftarrow q}(\hat{z})\big(\ln\big[\hat{z}(1-\hat{z})\big]+1\big)-C_\mathrm{F}\,\hat{z}\,,
\end{align}
with $\Delta C_q$ of Eq.\,\eqref{eq:q:DeltaCq}. The term with $\delta(1-\hat{z})$ is the quark channel contribution~\eqref{eq:q:Delta fqq} to the shift  of the quark distribution~\eqref{eq:q:PDF shift}, which enters SIDIS through the leading-order quark channel coefficient function $\delta(1-\hat{x})\delta(1-\hat{z})$, as in the gluon channel. 

The terms with $\delta(1-\hat{x})$ are related in the same way to the fragmentation functions of the quark into the off-shell outgoing quark and gluon, which we derive in the $\overline{\mathrm{MS}}$ scheme in Appendix~\ref{sec:app:FF}. Recalling the expressions for the scheme-independent partonic structure functions from Eq.\,\eqref{eq:F from C}, the terms with $\delta(1-\hat{x})$ of $\mathcal F_{qq}$ and $\mathcal F_{gq}$ are those of $C^{\rm S}_{qq}$ and $C^{\rm S}_{gq}$ plus $(2\pi/\alpha_\mathrm{s})\big[D^{\rm S}_{q^*(\mu_{\mathrm{out}})/q}(\hat{z},\mu_D^2)-\delta(1-\hat{z})\big]$ and $(2\pi/\alpha_\mathrm{s})\,D^{\rm S}_{g^*(\mu_{\mathrm{out}})/q}(\hat{z},\mu_D^2)$. In the  off-shell scheme $D^{\rm off}$ is given in Eq.\,\eqref{eq:offPDF}. The $\overline{\mathrm{MS}}$ scheme partonic fragmentation functions derived, similarly as the corresponding result for the parton distribution in  Eq.\,\eqref{eq:q:fqq}, from an UV renormalization of the operator definition in Appendix~\ref{sec:app:FF} are
\begin{align}
\label{eq:q:Dqq}
    D^{\overline{\mathrm{MS}}}_{q^*(\mu_{\mathrm{out}})/q}(\hat{z},\mu_D^2)&=\delta(1-\hat{z})+\frac{\alpha_\mathrm{s}}{2\pi}\bigg\{P_{q\leftarrow q}(\hat{z})\bigg[\ln\frac{\mu_D^2}{\mu_{\mathrm{out}}^2\hat{z}(1-\hat{z})}-1\bigg]-C_\mathrm{F}(1-\hat{z})\bigg\}_+\,,\nonumber\\
    D^{\overline{\mathrm{MS}}}_{g^*(\mu_{\mathrm{out}})/q}(\hat{z},\mu_D^2)&=\frac{\alpha_\mathrm{s}}{2\pi}\bigg\{P_{g\leftarrow q}(\hat{z})\bigg[\ln\frac{\mu_D^2}{\mu_{\mathrm{out}}^2\hat{z}(1-\hat{z})}-1\bigg]-C_\mathrm{F}\,\hat{z}\bigg\}\,,
\end{align}
The scheme independence of $\mathcal F_{qq}$ and $\mathcal F_{gq}$ is ensured by the relation between the transformations of the coefficient functions and fragmentation functions
\begin{align}
\label{eq:q:Delta D}
    \Delta D_{q^*/q}(\hat{z})&\equiv D^{\rm off}_{q^*(\mu_{\mathrm{out}})/q}(\hat{z},\mu_D^2)-D^{\overline{\mathrm{MS}}}_{q^*(\mu_{\mathrm{out}})/q}(\hat{z},\mu_D^2)=-\frac{\alpha_\mathrm{s}}{2\pi}\,\Delta C_q(\hat{z})\,,\nonumber\\
    \Delta D_{g^*/q}(\hat{z})&\equiv D^{\rm off}_{g^*(\mu_{\mathrm{out}})/q}(\hat{z},\mu_D^2)-D^{\overline{\mathrm{MS}}}_{g^*(\mu_{\mathrm{out}})/q}(\hat{z},\mu_D^2)=-\frac{\alpha_\mathrm{s}}{2\pi}\,\Delta C_{gq}(\hat{z})\,,
\end{align}
which Eqs.~\eqref{eq:offPDF}, \eqref{eq:q:DeltaCq}, \eqref{eq:q:DeltaCqq}
and \eqref{eq:q:Dqq} indeed satisfy, independently of $\mu_{\mathrm{out}}$ and $\mu_D$. This confirms that we reproduce the terms with $\delta(1-\hat{x})$ in Eq.\,\eqref{eq:q:DeltaCqq}, obtained from the $\overline{\mathrm{MS}}$ coefficient functions of Ref.~\cite{deFlorian:1997zj}, by a direct calculation of the fragmentation functions into the off-shell partons. At the hadronic level, every fragmenting parton has a contribution weighted with the fragmentation functions $D_{H/q}$ and $D_{H/g}$ into the hadron, so that the fragmentation functions of the hadron in the two schemes are related by
\begin{align}
\label{eq:q:FF shift}
    D^{\rm off}_{H/q}(z_H,\mu_{D}^2)&=D^{\overline{\mathrm{MS}}}_{H/q}(z_H,\mu_{D}^2)\nonumber\\
    &\quad-\frac{\alpha_\mathrm{s}}{2\pi}\int_{z_H}^1\frac{\mathrm{d}\hat{z}}{\hat{z}}\bigg[\Delta C_q(\hat{z})\,D_{H/q}\Big(\frac{z_H}{\hat{z}},\mu_{D}^2\Big)+\Delta C_{gq}(\hat{z})\,D_{H/g}\Big(\frac{z_H}{\hat{z}},\mu_{D}^2\Big)\bigg]\,,
\end{align}
and likewise for $\bar q$. The shift of $D_{H/q}$ has the same form as that of the quark distribution, since the outgoing fluctuation regulated by $\mu_{\mathrm{out}}$ has the same form as the incoming one regulated by $\mu_{\mathrm{in}}$. The other contribution $\Delta C_{gq}$ is its counterpart for the observed gluon, with $\zeta\to1-\hat{z}$. Consistently, $\int_0^1\mathrm{d}\hat{z}\,\Delta C_q(\hat{z})=0$ and $\int_0^1\mathrm{d}\hat{z}\,\hat{z}\,\big[\Delta C_q(\hat{z})+\Delta C_{gq}(\hat{z})\big]=0$, so that the shift conserves the quark number and the momentum sum rule of the fragmentation functions.

\section{Summary and outlook}
\label{sec:summary}

We have calculated the $\mathcal O(\alpha_\mathrm{s})$ quark- and gluon-initiated contributions to inclusive and semi-inclusive DIS in a picture in which the partons of the proton scatter off the photon field of the electron, treated as an abelian shockwave in the same way as the target color field in CGC calculations. The partonic cross sections were computed in light-cone perturbation theory, with the incoming and outgoing partons taken off shell with virtualities $\mu_{\mathrm{in}}^2$ and $\mu_{\mathrm{out}}^2$.  

The resulting coefficient functions, Eqs.~\eqref{eq:g:Cgoff}, \eqref{eq:q:Cqoff}, \eqref{eq:g:Coff result}, \eqref{eq:q:Cqq result} and \eqref{eq:q:Cgq result}, have the same collinear logarithms as the $\overline{\mathrm{MS}}$ ones \cite{Ellis:1996mzs, deFlorian:1997zj}. The differences are target independent and correspond to a finite scheme transformation of the quark distributions, Eq.\,\eqref{eq:q:PDF shift}, and of the fragmentation functions, Eq.\,\eqref{eq:q:FF shift}. We obtained the same transformation independently, by calculating the parton distributions and fragmentation functions of the off-shell partons in the $\overline{\mathrm{MS}}$ scheme from the light-cone wave functions, as in the derivation of the splitting functions in Ref.~\cite{Lappi:2026xaa}. The quark-number and momentum sum rules of the coefficient functions provide further checks of the calculation. 

We propose the method off-shell external partons as a method for higher order calculations of dilute-dense processes with colored projectiles, such as forward particle production in proton-nucleus collisions. Making the factorization scale appear as a physical scale in the calculation should also be useful in multiscale processes, such as dijet or massive quark production in DIS, where the natural scale of the parton distributions is not obvious from the kinematics.
The virtualities enter only through the energy denominators of the light-cone wave functions, where they regulate the collinear divergences. The numerators, i.e. the interaction vertices, are unchanged, preserving the fact that in Hamiltonian perturbation theory in light cone gauge one is always working only in terms of physical polarization states.  The cross sections calculated with these wavefunctions are collinear finite, and no separate subtraction is needed.

 If we want to work purely in the leading twist limit, we take the virtualities $\mu_{\mathrm{in}}^2$ and $\mu_{\mathrm{out}}^2$ to be small compared to the hard scale $Q^2$. Then they matter only in the collinear regions, and elsewhere they are power corrections in $\mu_{\mathrm{in}}^2/Q^2$ and $\mu_{\mathrm{out}}^2/Q^2$ which we neglect here.  We defined the off-shell factorization scheme by the condition that the parton distributions and fragmentation functions of the off-shell partons are trivial when the factorization scale equals the virtuality, Eq.\,\eqref{eq:rencond}. In this scheme, the partonic cross sections of the off-shell partons, with the virtualities taken to be equal to the factorization scales, are directly the coefficient functions,  see Eq.\,\eqref{eq:C from F}. 
 
The absence of collinear divergences in the cross section also suggests the possibility of another calculational scheme. One could also just set the off-shellness directly to the hard scale of the problem without expanding to keep just the leading power contribution. In terms of the renormalization condition in the off-shell scheme it would then be quite consistent to use these cross sections with PDFs evolved to this same hard scale. The advantage of this approach is that the resulting hard cross sections, without collinear logarithms or subtractions, could be more amenable to numerical integration. The downside, on the other hand, is that this would not be a consistent leading twist calculation, but include a series of power corrections in the cross section. It would be interesting to explore this approach more also quantitatively, but we have not done so in this paper.

One of the advantages of this approach is that all the way through the calculation the difference between the initial and final state collinear regions is clear, and they are explicitly regulated by different scales. In particular, the final state virtuality $\mu_{\mathrm{out}}^2$, which is associated with DGLAP evolution of the fragmentation function, cancels from the DIS cross section; although for the quark channel seeing this this requires care since it results from a cancellation between real and virtual contributions at the endpoint $\hat{x}=1$.
In general one might doubt that using external off-shell partons could cause issues with gauge invariance. However here, as usually in the high energy saturation regime, it is by most natural to work in light cone gauge and this is not an issue. Problems with regulating the additional gauge pole in covariant theory light cone gauge calculations, on the other hand, are avoided  or at least given a clear prescription by the use of Hamiltonian LCPT. 

The second salient feature of our calculation, relevant for dilute-dense processes in the CGC framework, is the consistent and precise formulation in terms of target momentum fractions. In the case of DIS with an electron target this is needed in order to reconstruct the experimental variable $x_\text{Bj}$, which in fact is an (inverse) target momentum fraction in the photon shockwave picture. While the presence of the collinear logarithms is clearly manifest already in terms of the projectile momentum fractions, our calculation shows that the change of variables  to target momentum fraction is essential for reproducing the correct finite parts, the coefficient functions. This outlines a method that we believe would give the correct NLO result also for the part corresponding to DGLAP evolution in processes with colored incoming our measured particles in CGC calculations, such as forward inclusive hadron and jet production in proton-nucleus collisions. Conversely, we argue that the Abelian limit, where one should recover known DIS coefficient functions, provides a test of different ways of performing such calculations suggested in the literature. 

In the case of an Abelian shockwave, there are no large logarithms corresponding to BK~\cite{Balitsky:1995ub,Kovchegov:1999ua,Kovchegov:1999yj} or JIMWLK evolution of the target. They will only appear when the target is a nonabelian shockwave, and need to be subtracted from the cross section and resummed by the high energy evolution of the target Wilson lines. It is well known~\cite{Altinoluk:2014eka,Beuf:2014uia,Watanabe:2015tja,Lappi:2016fmu,Ducloue:2019ezk,Ducloue:2019jmy,Beuf:2020dxl} that doing this in way that results in a stable perturbative expansion requires formulating the subtraction in terms of a target, rather than projectile, momentum fraction. Often in the literature this coordinate transformation is treated in an approximate way that captures the correct leading logarithms, but leaves some uncertaintly about the finite part of the cross section, called in this case the NLO impact factor. We believe that an explicit and exact constraint on the target momentum fraction, as used here, would be the ideal way to also understand this issue, provided that one is able to carry through the calculation in a similar way as here. 

In summary, the photon shockwave provides an illuminating different way to think about traditional collinear factorization. The calculation here is simple enough to be done analytically, but its structure  is the same as for a colored target: the partons of the projectile fluctuate before and after the eikonal interaction with the shockwave, and the target momentum fraction $\hat{x}$ is reconstructed from the final state. We believe that the methods developed here could be useful for various NLO calculations in the CGC picture in the future. 

\acknowledgments

T.L. thanks G. Beuf for pointing out, a long time ago in a private communication,  that partons from a renormalized PDF are really off-shell. We thank C. Marquet for useful comments on our approach.

This work has been supported by the Research Council of Finland, the Centre of Excellence in Quark Matter (projects 346324 and 364191), the Centre of Excellence in Neutron Star Physics (projects 374062 and 374063) and projects 347499, 353772, 354533, and 354572, and 
and by the European Research Council (ERC, grant agreement No. ERC-2023-101123801 GlueSatLight).
The content of this article does not reflect the official opinion of the European Union and responsibility for the information and views expressed therein lies entirely with the authors.
P.G.E. acknowledges  support from the Finnish Cultural Foundation and M.S.  from the  Magnus Ehrnrooth foundation.
T.L. also acknowledges support from grant
NSF PHY-2309135 to the Kavli Institute for Theoretical Physics (KITP).

\appendix
\section{Light-cone wave functions and amplitudes}
\label{sec:app:partonic}

This appendix gives the details of the partonic calculation of Sec.~\ref{sec:NLO}: the light-cone wave functions, the amplitude and cross section of the gluon channel, and the virtual corrections and real emission of the quark channel.

\subsection{Light-cone wave functions}
\label{sec:app:g LCWF}
\label{sec:app:q LCWF}

The wave function of the $q\bar q$ fluctuation \eqref{eq:g:Fock} of the off-shell gluon follows from Eq.\,\eqref{eq: LCPT: Definition LCWF} with the parent momentum $p=p_g$ of Eq.\,\eqref{eq: off-shell regulator}. The phase-space sum \eqref{eq: LCPT: Definition phase space} over the pair reduces to
\begin{align}
    {\sum_{q\bar q}}'=\sum_{\substack{\mathrm{helicity}\\\mathrm{colour}}}\int_0^1\frac{\mathrm{d}z}{4\pi z(1-z)}\int\frac{\mathrm{d}^{D-2}\boldsymbol{\ell}}{(2\pi)^{D-2}}\,.
\end{align}
The energy denominator $\mathrm{ED}^{\mathrm{in}}_{g\to q\bar q}$ is Eq.\,\eqref{eq:ED regulated}. The vertex is $g\,t^a_{ij}\,\bar u_{h_q}(k_q')\,\slashed\varepsilon_\lambda\,v_{h_{\bar q}}(k_{\bar q}')$, with $\varepsilon^\mu_\lambda=(0,0,\boldsymbol{\varepsilon}_\lambda)$ for a gluon without transverse momentum in light-cone gauge, and its spinor matrix element decomposes in $D$ dimensions as \cite{Beuf:2021qqa}
\begin{align}
    \bar u_{h_q}(k_q')\,\slashed\varepsilon_\lambda\,v_{h_{\bar q}}(k_{\bar q}')=\frac{\varepsilon^n_\lambda\,\boldsymbol{\ell}^m}{4p_g^+z(1-z)}\,\bar u_{h_q}\gamma^+\Big[-2(1-2z)\,\delta^{mn}+[\gamma^m,\gamma^n]\Big]v_{h_{\bar q}}\,.
\end{align}
Dividing by $2p_g^+\mathrm{ED}^{\mathrm{in}}_{g\to q\bar q}$ gives Eq.\,\eqref{eq:g:LCWF}.

The wave functions of the quark channel follow in the same way, with the phase-space sum
\begin{align}
    {\sum_{qg}}'=\sum_{\substack{\mathrm{helicity}\\\mathrm{colour}}}\int_0^1\frac{\mathrm{d}\zeta}{4\pi\zeta(1-\zeta)}\int\frac{\mathrm{d}^{D-2}\boldsymbol{\ell}}{(2\pi)^{D-2}}\,.
\end{align}
For the incoming splitting, the energy denominator is Eq.\,\eqref{eq:ED regulated} with $z\to\zeta$, and the vertex is $g\,t^a_{ij}\,\bar u_{h_q}(k_q')\,\slashed\varepsilon^*_\sigma(k_g')\,u_h(p_q)$, where in light-cone gauge the polarization vector of the gluon, $\varepsilon^\mu_\sigma(k_g')=\big(0,\,\boldsymbol\varepsilon_\sigma\cdot\boldsymbol{\mathrm{k}}_g'/{k_g'}^+,\,\boldsymbol\varepsilon_\sigma\big)$, has a nonvanishing minus component. In $D$ dimensions \cite{Beuf:2021qqa}
\begin{align}
\label{eq:q:spinor ME}
    \bar{u}_{h_q}(k_q')\,\slashed{\varepsilon}_\sigma^*(k_g')\,u_{h}(p_q)=-\frac{\varepsilon^{*n}_\sigma\,\boldsymbol{\ell}^m}{4\zeta(1-\zeta)p_q^+}\,\bar u_{h_q}\gamma^+\Big[2(1+\zeta)\,\delta^{mn}-(1-\zeta)\,[\gamma^m,\gamma^n]\Big]u_h\,,
\end{align}
where the $\delta^{mn}$ term receives contributions from both $\gamma^+\varepsilon^{*-}_\sigma$ and the transverse components. Dividing by $2p_q^+\mathrm{ED}^{\mathrm{in}}_{q\to qg}$ gives Eq.\,\eqref{eq:q:LCWF}. The outgoing splitting and merging have the same vertex and spinor decomposition, since the Dirac structure projects out only the ``good component'' \cite{Beuf:2016wdz} of the spinor and thus depends only on the plus momenta and the helicities, Their energy denominators are Eq.\,\eqref{eq:ED regulated} with the relative momentum of the outgoing pair and the virtuality $\mu_{\mathrm{out}}^2$ of the single quark, which is the outgoing quark $k_q$ in the virtual correction and the quark $k_q''$ after the shockwave in the real emission. For the merging, $\mathrm{ED}^{\mathrm{out}}_{qg\to q}=E_{qg}-E_{q''}$ has the opposite sign, and since the vertex $\langle q''|\hat V|qg\rangle$ is the complex conjugate of the splitting vertex, Eq.\,\eqref{eq:LCPT:merging} gives
\begin{align}
\label{eq:q:app LCWF out R}
    \Psi^{qg\to q}_{h;h_q,\sigma}(\boldsymbol{\mathrm{m}},z;\mu_{\mathrm{out}}^2)=-\Big[\Psi^{q\to qg}_{h_q,\sigma;h}(\boldsymbol{\mathrm{m}},z;\mu_{\mathrm{out}}^2)\Big]^*\,,
\end{align}
This gives Eq.\,\eqref{eq:q:LCWF cases}. In the bras of the amplitudes the outgoing splitting enters as $(\Psi^{q\to qg})^\dagger$ and the merging as $(\Psi^{qg\to q})^*=-\Psi^{q\to qg}$, which gives the relative sign in Eq.\,\eqref{eq:q:real amplitude}.

\subsection{Gluon channel}
\label{sec:app:gluon}

\begin{figure}[tbp!]
\centering
\includegraphics[width=0.8\linewidth,trim=0 14 0 0,clip]{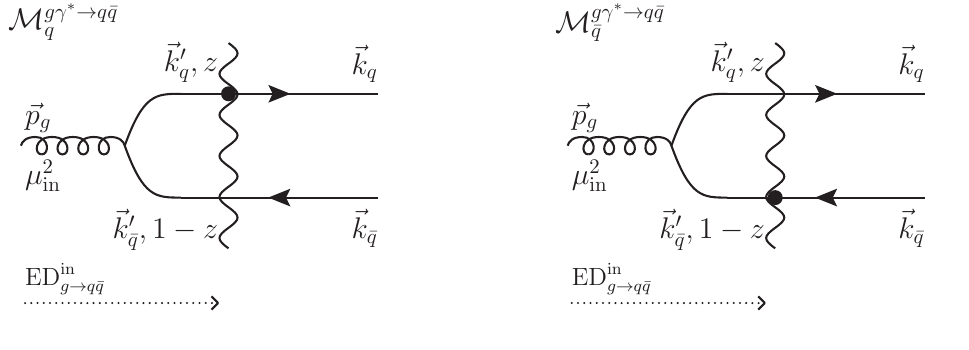}
\caption{Tree-level LCPT amplitudes for $g+\gamma^*\to q\bar q$. The off-shell gluon fluctuates into $q(k_q')\,\bar q(k_{\bar q}')$ before the shockwave at $x^+=0$ (wavy line). The photon, which transfers the transverse momentum $\boldsymbol{\mathrm{q}}=\boldsymbol{\mathrm{k}}_q+\boldsymbol{\mathrm{k}}_{\bar q}$, is absorbed (dot) by the quark (left) or by the antiquark (right), while the other parton is a spectator. The label $\mu_{\mathrm{in}}^2$ indicates the virtuality $-\mu_{\mathrm{in}}^2$ of the incoming gluon, and the dotted arrow the intermediate state with the energy denominator \eqref{eq:ED regulated}.}
\label{fig:gluon amplitude}
\end{figure}

Only the $q\bar q$ component of the incoming state \eqref{eq:g:Fock} contributes at this order, and the outgoing $q\bar q$ state is free (Fig.\,\ref{fig:gluon amplitude}). With $z=k_q^+/p_g^+$, inserting complete sets of transverse position states in Eq.\,\eqref{eq: LCPT: Definition scattering amplitude} gives
\begin{align}
\label{eq:g:amplitude position}
    i\mathcal M^{g\gamma^*\to q\bar q}_{h_q,h_{\bar q};\lambda}&=-\int\frac{\mathrm{d}^{D-2}\boldsymbol{\ell}}{(2\pi)^{D-2}}\,\Psi^{g\to q\bar q}_{h_q,h_{\bar q};\lambda}(\boldsymbol{\ell},z;\mu_{\mathrm{in}}^2)\nonumber\\
    &\quad\times\int\mathrm{d}^{D-2}\boldsymbol{\mathrm{x}}\,\mathrm{d}^{D-2}\boldsymbol{\mathrm{y}}\,e^{-i\boldsymbol{\mathrm{k}}_q\cdot\boldsymbol{\mathrm{x}}-i\boldsymbol{\mathrm{k}}_{\bar q}\cdot\boldsymbol{\mathrm{y}}}\,e^{i\boldsymbol{\ell}\cdot(\boldsymbol{\mathrm{x}}-\boldsymbol{\mathrm{y}})}\Big[1-V_q(\boldsymbol{\mathrm{x}})V_{\bar q}^\dagger(\boldsymbol{\mathrm{y}})\Big]\,,
\end{align}
where $\boldsymbol{\mathrm{x}}$ and $\boldsymbol{\mathrm{y}}$ are the transverse positions of the quark and the antiquark. With the dipole factor evaluated at leading order in the electromagnetic coupling \eqref{eq: LCPT: amplitude dipole}, the coordinate integrals set $\boldsymbol{\ell}=-\boldsymbol{\mathrm{k}}_{\bar q}$ when the quark absorbs the photon and $\boldsymbol{\ell}=\boldsymbol{\mathrm{k}}_q$ when the antiquark absorbs it, with $\boldsymbol{\mathrm{q}}=\boldsymbol{\mathrm{k}}_q+\boldsymbol{\mathrm{k}}_{\bar q}$ in both cases, so that
\begin{align}
    \mathcal M^{g\gamma^*\to q\bar q}_{h_q,h_{\bar q};\lambda}=\frac{e_fe^2}{\boldsymbol{\mathrm{q}}^2}\Big[\Psi^{g\to q\bar q}_{h_q,h_{\bar q};\lambda}(\boldsymbol{\ell}_1,z;\mu_{\mathrm{in}}^2)-\Psi^{g\to q\bar q}_{h_q,h_{\bar q};\lambda}(\boldsymbol{\ell}_2,z;\mu_{\mathrm{in}}^2)\Big]\,.
\end{align}
The two wave functions have the same Dirac structure and differ only in the transverse vector, which combine into $\boldsymbol{\mathrm{V}}$; this gives Eqs.~\eqref{eq:g:amplitude} and \eqref{eq:g:V}.

For the squared amplitude we average over the color and the $D-2$ transverse polarizations of the gluon and sum over the final state, with $\frac{1}{N_\mathrm{c}^2-1}\sum t^a_{ij}(t^a_{ij})^*=T_\mathrm{R}$ and $\sum_\lambda\varepsilon^n_\lambda\varepsilon^{*n'}_\lambda=\delta^{nn'}$. The helicity sum is performed in $D$ dimensions: with $\sum_hu_h\bar u_h=\sum_hv_h\bar v_h=\slashed k$ and $\gamma^+\slashed k\gamma^+=2k^+\gamma^+$, for products $\Gamma$, $\Gamma'$ of an even number of transverse $\gamma$ matrices
\begin{align}
\label{eq:helicity trace}
    \sum_{h_q,h_{\bar q}}\big[\bar u_{h_q}(k_q)\gamma^+\Gamma\,v_{h_{\bar q}}(k_{\bar q})\big]\big[\bar u_{h_q}(k_q)\gamma^+\Gamma'v_{h_{\bar q}}(k_{\bar q})\big]^*=2k_q^+k_{\bar q}^+\,\mathrm{Tr}\big[\Gamma\bar\Gamma'\big]\,,
\end{align}
with $\bar\Gamma'=\gamma^0\Gamma'^\dagger\gamma^0$. With $\mathrm{Tr}\,[\gamma^m,\gamma^n]=0$ and
\begin{align}
\label{eq:commutator trace}
    \mathrm{Tr}\Big([\gamma^m,\gamma^n][\gamma^{n'},\gamma^{m'}]\Big)=16\big(\delta^{mm'}\delta^{nn'}-\delta^{mn'}\delta^{nm'}\big)\,,
\end{align}
and $\delta^{nn}=D-2$,
\begin{align}
\label{eq:g:helicity sum D}
    &\sum_{\lambda,h_q,h_{\bar q}}\Big|\varepsilon^n_\lambda\,\bar u_{h_q}\gamma^+\Big[-2(1-2z)\,\delta^{mn}+[\gamma^m,\gamma^n]\Big]v_{h_{\bar q}}\,\boldsymbol{\mathrm{V}}^m\Big|^2\nonumber\\
    &=32\,z(1-z)\,(p_g^+)^2\Big[(1-2z)^2+D-3\Big]\boldsymbol{\mathrm{V}}^2\,,
\end{align}
which with the prefactors of Eq.\,\eqref{eq:g:amplitude} gives Eq.\,\eqref{eq:g:Msq}. In $D=4$ the bracket is $2[z^2+(1-z)^2]$, and with the two-body phase space of Sec.~\ref{sec:gluon channel partonic}, $\mathrm{d}^2\boldsymbol{\mathrm{k}}_{\bar q}=\mathrm{d}^2\boldsymbol{\mathrm{q}}\to\pi\,\mathrm{d}Q^2$ at fixed $\boldsymbol{\mathrm{k}}_q$, this gives Eq.\,\eqref{eq:g:partonic xs}.

\subsection{Virtual corrections in the quark channel}
\label{sec:app:q virtual}

\begin{figure*}[t]
    \centering
    \includegraphics[width=0.8\linewidth]{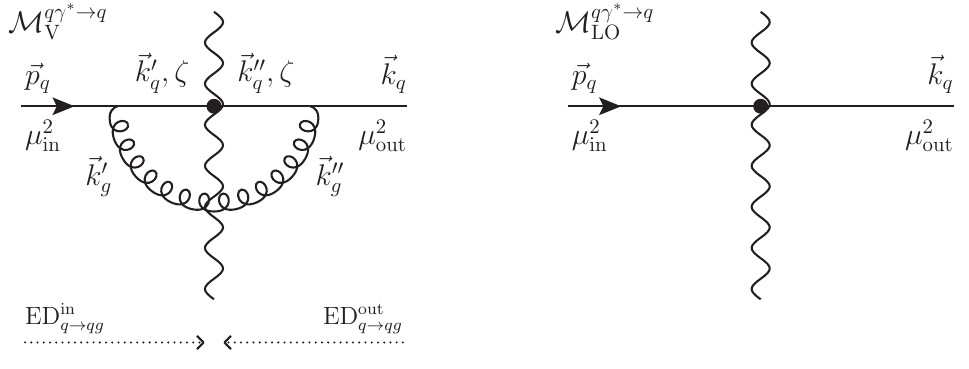}
    \caption{Loop amplitude (left) and leading-order amplitude (right) entering the virtual contribution. The wavy line denotes the photon shockwave at $x^+=0$, the dot the absorption of the photon by the quark, $\mu_{\mathrm{in}}^2$, $\mu_{\mathrm{out}}^2$ indicate the virtualities $-\mu_{\mathrm{in}}^2$, $-\mu_{\mathrm{out}}^2$ of the off-shell quark lines, and the dotted arrows the intermediate states, whose energy denominators are of the form \eqref{eq:ED regulated}.}
    \label{fig:virtual_amplitude}
\end{figure*}

In the loop amplitude $\mathcal M^{q\gamma^*\to q}_{\rm V}$ (left in Fig.\,\ref{fig:virtual_amplitude}) the quark of the incoming $qg$ fluctuation absorbs the photon, while the gluon is a spectator, $\boldsymbol{\mathrm{k}}_g''=\boldsymbol{\mathrm{k}}_g'=-\boldsymbol{\ell}$. Hence $\boldsymbol{\mathrm{k}}_q''=\boldsymbol{\mathrm{k}}_q+\boldsymbol{\ell}$, the relative momentum of the pair after the shockwave is $\boldsymbol{\ell}+(1-\zeta)\boldsymbol{\mathrm{q}}$ with $\boldsymbol{\mathrm{q}}=\boldsymbol{\mathrm{k}}_q$. The amplitude is a product of splitting wavefunctions with arguments $\boldsymbol{\ell}$ before the shockwave and $\boldsymbol{\ell}+(1-\zeta)\boldsymbol{\mathrm{q}}$ after it, which gives Eq.\,\eqref{eq:q:MV}. In addition to this diagram, one needs the  normalization factors for the dressed quark states $g^2Z_q^{(1)}=-{\sum}'|\Psi^{q\to qg}|^2$ from the fluctuation into the bare  $qg$ component, i.e.\ Eq.\,\eqref{eq:q:Z}.

Both $\mathcal M^{q\gamma^*\to q}_{\rm V}$ and $Z_q$ contain the product of an outgoing and an incoming wave function, summed over the colors $i$, $a$, the helicity $h_q$ and the polarization $\sigma$ of the intermediate $qg$ pair. These sums run over the internal states of the loop and are part of the amplitude; they are not the sums over external states of a squared amplitude. The color sum is $\sum_{i,a}(t^a_{ij'})^*t^a_{ij}=C_\mathrm{F}\,\delta_{jj'}$, diagonal in the colors $j$, $j'$ of the incoming and outgoing quark. The Dirac structure, with an even number of transverse $\gamma$ matrices, conserves the helicity $h$ of the external quark. Since the leading-order amplitude is diagonal in and independent of color and helicity, the average over $h$ of the interference $2\,{\rm Re}\big[\mathcal M^{q\gamma^*\to q}_{\rm V}\big(\mathcal M^{q\gamma^*\to q}_{\rm LO}\big)^*\big]$ acts only on the product of wave functions in $\mathcal M^{q\gamma^*\to q}_{\rm V}$. This average allows us to use the trace identity \eqref{eq:helicity trace} of the gluon channel with $v\to u$, where now one of the two summed helicities is that of the incoming quark. With the polarization sum, the trace $\mathrm{Tr}\big(\big[2(1+\zeta)\delta^{mn}-(1-\zeta)[\gamma^m,\gamma^n]\big]\big[2(1+\zeta)\delta^{m'n}-(1-\zeta)[\gamma^n,\gamma^{m'}]\big]\big)=16\big[(1+\zeta)^2+(1-\zeta)^2(D-3)\big]\delta^{mm'}$ obtained from Eq.\,\eqref{eq:commutator trace}, and $2{k_q'}^+p_q^+=2\zeta(p_q^+)^2$, one finds
\begin{align}
\label{eq:q:LCWF squared}
    &\frac12\sum_{h}\sum_{i,a,h_q,\sigma}\Big[\Psi^{q\to qg}_{h_q,\sigma;h}(\boldsymbol{\rho}_2,\zeta;\mu_2^2)\Big]^\dagger\Psi^{q\to qg}_{h_q,\sigma;h}(\boldsymbol{\rho}_1,\zeta;\mu_1^2)\nonumber\\
    &=g^2C_\mathrm{F}\,\zeta\Big[(1+\zeta)^2+(1-\zeta)^2(D-3)\Big]\frac{\boldsymbol{\rho}_2\cdot\boldsymbol{\rho}_1}{\big[\boldsymbol{\rho}_2^2+\tilde\mu_2^2\big]\big[\boldsymbol{\rho}_1^2+\tilde\mu_1^2\big]}\,,
\end{align}
with $\tilde\mu_{1,2}^2=\zeta(1-\zeta)\mu_{1,2}^2$. Here $\boldsymbol{\rho}_1$, $\boldsymbol{\rho}_2$ are the relative transverse momenta of the two $qg$ pairs and $\mu_1$, $\mu_2$ the virtualities of their parents. In $\mathcal M^{q\gamma^*\to q}_{\rm V}$, the incoming wave function \eqref{eq:q:LCWF} has $\boldsymbol{\rho}_1=\boldsymbol{\ell}$, $\mu_1=\mu_{\mathrm{in}}$, and the outgoing one \eqref{eq:q:LCWF} has $\boldsymbol{\rho}_2=\boldsymbol{\mathrm{k}}_q''-\zeta\boldsymbol{\mathrm{k}}_q=\boldsymbol{\ell}+(1-\zeta)\boldsymbol{\mathrm{q}}$, $\mu_2=\mu_{\mathrm{out}}$. In $Z_q^{\rm in}$ both are the incoming wave function, $\boldsymbol{\rho}_1=\boldsymbol{\rho}_2=\boldsymbol{\ell}$ and $\mu_1=\mu_2=\mu_{\mathrm{in}}$, and in $Z_q^{\rm out}$ both are the outgoing one, $\boldsymbol{\rho}_1=\boldsymbol{\rho}_2=\boldsymbol{\mathrm{k}}_q''-\zeta\boldsymbol{\mathrm{k}}_q$ and $\mu_1=\mu_2=\mu_{\mathrm{out}}$. For fixed $h$, the product of wave functions contains, in addition, a term proportional to $\bar u_h\gamma^+[\gamma^m,\gamma^{m'}]u_h\,\boldsymbol{\rho}_1^m\boldsymbol{\rho}_2^{m'}$, which is odd in $h$ and is removed by the average. It does not contribute to either quantity: in $Z_q$ one has $\boldsymbol{\rho}_1=\boldsymbol{\rho}_2$, and in $\mathcal M^{q\gamma^*\to q}_{\rm V}$, where $\boldsymbol{\rho}_1=\boldsymbol{\ell}$ and $\boldsymbol{\rho}_2=\boldsymbol{\ell}+(1-\zeta)\boldsymbol{\mathrm{q}}$, it is proportional to $\boldsymbol{\ell}^m\boldsymbol{\mathrm{q}}^{m'}-\boldsymbol{\ell}^{m'}\boldsymbol{\mathrm{q}}^m$ and vanishes after the integration over $\boldsymbol{\ell}$, since the rest of the integrand depends on $\boldsymbol{\ell}$ only through $\boldsymbol{\ell}^2$ and $\boldsymbol{\ell}\cdot\boldsymbol{\mathrm{q}}$. Both $Z_q$ and $\mathcal M^{q\gamma^*\to q}_{\rm V}$ are therefore independent of $h$, and they may be evaluated with the averaged product \eqref{eq:q:LCWF squared}.

At $\mathcal O(g^2)$ the amplitude is $\mathcal M^{q\gamma^*\to q}=\sqrt{Z_q^{\rm in}Z_q^{\rm out}}\,\mathcal M^{q\gamma^*\to q}_{\rm LO}+\mathcal M^{q\gamma^*\to q}_{\rm V}$, so that the virtual cross section contains $|\mathcal M^{q\gamma^*\to q}_{\rm LO}|^2(Z_q^{\rm in}+Z_q^{\rm out}-2)+2\,{\rm Re}[\mathcal M^{q\gamma^*\to q}_{\rm V}(\mathcal M^{q\gamma^*\to q}_{\rm LO})^*]$. With Eq.\,\eqref{eq:q:LCWF squared}, writing $\boldsymbol{\mathrm{k}}_q''=\boldsymbol{\mathrm{k}}_q+\boldsymbol{\ell}$ in $Z_q^{\rm out}$, the three terms combine into Eq.\,\eqref{eq:q:virtual square}, or
\begin{align}
    \frac{\mathrm{d}\hat\sigma^{q\gamma^*\to q}_{\rm V}}{\mathrm{d}z\,\mathrm{d}Q^2}&=\sigma_0\,\alpha_\mathrm{s}C_\mathrm{F}\,\delta(1-z) \int _0^1\mathrm{d}\zeta\, \frac{(1+\zeta)^2+(1-\zeta)^2(D-3)}{1-\zeta}\nonumber\\
    &\quad\times \Bigg[-\int\frac{\mathrm{d}^{D-2}\boldsymbol{\ell}}{(2\pi)^{D-2}}\frac{\boldsymbol{\ell}^2}{\left[\boldsymbol{\ell}^2+\tilde{\mu}_{\mathrm{in}}^2\right]^2}-\int\frac{\mathrm{d}^{D-2}\boldsymbol{\ell}}{(2\pi)^{D-2}}\frac{\boldsymbol{\ell}^2}{\left[\boldsymbol{\ell}^2+\tilde{\mu}_{\mathrm{out}}^2\right]^2}\nonumber\\
    &\qquad+ 2 \int\frac{\mathrm{d}^{D-2}\boldsymbol{\ell}}{(2\pi)^{D-2}}\frac{\boldsymbol{\ell}^2+\boldsymbol{\ell}\cdot\tilde{\boldsymbol{\mathrm{q}}}}{\left[\boldsymbol{\ell}^2+\tilde{\mu}_{\mathrm{in}}^2\right]\left[(\boldsymbol{\ell}+\tilde{\boldsymbol{\mathrm{q}}})^2+\tilde{\mu}_{\mathrm{out}}^2\right]}\Bigg]\,.\label{eq:q:virtual integrals}
\end{align}
Each of the three integrals is ultraviolet divergent. They can be  expressed through the one-loop master integrals
\begin{align}
    \mathcal{A}_0(\Delta) &\equiv \bar\mu^{4-D}\int\frac{\mathrm{d}^{D-2} \boldsymbol{\ell}}{(2\pi)^{D-2}}\frac{1}{\boldsymbol{\ell}^2+\Delta}=\frac{\Gamma(\varepsilon)}{4\pi}\left(\frac{4\pi \bar\mu^2}{\Delta}\right)^\varepsilon, \\
    \mathcal{B}_0(\Delta_1, \Delta_2, \boldsymbol{\mathrm{q}}) & \equiv \bar\mu^{4-D}\int\frac{\mathrm{d}^{D-2} \boldsymbol{\ell}}{(2\pi)^{D-2}}\frac{1}{\left[\boldsymbol{\ell}^2+\Delta_1\right]\left[(\boldsymbol{\ell}+\boldsymbol{\mathrm{q}})^2+\Delta_2\right]}\nonumber\\
    &=\frac{\varepsilon \Gamma(\varepsilon)}{4\pi}\left(4\pi\bar \mu^2\right)^\varepsilon \int_0^1\mathrm{d}y\left[y(1-y)\boldsymbol{\mathrm{q}}^2+(1-y)\Delta_1+y\Delta_2\right]^{-1-\varepsilon},
\end{align}
with $D=4-2\varepsilon$. Writing the numerators in terms of the denominators, one finds
\begin{align}
    \int\frac{\mathrm{d}^{D-2}\boldsymbol{\ell}}{(2\pi)^{D-2}}\frac{\boldsymbol{\ell}^2}{\left[\boldsymbol{\ell}^2+\tilde{\mu}_{\mathrm{in}}^2\right]^2} &= (1-\varepsilon)\mathcal A_0(\tilde{\mu}_{\mathrm{in}}^2),\\
    \int\frac{\mathrm{d}^{D-2}\boldsymbol{\ell}}{(2\pi)^{D-2}}\frac{\boldsymbol{\ell}^2}{\left[\boldsymbol{\ell}^2+\tilde{\mu}_{\mathrm{in}}^2\right]\left[(\boldsymbol{\ell}+\tilde{\boldsymbol{\mathrm{q}}})^2+\tilde{\mu}_{\mathrm{out}}^2\right]}&=\mathcal A_0(\tilde{\mu}_{\mathrm{out}}^2)-\tilde{\mu}_{\mathrm{in}}^2\mathcal{B}_0(\tilde{\mu}_{\mathrm{in}}^2, \tilde{\mu}_{\mathrm{out}}^2,\tilde{\boldsymbol{\mathrm{q}}}),\\
    \int\frac{\mathrm{d}^{D-2}\boldsymbol{\ell}}{(2\pi)^{D-2}}\frac{\boldsymbol{\ell}\cdot \tilde{\boldsymbol{\mathrm{q}}}}{\left[\boldsymbol{\ell}^2+\tilde{\mu}_{\mathrm{in}}^2\right]\left[(\boldsymbol{\ell}+\tilde{\boldsymbol{\mathrm{q}}})^2+\tilde{\mu}_{\mathrm{out}}^2\right]}&=\frac{1}{2}\Big[\mathcal A_0(\tilde{\mu}_{\mathrm{in}}^2)-\mathcal A_0(\tilde{\mu}_{\mathrm{out}}^2)\nonumber\\
    &\quad+(\tilde{\mu}_{\mathrm{in}}^2-\tilde{\mu}_{\mathrm{out}}^2-\tilde{\boldsymbol{\mathrm{q}}}^2)\mathcal{B}_0(\tilde{\mu}_{\mathrm{in}}^2, \tilde{\mu}_{\mathrm{out}}^2,\tilde{\boldsymbol{\mathrm{q}}})\Big].
\end{align}
In the sum the ultraviolet poles of the $\mathcal A_0$ integrals cancel, and the square bracket in Eq.\,\eqref{eq:q:virtual integrals} becomes
\begin{align}
    \varepsilon\mathcal{A}_0(\tilde{\mu}_{\mathrm{in}}^2)+\varepsilon\mathcal{A}_0(\tilde{\mu}_{\mathrm{out}}^2)-(\tilde{\mu}_{\mathrm{in}}^2+\tilde{\mu}_{\mathrm{out}}^2+\tilde{\boldsymbol{\mathrm{q}}}^2)\,\mathcal{B}_0(\tilde{\mu}_{\mathrm{in}}^2, \tilde{\mu}_{\mathrm{out}}^2,\tilde{\boldsymbol{\mathrm{q}}})\,,
\end{align}
which is finite. Taking $\varepsilon\to0$, $\varepsilon\mathcal A_0\to1/(4\pi)$, and the Feynman-parameter integral in $\mathcal B_0$ gives
\begin{align}
    (\tilde{\mu}_{\mathrm{in}}^2+\tilde{\mu}_{\mathrm{out}}^2+\tilde{\boldsymbol{\mathrm{q}}}^2)\int_0^1\frac{\mathrm{d}y}{y(1-y)\tilde{\boldsymbol{\mathrm{q}}}^2+(1-y)\tilde{\mu}_{\mathrm{in}}^2+y\tilde{\mu}_{\mathrm{out}}^2}=\frac{2\,\operatorname{artanh}\beta}{\beta}\,,
\end{align}
with $\beta$ of Eq.\,\eqref{eq:q:beta}, which gives Eq.\,\eqref{eq:q:virtual result} with $(1+\zeta)^2+(1-\zeta)^2=2(1+\zeta^2)$ in $D=4$. Since the final contribution  to the cross section is finite, we could have kept $D=4$ in the polarization sum and gotten the correct result for the cross section. However, the derivation of the UV-renormalized PDF and FF in Appendices~\ref{sec:app:q PDF} and \ref{sec:app:FF} needs the squared wavefunctions in $D$ dimensions.

\subsection{Real emission in the quark channel}
\label{sec:app:q real}

\begin{figure}[tbp!]
    \centering
    \includegraphics[width=0.8\linewidth]{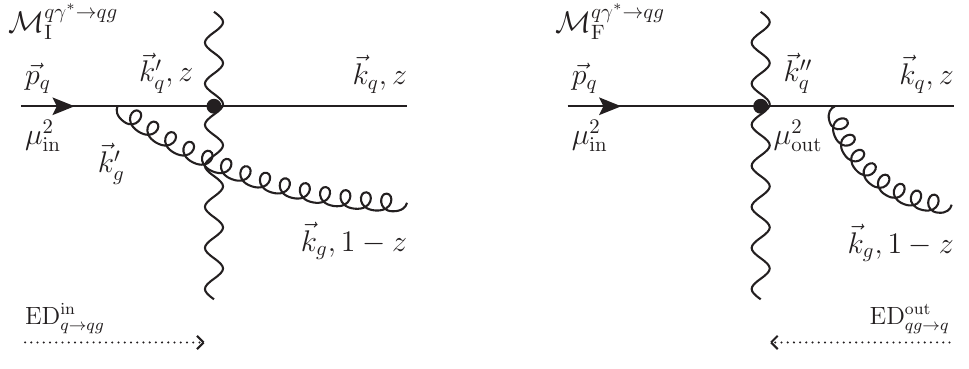}
    \caption{Real-emission amplitudes with the gluon emitted before (left) and after (right) the shockwave, with the energy denominators $\mathrm{ED}^{\mathrm{in}}_{q\to qg}$ and $\mathrm{ED}^{\mathrm{out}}_{qg\to q}$ of Appendix~\ref{sec:app:q LCWF}. Notation as in Fig.\,\ref{fig:virtual_amplitude}.}
    \label{fig:mixed_amplitude}
\end{figure}

In $\mathcal M^{q\gamma^*\to qg}_{\rm I}$ (left in Fig.\,\ref{fig:mixed_amplitude}) the quark of the incoming fluctuation absorbs the photon and the gluon is a spectator. As in Eq.\,\eqref{eq:g:amplitude position}, the coordinate integrals set $\boldsymbol{\ell}=-\boldsymbol{\mathrm{k}}_g$ and $\boldsymbol{\mathrm{q}}=\boldsymbol{\mathrm{k}}_q+\boldsymbol{\mathrm{k}}_g$, so that $\mathcal M^{q\gamma^*\to qg}_{\rm I}=(e_fe^2/\boldsymbol{\mathrm{q}}^2)\,\Psi^{q\to qg}_{h_q,\sigma;h}(\boldsymbol{\ell},z;\mu_{\mathrm{in}}^2)$. In $\mathcal M^{q\gamma^*\to qg}_{\rm F}$ (right in Fig.\,\ref{fig:mixed_amplitude}) the bare quark absorbs the photon, becoming the quark $k_q''$ with $\boldsymbol{\mathrm{k}}_q''=\boldsymbol{\mathrm{q}}$, which is projected on the single-quark component of the bra ${}^{\rm out}_{\ \,\rm D}\langle qg|$. By Eq.\,\eqref{eq:q:app LCWF out R} this component enters with $-\Psi^{q\to qg}_{h_q,\sigma;h}(\boldsymbol{\mathrm{m}},z;\mu_{\mathrm{out}}^2)$, whose Dirac structure may be evaluated with $u_h(p_q)$, since ${k_q''}^+=p_q^+$. The two amplitudes have the same Dirac structure, and their sum gives Eqs.~\eqref{eq:q:real amplitude} and \eqref{eq:q:real amplitude W}.

For the squared amplitude we average over $h$ and the color of the incoming quark and sum over the final state. With $\frac{1}{N_\mathrm{c}}{\rm Tr}(t^at^a)=C_\mathrm{F}$, the polarization sum, and the trace identity \eqref{eq:helicity trace} with $v\to u$ and $2k_q^+p_q^+=2z(p_q^+)^2$,
\begin{align}
    &\frac12\sum_{h,h_q,\sigma}\Big|\varepsilon^{*n}_\sigma\,\bar u_{h_q}\gamma^+\Big[2(1+z)\,\delta^{mn}-(1-z)\,[\gamma^m,\gamma^n]\Big]u_h\,\boldsymbol{\mathrm{W}}^m\Big|^2\nonumber\\
    &=16\,z\,(p_q^+)^2\Big[(1+z)^2+(1-z)^2(D-3)\Big]\boldsymbol{\mathrm{W}}^2\,,
\end{align}
which gives Eq.\,\eqref{eq:q:Msq}, and in $D=4$, with the two-body phase space and $\mathrm{d}^2\boldsymbol{\mathrm{k}}_g=\mathrm{d}^2\boldsymbol{\mathrm{q}}\to\pi\,\mathrm{d}Q^2$ as in the gluon channel, Eq.\,\eqref{eq:q:real result}.

At fixed $\boldsymbol{\mathrm{q}}$ one has $\boldsymbol{\mathrm{m}}=\boldsymbol{\ell}+(1-z)\boldsymbol{\mathrm{q}}$, so that
\begin{align}
    \boldsymbol{\mathrm{W}}=-\bigg[\frac{\boldsymbol{\ell}}{\boldsymbol{\ell}^2+\tilde{\mu}_{\mathrm{in}}^2}-\frac{\boldsymbol{\ell}+\tilde{\boldsymbol{\mathrm{q}}}}{(\boldsymbol{\ell}+\tilde{\boldsymbol{\mathrm{q}}})^2+\tilde{\mu}_{\mathrm{out}}^2}\bigg]\,,
\end{align}
with $\tilde{\mu}_{\mathrm{in,out}}^2=z(1-z)\mu_{\mathrm{in,out}}^2$ and $\tilde{\boldsymbol{\mathrm{q}}}=(1-z)\boldsymbol{\mathrm{q}}$, i.e.\ minus the square bracket of Eq.\,\eqref{eq:q:virtual square} at $\zeta=z$. Its square falls off as $1/\boldsymbol{\ell}^4$, and by Eq.\,\eqref{eq:q:virtual integrals}
\begin{align}
    \int\frac{\mathrm{d}^2\boldsymbol{\ell}}{(2\pi)^2}\,\boldsymbol{\mathrm{W}}^2=\frac{1}{4\pi}\bigg[\frac{2\operatorname{artanh}\beta}{\beta}-2\bigg]\,,
\end{align}
with $\beta$ of Eq.\,\eqref{eq:q:beta} at $\zeta=z$, which gives Eq.\,\eqref{eq:q:real integrated}. The virtual result \eqref{eq:q:virtual result} contains the same function with the opposite sign, multiplied by $\delta(1-z)$ and integrated over $z$, so that the sum is the plus distribution \eqref{eq:q:R+V}.

\section{Calculational details: Inclusive DIS}
\label{sec:app:DIS all}

This appendix gives the details of the inclusive DIS calculation of Sec.~\ref{sec:DIS}: the reduced partonic structure functions of the gluon and quark channels, and the $\overline{\mathrm{MS}}$ parton distributions of the off-shell partons.

\subsection{Gluon channel}
\label{sec:app:DIS}

We evaluate the reduced partonic structure function $\mathcal F_g$ of Eq.\,\eqref{eq:g:Cg def}. In terms of $\boldsymbol{\mathrm{m}}$, the vector $\boldsymbol{\mathrm{V}}$ reads
\begin{align}
    \boldsymbol{\mathrm{V}}&=\frac{\boldsymbol{\ell}_1}{\tilde\ell_1^2}-\frac{\boldsymbol{\ell}_2}{\tilde\ell_2^2}\,,\qquad\tilde\ell_{1,2}^2\equiv\boldsymbol{\ell}_{1,2}^2+\tilde{\mu}_{\mathrm{in}}^2\,,\nonumber\\
    \boldsymbol{\ell}_1&=\boldsymbol{\mathrm{m}}-(1-z)\boldsymbol{\mathrm{q}}\,,\qquad\boldsymbol{\ell}_2=\boldsymbol{\mathrm{m}}+z\boldsymbol{\mathrm{q}}\,,
\end{align}
with $\tilde{\mu}_{\mathrm{in}}^2=z(1-z)\mu_{\mathrm{in}}^2$. Using $2\boldsymbol{\ell}_1\cdot\boldsymbol{\ell}_2=\boldsymbol{\ell}_1^2+\boldsymbol{\ell}_2^2-Q^2$, since $\boldsymbol{\ell}_2-\boldsymbol{\ell}_1=\boldsymbol{\mathrm{q}}$,
\begin{align}
\label{eq:g:V2 D}
    \boldsymbol{\mathrm{V}}^2=-\frac{\tilde{\mu}_{\mathrm{in}}^2}{\tilde\ell_1^4}-\frac{\tilde{\mu}_{\mathrm{in}}^2}{\tilde\ell_2^4}+\frac{Q^2+2\tilde{\mu}_{\mathrm{in}}^2}{\tilde\ell_1^2\tilde\ell_2^2}\,.
\end{align}
On the constraint on $m_\perp$, the combination $C_0\equiv z\tilde\ell_1^2+(1-z)\tilde\ell_2^2=z(1-z)Q^2/\hat{x}+\tilde{\mu}_{\mathrm{in}}^2$ is independent of the angle $\phi$ between $\boldsymbol{\mathrm{m}}$ and $\boldsymbol{\mathrm{q}}$, which allows the partial-fraction decomposition
\begin{align}
    \frac{1}{\tilde\ell_1^2\tilde\ell_2^2}=\frac{1}{C_0}\bigg[\frac{1-z}{\tilde\ell_1^2}+\frac{z}{\tilde\ell_2^2}\bigg]\,.
\end{align}
The denominators are linear in $\cos\phi$,
\begin{align}
\label{eq:g:app D}
    \tilde\ell_1^2&=A_1-B_1\cos\phi\,,\qquad A_1\equiv m_\perp^2+(1-z)^2Q^2+\tilde{\mu}_{\mathrm{in}}^2\,,\qquad B_1\equiv2(1-z)m_\perp Q\,,\nonumber\\
    \tilde\ell_2^2&=A_2+B_2\cos\phi\,,\qquad A_2\equiv m_\perp^2+z^2Q^2+\tilde{\mu}_{\mathrm{in}}^2\,,\qquad B_2\equiv2zm_\perp Q\,,
\end{align}
and with $\int_0^{2\pi}\frac{\mathrm{d}\phi}{2\pi}\frac{1}{A-B\cos\phi}=\frac{1}{\sqrt{A^2-B^2}}$ and $\int_0^{2\pi}\frac{\mathrm{d}\phi}{2\pi}\frac{1}{(A-B\cos\phi)^2}=\frac{A}{(A^2-B^2)^{3/2}}$,
\begin{align}
\label{eq:g:Iphi}
    \int_0^{2\pi}\frac{\mathrm{d}\phi}{2\pi}\,\boldsymbol{\mathrm{V}}^2=-\frac{\tilde{\mu}_{\mathrm{in}}^2A_1}{(A_1^2-B_1^2)^{3/2}}-\frac{\tilde{\mu}_{\mathrm{in}}^2A_2}{(A_2^2-B_2^2)^{3/2}}+\frac{Q^2+2\tilde{\mu}_{\mathrm{in}}^2}{C_0}\bigg[\frac{1-z}{\sqrt{A_1^2-B_1^2}}+\frac{z}{\sqrt{A_2^2-B_2^2}}\bigg]\,,
\end{align}
where
\begin{align}
\label{eq:g:A2B2}
    A_1^2-B_1^2=\frac{(1-z)^2Q^4}{\hat{x}^2}(z-\hat{x})^2+2\tilde{\mu}_{\mathrm{in}}^2\big[m_\perp^2+(1-z)^2Q^2\big]+\tilde{\mu}_{\mathrm{in}}^4\,,
\end{align}
and $A_2^2-B_2^2$ follows by $z\to1-z$. Under $z\to1-z$ the terms with $A_1,B_1$ and $A_2,B_2$ are exchanged, while $P_{q\leftarrow g}(z)$, $z(1-z)$, $\tilde{\mu}_{\mathrm{in}}^2$ and $C_0$ are invariant. Inserting Eq.\,\eqref{eq:g:Iphi} into Eq.\,\eqref{eq:g:Cg def} therefore gives
\begin{align}
\label{eq:g:Cg I}
    \mathcal F_g(\hat{x};r_\mathrm{in})&=2\big(I^g_1+I^g_2\big)\,,\nonumber\\
    I^g_1&\equiv-\int_0^1\mathrm{d}z\,P_{q\leftarrow g}(z)\,\frac{z(1-z)Q^2}{\hat{x}^2}\,\frac{\tilde{\mu}_{\mathrm{in}}^2A_1}{\big(A_1^2-B_1^2\big)^{3/2}}\,,\nonumber\\
    I^g_2&\equiv\int_0^1\mathrm{d}z\,P_{q\leftarrow g}(z)\,\frac{z(1-z)Q^2}{\hat{x}^2}\,\frac{Q^2+2\tilde{\mu}_{\mathrm{in}}^2}{C_0}\,\frac{1-z}{\sqrt{A_1^2-B_1^2}}\,,
\end{align}
which we evaluate for $r_\mathrm{in}\to0$, neglecting corrections of $\mathcal O(r_\mathrm{in}\ln(1/r_\mathrm{in}))$, keeping $\tilde{\mu}_{\mathrm{in}}^2$ only in the collinear regions $z\simeq\hat{x}$ and $z\simeq1-\hat{x}$ of Sec.~\ref{sec:gluon channel DIS}.

By Eq.\,\eqref{eq:g:A2B2}, at $r_\mathrm{in}=0$ the combination $A_1^2-B_1^2$ vanishes only at $z=\hat{x}$, the collinear region $\boldsymbol{\mathrm{k}}_{\bar q}\to0$, where $\tilde{\mu}_{\mathrm{in}}^2$ is kept. Near this point
\begin{align}
    \tilde{\mu}_{\mathrm{in}}^2A_1&\simeq2\hat{x}(1-\hat{x})^3Q^4r_\mathrm{in}\,,\qquad\kappa^2\equiv4\hat{x}^3(1-\hat{x})\,,\nonumber\\
    A_1^2-B_1^2&\simeq\frac{(1-\hat{x})^2Q^4}{\hat{x}^2}\Big[(z-\hat{x})^2+\kappa^2r_\mathrm{in}\Big]\,.
\end{align}
Note that since the $z$-integral in $I^g_1$ has a power law divergence regulated by $\tilde{\mu}_{\mathrm{in}}^2$, we cannot neglect it even if it is superficially proportional to $\tilde{\mu}_{\mathrm{in}}^2$, while in
$I^g_2$ which is only logarithmically divergent we can neglect the $\tilde{\mu}_{\mathrm{in}}^2$ in the numerator. 
Since $\tilde{\mu}_{\mathrm{in}}^2\propto r_\mathrm{in}$, the integrand of $I^g_1$ is $\mathcal O(r_\mathrm{in})$ away from $z=\hat{x}$, and with $u=z-\hat{x}$
\begin{align}
    I^g_1=-\frac12P_{q\leftarrow g}(\hat{x})\int_{-\hat{x}}^{1-\hat{x}}\frac{\kappa^2r_\mathrm{in}\,\mathrm{d}u}{(u^2+\kappa^2r_\mathrm{in})^{3/2}}=-P_{q\leftarrow g}(\hat{x})\,.
\end{align}
Dropping the part $2\tilde{\mu}_{\mathrm{in}}^2$ of the numerator in $I^g_2$ and keeping the dependence of the square root on $r_\mathrm{in}$, Eq.\,\eqref{eq:g:A2B2} gives
\begin{align}
    I^g_2=\int_0^1\mathrm{d}z\,\frac{P_{q\leftarrow g}(z)}{\sqrt{(z-\hat{x})^2+\kappa^2r_\mathrm{in}}}\,.
\end{align}
Writing $P_{q\leftarrow g}(z)=P_{q\leftarrow g}(\hat{x})+(z-\hat{x})(z+\hat{x}-1)$, the first term gives the collinear logarithm, while the second is regular at $z=\hat{x}$ and can be evaluated at $r_\mathrm{in}=0$:
\begin{align}
    I^g_2&=P_{q\leftarrow g}(\hat{x})\ln\frac{4\hat{x}(1-\hat{x})}{\kappa^2r_\mathrm{in}}+\int_0^1\mathrm{d}z\,{\rm sgn}(z-\hat{x})(z+\hat{x}-1)\nonumber\\
    &=P_{q\leftarrow g}(\hat{x})\Big[\ln\frac{Q^2}{\mu_{\mathrm{in}}^2}-2\ln\hat{x}\Big]+3\hat{x}(1-\hat{x})-\frac12\,.
\end{align}
Inserting $I^g_1$ and $I^g_2$ into Eq.\,\eqref{eq:g:Cg I}, with $-2P_{q\leftarrow g}(\hat{x})=2\hat{x}(1-\hat{x})-1$, gives Eq.\,\eqref{eq:g:Fg}.

\subsection{Quark channel}
\label{sec:app:q DIS}

We evaluate the real part $H_q$ of Eq.\,\eqref{eq:q:H def}, with the two regulators independent. In terms of $\boldsymbol{\mathrm{m}}$, the vector $\boldsymbol{\mathrm{W}}$ of Eq.\,\eqref{eq:q:W} reads
\begin{align}
    \boldsymbol{\mathrm{W}}&=\frac{\boldsymbol{\mathrm{m}}}{\tilde m^2}-\frac{\boldsymbol{\ell}}{\tilde\ell^2}\,,\qquad\tilde\ell^2\equiv\boldsymbol{\ell}^2+\tilde{\mu}_{\mathrm{in}}^2\,,\qquad\tilde m^2\equiv m_\perp^2+\tilde{\mu}_{\mathrm{out}}^2\,,\qquad\boldsymbol{\ell}=\boldsymbol{\mathrm{m}}-(1-z)\boldsymbol{\mathrm{q}}\,,\nonumber\\
    \tilde{\mu}_{\mathrm{in}}^2&\equiv z(1-z)\mu_{\mathrm{in}}^2\,,\qquad\tilde{\mu}_{\mathrm{out}}^2\equiv z(1-z)\mu_{\mathrm{out}}^2\,,
\end{align}
where $\tilde\ell^2=\tilde\ell_1^2$ of Appendix~\ref{sec:app:DIS}, since $\boldsymbol{\ell}=\boldsymbol{\ell}_1$, while $\tilde m^2$ does not depend on $\phi$. Using $2\boldsymbol{\ell}\cdot\boldsymbol{\mathrm{m}}=\boldsymbol{\ell}^2+m_\perp^2-c^2$ with $c^2\equiv(\boldsymbol{\ell}-\boldsymbol{\mathrm{m}})^2=(1-z)^2Q^2$,
\begin{align}
\label{eq:q:W2 D}
    \boldsymbol{\mathrm{W}}^2=-\frac{\tilde{\mu}_{\mathrm{in}}^2}{\tilde\ell^4}+\frac{c^2+\tilde{\mu}_{\mathrm{in}}^2+\tilde{\mu}_{\mathrm{out}}^2}{\tilde\ell^2\tilde m^2}-\frac{\tilde{\mu}_{\mathrm{out}}^2}{\tilde m^4}\,,
\end{align}
the analog of Eq.\,\eqref{eq:g:V2 D}. Unlike the gluon channel, the final-state denominator $\tilde m^2=m_\perp^2+\tilde{\mu}_{\mathrm{out}}^2$ is fixed by the constraint on $m_\perp$, so that only $\tilde\ell^2=A_1-B_1\cos\phi$ of Eq.\,\eqref{eq:g:app D} depends on $\phi$, and the angular integrals of Appendix~\ref{sec:app:DIS} give
\begin{align}
\label{eq:q:Iphi}
    \int_0^{2\pi}\frac{\mathrm{d}\phi}{2\pi}\,\boldsymbol{\mathrm{W}}^2=-\frac{\tilde{\mu}_{\mathrm{in}}^2A_1}{(A_1^2-B_1^2)^{3/2}}+\frac{c^2+\tilde{\mu}_{\mathrm{in}}^2+\tilde{\mu}_{\mathrm{out}}^2}{(m_\perp^2+\tilde{\mu}_{\mathrm{out}}^2)\sqrt{A_1^2-B_1^2}}-\frac{\tilde{\mu}_{\mathrm{out}}^2}{(m_\perp^2+\tilde{\mu}_{\mathrm{out}}^2)^2}\,,
\end{align}
where $A_1^2-B_1^2$  in the presence of the constraint on $m_\perp$ is given by Eq.\,\eqref{eq:g:A2B2}. At $r_\mathrm{in}=0$ it vanishes only at $z=\hat{x}$, i.e.\ at $\boldsymbol{\mathrm{k}}_g=0$, which is the initial-state collinear singularity.

\paragraph{Bulk region $\hat{x}<1$.} For fixed $\hat{x}<1$ the final-state collinear region of Eq.\,\eqref{eq:collinear} is avoided and $\mu_{\mathrm{out}}$ is dropped, while $\tilde{\mu}_{\mathrm{in}}^2$ is kept in the initial-state collinear region $z\simeq\hat{x}$. Then
\begin{align}
    H_q(\hat{x};r_\mathrm{in},r_\mathrm{out})&=I^q_1+I^q_2+\dots\,,\nonumber\\
    I^q_1&\equiv-\int_0^1\mathrm{d}z\,P_{q\leftarrow q}(z)\,\frac{z(1-z)Q^2}{\hat{x}^2}\,\frac{\tilde{\mu}_{\mathrm{in}}^2A_1}{\big(A_1^2-B_1^2\big)^{3/2}}\,,\nonumber\\
    I^q_2&\equiv\int_0^1\mathrm{d}z\,P_{q\leftarrow q}(z)\,\frac{z(1-z)Q^2}{\hat{x}^2}\,\frac{c^2+\tilde{\mu}_{\mathrm{in}}^2}{m_\perp^2\sqrt{A_1^2-B_1^2}}\,,
\end{align}
the analogs of $I^g_1$ and $I^g_2$ in Eq.\,\eqref{eq:g:Cg I}, which we evaluate for $r_\mathrm{in}\to0$.

Since $A_1$ and $B_1$ are the same as in the gluon channel, $I^q_1$ is $I^g_1$ of Eq.\,\eqref{eq:g:Cg I} with $P_{q\leftarrow g}\to P_{q\leftarrow q}$. Its integrand is $\mathcal O(r_\mathrm{in})$ except near $z=\hat{x}$, so that the evaluation in Appendix~\ref{sec:app:DIS} applies and gives
\begin{align}
    I^q_1=-P_{q\leftarrow q}(\hat{x})\,.
\end{align}
$I^q_2$ is simplified in the same way as $I^g_2$: the part $\tilde{\mu}_{\mathrm{in}}^2$ of the numerator contributes only at $\mathcal O(r_\mathrm{in}\ln(1/r_\mathrm{in}))$, and with the prefactor $z(1-z)Q^2/(\hat{x}^2m_\perp^2)=1/[\hat{x}(1-\hat{x})]$,
\begin{align}
\label{eq:q:F def}
    I^q_2&=\int_0^1\mathrm{d}z\,\frac{F(z)}{\sqrt{(z-\hat{x})^2+\kappa^2r_\mathrm{in}}}\,,\nonumber\\
    F(z)&\equiv\frac{(1-z)P_{q\leftarrow q}(z)}{1-\hat{x}}=C_\mathrm{F}\,\frac{1+z^2}{1-\hat{x}}\,,\qquad F(\hat{x})=P_{q\leftarrow q}(\hat{x})\,.
\end{align}
Writing $F(z)=F(\hat{x})+[F(z)-F(\hat{x})]$, the first term gives the collinear logarithm, while the second is regular at $z=\hat{x}$ and can be evaluated at $r_\mathrm{in}=0$:
\begin{align}
    I^q_2&=P_{q\leftarrow q}(\hat{x})\ln\frac{4\hat{x}(1-\hat{x})}{\kappa^2r_\mathrm{in}}+\frac{C_\mathrm{F}}{1-\hat{x}}\int_0^1\mathrm{d}z\,{\rm sgn}(z-\hat{x})(z+\hat{x})\nonumber\\
    &=P_{q\leftarrow q}(\hat{x})\bigg[\ln\frac{Q^2}{\mu_{\mathrm{in}}^2}-2\ln\hat{x}\bigg]+C_\mathrm{F}\,\frac{\frac12+\hat{x}-3\hat{x}^2}{1-\hat{x}}\,.
\end{align}
Adding $I^q_1$ gives Eq.\,\eqref{eq:q:Hbulk}.

\paragraph{Cancellation check.} We check that the cancellation \eqref{eq:q:R+V} between the real emission and the virtual correction is recovered from $H_q$. The virtual correction contributes at $\hat{x}=1$ with the coefficient $V_q$ of Eq.\,\eqref{eq:q:V}. Integrating $H_q$ over $\hat{x}$ removes the delta function in Eq.\,\eqref{eq:q:H def}, and with Eq.\,\eqref{eq:q:real integrated}
\begin{align}
\label{eq:q:H xhat integral}
    \int_0^1\mathrm{d}\hat{x}\,H_q(\hat{x};r_\mathrm{in},r_\mathrm{out})=\frac1\pi\int_0^1\mathrm{d}z\,P_{q\leftarrow q}(z)\int\mathrm{d}^2\boldsymbol{\mathrm{k}}_q\,\boldsymbol{\mathrm{W}}^2=C_\mathrm{F}\int_0^1\mathrm{d}z\,\frac{1+z^2}{1-z}\bigg[\frac{2\operatorname{artanh}\beta}{\beta}-2\bigg]\,,
\end{align}
so that, for any $r_\mathrm{in}$ and $r_\mathrm{out}$,
\begin{align}
\label{eq:q:endpoint cancel}
    \int_0^1\mathrm{d}\hat{x}\,H_q(\hat{x};r_\mathrm{in},r_\mathrm{out})+V_q(r_\mathrm{in},r_\mathrm{out})=0\,.
\end{align}
The same holds for the soft pole. At $z\to1$, $\beta\to|r_\mathrm{in}-r_\mathrm{out}|/(r_\mathrm{in}+r_\mathrm{out})$, and the integrand of Eq.\,\eqref{eq:q:H xhat integral} has the pole
\begin{align}
\label{eq:q:soft poles}
    \frac{2C_\mathrm{F}}{1-z}\bigg[\frac{r_\mathrm{in}+r_\mathrm{out}}{r_\mathrm{in}-r_\mathrm{out}}\ln\frac{r_\mathrm{in}}{r_\mathrm{out}}-2\bigg]\,,
\end{align}
which cancels against $V_q$ at $\zeta=z\to1$. Before the integration over $\hat{x}$, the integrand of $H_q$, Eq.\,\eqref{eq:q:H def}, at $z\to1$ follows from $\boldsymbol{\mathrm{k}}_g\simeq-\boldsymbol{\mathrm{m}}$, $\boldsymbol{\mathrm{W}}\simeq-\boldsymbol{\mathrm{m}}/(m_\perp^2+\tilde{\mu}_{\mathrm{in}}^2)+\boldsymbol{\mathrm{m}}/(m_\perp^2+\tilde{\mu}_{\mathrm{out}}^2)$, and $m_\perp^2,\tilde{\mu}_{\mathrm{in}}^2,\tilde{\mu}_{\mathrm{out}}^2\propto z(1-z)$,
\begin{align}
\label{eq:q:soft xhat}
    P_{q\leftarrow q}(z)\,\frac{z(1-z)Q^2}{\hat{x}^2}\int_0^{2\pi}\frac{\mathrm{d}\phi}{2\pi}\,\boldsymbol{\mathrm{W}}^2\simeq\frac{2C_\mathrm{F}}{1-z}\,\frac{\hat{x}(1-\hat{x})\,(r_\mathrm{in}-r_\mathrm{out})^2}{(1-\hat{x}+\hat{x}r_\mathrm{in})^2(1-\hat{x}+\hat{x}r_\mathrm{out})^2}\,,\qquad z\to1\,,
\end{align}
and with $u=(1-\hat{x})/\hat{x}$ its integral over $\hat{x}$,
\begin{align}
    \frac{2C_\mathrm{F}}{1-z}\int_0^\infty\mathrm{d}u\,\frac{u\,(r_\mathrm{in}-r_\mathrm{out})^2}{(u+r_\mathrm{in})^2(u+r_\mathrm{out})^2}=\frac{2C_\mathrm{F}}{1-z}\bigg[\frac{r_\mathrm{in}+r_\mathrm{out}}{r_\mathrm{in}-r_\mathrm{out}}\ln\frac{r_\mathrm{in}}{r_\mathrm{out}}-2\bigg]\,,
\end{align}
is the pole \eqref{eq:q:soft poles}.

\paragraph{Endpoint $\hat{x}=1$.} Acting with $\mathcal F_q=H_q+V_q\,\delta(1-\hat{x})$ on a smooth test function $\varphi(\hat{x})$,
\begin{align}
\label{eq:q:endpoint test}
    \int_0^1\mathrm{d}\hat{x}\,\mathcal F_q\,\varphi(\hat{x})&=\varphi(1)\bigg[\int_0^1\mathrm{d}\hat{x}\,H_q+V_q\bigg]+\int_0^1\mathrm{d}\hat{x}\,H_q\big[\varphi(\hat{x})-\varphi(1)\big]\nonumber\\
    &=\int_0^1\mathrm{d}\hat{x}\,H_q^{\rm bulk}\big[\varphi(\hat{x})-\varphi(1)\big]+\int_0^1\mathrm{d}\hat{x}\,\big(H_q-H_q^{\rm bulk}\big)\big[\varphi(\hat{x})-\varphi(1)\big]\,.
\end{align}
The first bracket vanishes by Eq.\,\eqref{eq:q:endpoint cancel}: the virtual correction and the endpoint contributions of the real emission cancel there, including their dependence on $\mu_{\mathrm{out}}$. The endpoint contributions of the real emission remain in $H_q-H_q^{\rm bulk}$, which is confined to $1-\hat{x}\lesssim r_{\rm in,out}$ and thus acts as a multiple of $\delta(1-\hat{x})$ for $r_{\rm in,out}\to0$. Since this term is multiplied by $\varphi(\hat{x})-\varphi(1)=\mathcal O(1-\hat{x})$, the contribution of the last term is $\mathcal O(r_{\rm in,out}\ln r_{\rm in,out})$. Hence
\begin{align}
    \mathcal F_q=\big[H_q^{\rm bulk}\big]_+\,,
\end{align}
which is Eq.\,\eqref{eq:q:Fq}. In terms of standard distributions, with $\int_0^1\mathrm{d}\hat{x}\,\frac{1+\hat{x}^2}{1-\hat{x}}\ln\hat{x}=\frac54-\frac{\pi^2}{3}$ for the coefficient of $\delta(1-\hat{x})$, this gives the coefficient function Eq.\,\eqref{eq:q:Cqoff} with $\mu_F\to\mu_{\mathrm{in}}$. Here  $\mu_{\mathrm{in}}$ enters through the initial-state collinear region $z\simeq\hat{x}$, present at every $\hat{x}$, while $\mu_{\mathrm{out}}$ drops out.

\subsection{Parton distributions of the off-shell partons}
\label{sec:app:PDF}
\label{sec:app:q PDF}

Here we derive the $\overline{\textrm{MS}}$ quark distribution in a gluon and a quark, Eqs.~\eqref{eq:g:fqg} and \eqref{eq:q:fqq}. Following Ref.~\cite{Lappi:2026xaa}, the light-cone gauge distribution of partons $j$ in the off-shell parton $i^*$ with plus momentum $p^+$ is the expectation value of the bare parton $j$ number operator in the dressed incoming state $i^*$,
\begin{align}
\label{eq:app:PDF def}
    f^0_{j/i^*}(x)\equiv\frac{1}{4\pi x}\int\frac{\mathrm{d}^{D-2}\boldsymbol{\mathrm{k}}}{(2\pi)^{D-2}}\,\frac{{}^{\rm in}_{\ \rm D}\big\langle i(p^+)\big|a_j^\dagger(xp^+,\boldsymbol{\mathrm{k}})\,a_j(xp^+,\boldsymbol{\mathrm{k}})\big|i(p^+)\big\rangle^{\rm in}_{\rm D}}{{}^{\rm in}_{\ \rm D}\big\langle i(p^+)\big|i(p^+)\big\rangle^{\rm in}_{\rm D}}\,,
\end{align}
where $a_j^\dagger$ ($a_j$) creates (annihilates) a bare parton $j$ with momentum $(xp^+,\boldsymbol{\mathrm{k}})$, normalized as the states in Eq.\,\eqref{eq: LCPT: orthonormalization states}, and helicities and colors are summed over. In light-cone gauge the Wilson line in the usual gauge-invariant operator definition of the PDF is unity, and with a hadron in place of $|i\rangle^{\rm in}_{\rm D}$ this is the usual operator definition, normalized as the number density in $x$. The matrix element is easily computed with the light-cone wave functions, as in Ref.~\cite{Lappi:2026xaa}: the one-particle component contributes $Z_q^{\rm in}\,\delta(1-x)$ for $i=q$, and the two-particle components, with the phase space \eqref{eq: LCPT: Definition phase space} at $z=x$, give at $\mathcal O(\alpha_\mathrm{s})$
\begin{align}
\label{eq:app:PDF LCPT}
    f^0_{q/i^*(\mu_{\mathrm{in}})}(x)=\delta_{iq}\,Z_q^{\rm in}\,\delta(1-x)+\frac{1}{4\pi x(1-x)}\int\frac{\mathrm{d}^{D-2}\boldsymbol{\ell}}{(2\pi)^{D-2}}\,\overline{\sum}\big|\Psi^{i\to q\,X}(\boldsymbol{\ell},x;\mu_{\mathrm{in}}^2)\big|^2\,,
\end{align}
where the bar denotes the average over the color and polarization of $i$ and the sum over those of the pair. The transverse-momentum integral diverges logarithmically in the ultraviolet. In the $\overline{\mathrm{MS}}$ scheme it is regulated in $D=4-2\varepsilon$ dimensions, with $g\to\bar\mu^{\varepsilon}g$ and $\bar\mu^2=e^{\gamma_E}\mu^2/(4\pi)$, where $\mu$ is the regularization scale, and the pole is subtracted together with $\ln\mu^2$, which replaces $\mu$ by the factorization scale $\mu_F$ \cite{Lappi:2026xaa}.

For $i=g$ only the $q\bar q$ fluctuation \eqref{eq:g:Fock} contributes, averaged over the color and the $D-2$ polarizations of the gluon and summed over the colors and helicities of the pair as in Eq.\,\eqref{eq:g:helicity sum D}. For $i=q$ the $qg$ fluctuation \eqref{eq:q:Fock} contributes, by Eq.\,\eqref{eq:q:LCWF squared} at $\boldsymbol{\rho}_1=\boldsymbol{\rho}_2=\boldsymbol{\ell}$, with $\zeta=x$ for the observed quark and $\zeta=1-x$ for the observed gluon. In all cases
\begin{align}
\label{eq:app:Psi2}
    \overline{\sum}\big|\Psi^{i\to jX}(\boldsymbol{\ell},x;\mu_0^2)\big|^2=2g^2\,x(1-x)\,P^{(D)}_{j\leftarrow i}(x)\,\frac{\boldsymbol{\ell}^2}{(\boldsymbol{\ell}^2+\Delta)^2}\,,\qquad\Delta\equiv x(1-x)\,\mu_0^2\,,
\end{align}
where $x$ is the plus-momentum fraction of $j$, $\mu_0$ the virtuality of $i$, and $P^{(D)}_{j\leftarrow i}$ the splitting functions in $D$ dimensions,
\begin{gather}
    P^{(D)}_{j\leftarrow i}(x)=P_{j\leftarrow i}(x)+\varepsilon\,P^{\varepsilon}_{j\leftarrow i}(x)+\mathcal O(\varepsilon^2)\,,\nonumber\\
    P^{\varepsilon}_{q\leftarrow g}(x)=-2T_\mathrm{R}\,x(1-x)\,,\qquad P^{\varepsilon}_{q\leftarrow q}(x)=-C_\mathrm{F}(1-x)\,,\qquad P^{\varepsilon}_{g\leftarrow q}(x)=-C_\mathrm{F}\,x\,.\label{eq:app:Peps}
\end{gather}
With $\int\frac{\mathrm{d}^{D-2}\boldsymbol{\ell}}{(2\pi)^{D-2}}\frac{\boldsymbol{\ell}^2}{(\boldsymbol{\ell}^2+\Delta)^2}=(1-\varepsilon)\,\frac{\Gamma(\varepsilon)}{4\pi}\Big(\frac{4\pi\bar\mu^2}{\Delta}\Big)^\varepsilon$ and $\Gamma(\varepsilon)(4\pi\bar\mu^2/\Delta)^\varepsilon=1/\varepsilon+\ln(\mu^2/\Delta)+\mathcal O(\varepsilon)$, the two-particle components of Eq.\,\eqref{eq:app:PDF LCPT} are
\begin{align}
\label{eq:app:PDF PD}
    f^0_{j/i^*(\mu_{\mathrm{in}})}(x)\big|_{jX}&=\frac{\alpha_\mathrm{s}}{2\pi}\,P^{(D)}_{j\leftarrow i}(x)\,(1-\varepsilon)\,\Gamma(\varepsilon)\Big(\frac{4\pi\bar\mu^2}{x(1-x)\mu_{\mathrm{in}}^2}\Big)^\varepsilon\nonumber\\
    &=\frac{\alpha_\mathrm{s}}{2\pi}\bigg\{P_{j\leftarrow i}(x)\bigg[\frac1\varepsilon+\ln\frac{\mu^2}{\mu_{\mathrm{in}}^2x(1-x)}-1\bigg]+P^{\varepsilon}_{j\leftarrow i}(x)\bigg\}+\mathcal O(\varepsilon)\,.
\end{align}
The ultraviolet pole multiplies the factor $1-\varepsilon$ of the integral, which gives $-P_{j\leftarrow i}$, and the $\mathcal O(\varepsilon)$ part of $P^{(D)}_{j\leftarrow i}$, which gives $P^{\varepsilon}_{j\leftarrow i}$.

For $i=g$, $P^{\varepsilon}_{q\leftarrow g}-P_{q\leftarrow g}=-T_\mathrm{R}$, and
\begin{align}\label{eq:msbarbareqing}
    f^0_{q/g^*(\mu_{\mathrm{in}})}(x)=\frac{\alpha_\mathrm{s}}{2\pi}\bigg\{P_{q\leftarrow g}(x)\bigg[\frac1\varepsilon+\ln\frac{\mu^2}{\mu_{\mathrm{in}}^2x(1-x)}\bigg]-T_\mathrm{R}\bigg\}+\mathcal O(\varepsilon)\,.
\end{align}
In the $\overline{\mathrm{MS}}$ scheme this is renormalized by subtracting just the pole and the associated log, i.e.\ $P_{q\leftarrow g}(x)\big[1/\varepsilon+\ln(\mu^2/\mu_F^2)\big]$, which leads to the form in Eq.\,\eqref{eq:g:fqg}, with $T_\mathrm{R}=1/2$.

For $i=q$, Eq.\,\eqref{eq:app:PDF LCPT} also contains the one-particle component, $Z_q^{\rm in}\,\delta(1-x)$, where by Eq.\,\eqref{eq:q:Z} $Z_q^{\rm in}-1$ is minus the integral over $x$ of Eq.\,\eqref{eq:app:PDF PD} for $j=q$. This ensures quark-number conservation and turns the sum into a plus distribution,
\begin{align}
    f^0_{q/q^*(\mu_{\mathrm{in}})}(x)=\delta(1-x)+\frac{\alpha_\mathrm{s}}{2\pi}\bigg\{P_{q\leftarrow q}(x)\bigg[\frac1\varepsilon+\ln\frac{\mu^2}{\mu_{\mathrm{in}}^2x(1-x)}-1\bigg]-C_\mathrm{F}(1-x)\bigg\}_++\mathcal O(\varepsilon)\,.
\end{align}
Subtracting $\mathcal P_{q\leftarrow q}(x)\big[1/\varepsilon+\ln(\mu^2/\mu_F^2)\big]$ in the $\overline{\mathrm{MS}}$ scheme gives Eq.\,\eqref{eq:q:fqq}.

\section{Calculational details: SIDIS}
\label{sec:app:SIDIS all}

This appendix gives the details of the SIDIS calculation of Sec.~\ref{sec:SIDIS}: the reduced partonic structure functions of the gluon and quark channels, and the $\overline{\mathrm{MS}}$ fragmentation functions of the quark into off-shell partons.

\subsection{Gluon channel}
\label{sec:app:SIDIS}

We evaluate the reduced partonic structure function $\mathcal F_{qg}$ of Eq.\,\eqref{eq:g:Coff m}, with $\mathrm{d}^2\boldsymbol{\mathrm{m}}=\frac12\mathrm{d}m_\perp^2\,\mathrm{d}\phi$. The first delta function fixes $m_\perp$ with the Jacobian \eqref{eq:g:Jacobian}. With this  constraint $\hat{z}$ depends on the angle $\phi$ between $\boldsymbol{\mathrm{m}}$ and $\boldsymbol{\mathrm{q}}$ as in Eq.\,\eqref{eq:g:zhat phi}, $\hat{z}=\hat{z}_0+\Delta\cos\phi$, with $\hat{z}_0\equiv\hat{x}z+(1-\hat{x})(1-z)$ and $\Delta\equiv2\sqrt{\hat{x}(1-\hat{x})z(1-z)}$. Since $\boldsymbol{\mathrm{V}}^2$ depends only on $\cos\phi$, the two roots of the second delta function contribute equally,
\begin{align}
    \int_0^{2\pi}\mathrm{d}\phi\,\delta\big(\hat{z}-\hat{z}_0-\Delta\cos\phi\big)=\frac{2}{\sqrt{G}}\,,\qquad G\equiv\Delta^2-(\hat{z}-\hat{z}_0)^2=\delta^2-(z-z_0)^2\,,
\end{align}
where the second form, with $z_0=\hat{x}\hat{z}+(1-\hat{x})(1-\hat{z})$ and $\delta=2\sqrt{\hat{x}(1-\hat{x})\hat{z}(1-\hat{z})}$ as in Eq.\,\eqref{eq:g:Coff psi}, follows from the symmetry of $G$ under $z\leftrightarrow\hat{z}$. Thus
\begin{align}
    \mathcal F_{qg}(\hat{x},\hat{z};r_\mathrm{in})=\frac{1}{\pi\hat{x}^2}\int_{z_0-\delta}^{z_0+\delta}\frac{\mathrm{d}z}{\sqrt{\delta^2-(z-z_0)^2}}\,P_{q\leftarrow g}(z)\,z(1-z)Q^2\,\boldsymbol{\mathrm{V}}^2\,,
\end{align}
and the substitution $z=z_0+\delta\cos\psi$ gives Eq.\,\eqref{eq:g:Coff psi}.

When both constraints are imposed we have $\boldsymbol{\mathrm{k}}_q^2=z\hat{z}Q^2/\hat{x}$ and $\boldsymbol{\mathrm{k}}_{\bar q}^2=(1-z)(1-\hat{z})Q^2/\hat{x}$, so that $\tilde\ell_1^2=\boldsymbol{\mathrm{k}}_{\bar q}^2+\tilde{\mu}_{\mathrm{in}}^2$ and $\tilde\ell_2^2=\boldsymbol{\mathrm{k}}_q^2+\tilde{\mu}_{\mathrm{in}}^2$ in Eq.\,\eqref{eq:g:V2 D}. In terms of $d_1\equiv\tilde\ell_1^2/[(1-z)Q^2]$ and $d_2\equiv\tilde\ell_2^2/(zQ^2)$, the factor multiplying the splitting function in the $\psi$ integral is
\begin{align}
\label{eq:g:zV2}
    z(1-z)Q^2\,\boldsymbol{\mathrm{V}}^2&=\frac{1+2z(1-z)r_\mathrm{in}}{d_1d_2}-\frac{z^2r_\mathrm{in}}{d_1^2}-\frac{(1-z)^2r_\mathrm{in}}{d_2^2}\,,\nonumber\\
    d_1&=\frac{1-\hat{z}}{\hat{x}}+zr_\mathrm{in}\,,\qquad d_2=\frac{\hat{z}}{\hat{x}}+(1-z)r_\mathrm{in}\,.
\end{align}
By Eq.\,\eqref{eq:collinear}, the collinear region $\boldsymbol{\mathrm{k}}_{\bar q}\to0$ lies at $\hat{z}=1$, where $d_1\to zr_\mathrm{in}$, while the region $\boldsymbol{\mathrm{k}}_q\to0$ lies at $\hat{z}=0$, outside the current fragmentation region, and for $\hat{z}\geq z_H>0$ one has $d_2\geq z_H/\hat{x}$. Hence $r_\mathrm{in}$ is kept only in the endpoint region $1-\hat{z}=\mathcal O(r_\mathrm{in})$.

\paragraph{Regular part.} For fixed $\hat{z}<1$ we set $r_\mathrm{in}=0$. Then $z(1-z)Q^2\,\boldsymbol{\mathrm{V}}^2=1/(d_1d_2)=\hat{x}^2/[\hat{z}(1-\hat{z})]$ is independent of $z$, and the $\psi$ integral acts only on the splitting function,
\begin{align}
    \frac1\pi\int_0^\pi\mathrm{d}\psi\,P_{q\leftarrow g}(z_0+\delta\cos\psi)&=T_\mathrm{R}\big[z_0^2+(1-z_0)^2+\delta^2\big]\nonumber\\
    &=P_{q\leftarrow g}(\hat{x})\big[1-2\hat{z}(1-\hat{z})\big]+4\hat{x}(1-\hat{x})\hat{z}(1-\hat{z})\,,
\end{align}
so that
\begin{align}
\label{eq:g:Creg}
    \mathcal F^{\rm reg}_{qg}(\hat{x},\hat{z})&=\frac{1}{\hat{z}(1-\hat{z})}\,\frac1\pi\int_0^\pi\mathrm{d}\psi\,P_{q\leftarrow g}(z_0+\delta\cos\psi)\nonumber\\
    &=P_{q\leftarrow g}(\hat{x})\bigg[\frac{1}{1-\hat{z}}+\frac1{\hat{z}}-2\bigg]+4\hat{x}(1-\hat{x})\,.
\end{align}
\paragraph{Initial-state endpoint $\hat{z}\to1$.} The pole at $\hat{z}=1$ is not integrable, and its regularization is determined by the endpoint region. For $\eta_z\equiv1-\hat{z}\to0$ one has $\delta\to0$, so that $z\to\hat{x}$ and the integrand no longer depends on $\psi$. With $d_1\to(\eta_z+\hat{x}^2r_\mathrm{in})/\hat{x}$ and $d_2\to1/\hat{x}$, the term $2z(1-z)r_\mathrm{in}$ in the numerator of the first term of Eq.\,\eqref{eq:g:zV2} gives $2\hat{x}(1-\hat{x})\,r_\mathrm{in}/(\eta_z+\hat{x}^2r_\mathrm{in})$, whose integral over $\eta_z$ is $\mathcal O(r_\mathrm{in}\ln(1/r_\mathrm{in}))$, and the last term is $\mathcal O(r_\mathrm{in})$, since $d_2$ stays finite. Dropping both, the remaining terms give
\begin{align}
    \mathcal F^{\rm end}_{qg}(\hat{x},\hat{z};r_\mathrm{in})=P_{q\leftarrow g}(\hat{x})\bigg[\frac{1}{\eta_z+\hat{x}^2r_\mathrm{in}}-\frac{\hat{x}^2r_\mathrm{in}}{(\eta_z+\hat{x}^2r_\mathrm{in})^2}\bigg]=P_{q\leftarrow g}(\hat{x})\,\frac{\eta_z}{(\eta_z+\hat{x}^2r_\mathrm{in})^2}\,,
\end{align}
which reduces to the pole term of Eq.\,\eqref{eq:g:Creg} for $\eta_z\gg r_\mathrm{in}$. Its limit $r_\mathrm{in}\to0$ in the sense of distributions follows from acting on a smooth test function, $\varphi(\hat{z})=\varphi(1)+[\varphi(\hat{z})-\varphi(1)]$. The second term reproduces the plus prescription, while the first gives
\begin{align}
    \varphi(1)\int_0^1\mathrm{d}\eta_z\,\frac{\eta_z}{(\eta_z+\hat{x}^2r_\mathrm{in})^2}=\varphi(1)\bigg[\ln\frac{1}{\hat{x}^2r_\mathrm{in}}-1\bigg]+\mathcal O(r_\mathrm{in})\,.
\end{align}
Hence
\begin{align}
    \frac{\eta_z}{(\eta_z+\hat{x}^2r_\mathrm{in})^2}=\bigg[\ln\frac{Q^2}{\mu_{\mathrm{in}}^2}-2\ln\hat{x}-1\bigg]\delta(1-\hat{z})+\frac{1}{(1-\hat{z})_+}+\mathcal O\big(r_\mathrm{in}\ln(1/r_\mathrm{in})\big)\,.
\end{align}
Replacing the pole $P_{q\leftarrow g}(\hat{x})/(1-\hat{z})$ in Eq.\,\eqref{eq:g:Creg} by $P_{q\leftarrow g}(\hat{x})$ times this distribution gives $\mathcal F_{qg}$, which by Eq.\,\eqref{eq:C from F} is the gluon channel SIDIS coefficient function in Eq.\,\eqref{eq:g:Coff result} with $\mu_F\to\mu_{\mathrm{in}}$.

\subsection{Quark channel}
\label{sec:app:q SIDIS}

Let us then  derive the quark channel SIDIS coefficient functions in Eqs.~\eqref{eq:q:Cqq result} and \eqref{eq:q:Cgq result} following a similar procedure as  in Appendix~\ref{sec:app:SIDIS}. For the tagged quark, the first delta function in Eq.\,\eqref{eq:q:Cqq def} fixes $m_\perp$ as in Eq.\,\eqref{eq:q:H def}, and the second fixes $\phi$ through Eq.\,\eqref{eq:g:zhat phi}, with $z$ the plus-momentum fraction of the final-state quark. As in Appendix~\ref{sec:app:SIDIS},
\begin{align}
\label{eq:q:app Cqq psi}
    \mathcal F_{qq}(\hat{x},\hat{z};r_\mathrm{in},r_\mathrm{out})&=\frac{1}{\pi\hat{x}^2}\int_0^\pi\mathrm{d}\psi\,\Big[P_{q\leftarrow q}(z)\,z(1-z)Q^2\,\boldsymbol{\mathrm{W}}^2\Big]_{z=z_0+\delta\cos\psi}\nonumber\\
    &\quad+V_q(r_\mathrm{in},r_\mathrm{out})\,\delta(1-\hat{x})\,\delta(1-\hat{z})\,,
\end{align}
with $z_0=\hat{x}\hat{z}+(1-\hat{x})(1-\hat{z})$ and $\delta=2\sqrt{\hat{x}(1-\hat{x})\hat{z}(1-\hat{z})}$. The contribution where one tags the gluon, $\mathcal F_{gq}$, is obtained from the first term with $\hat{z}\to1-\hat{z}$. Due to the two constraints $\boldsymbol{\mathrm{k}}_q^2=z\hat{z}Q^2/\hat{x}$, $\boldsymbol{\mathrm{k}}_g^2=(1-z)(1-\hat{z})Q^2/\hat{x}$ and $m_\perp^2=z(1-z)Q^2(1-\hat{x})/\hat{x}$, so that Eq.\,\eqref{eq:q:W2 D} gives
\begin{align}
\label{eq:q:zW2 SIDIS}
    z(1-z)Q^2\,\boldsymbol{\mathrm{W}}^2&=z\bigg[\frac{1-z+zr_\mathrm{in}+zr_\mathrm{out}}{d_1d_2}-\frac{zr_\mathrm{in}}{d_1^2}-\frac{zr_\mathrm{out}}{d_2^2}\bigg]\,,\nonumber\\
    d_1&\equiv\frac{\tilde\ell^2}{(1-z)Q^2}=\frac{1-\hat{z}}{\hat{x}}+zr_\mathrm{in}\,,\qquad d_2\equiv\frac{\tilde m^2}{(1-z)Q^2}=z\bigg[\frac{1-\hat{x}}{\hat{x}}+r_\mathrm{out}\bigg]\,.
\end{align}

\paragraph{Regular part.} For $\hat{x}<1$ and $\hat{z}<1$ both collinear regions of Eq.\,\eqref{eq:collinear} are avoided, and the regulators are dropped. Then $z(1-z)Q^2\boldsymbol{\mathrm{W}}^2=z(1-z)/(d_1d_2)=\hat{x}^2(1-z)/[(1-\hat{x})(1-\hat{z})]$, so that
\begin{align}
\label{eq:q:Cqq reg}
    \mathcal F^{\rm reg}_{qq}(\hat{x},\hat{z})&=C_\mathrm{F}\,\frac{1+z_0^2+\delta^2/2}{(1-\hat{x})(1-\hat{z})}\nonumber\\
    &=\frac{P_{q\leftarrow q}(\hat{x})}{1-\hat{z}}+\frac{P_{q\leftarrow q}(\hat{z})}{1-\hat{x}}-\frac{2C_\mathrm{F}}{(1-\hat{x})(1-\hat{z})}+C_\mathrm{F}\big[2(1+\hat{x}\hat{z})+4\hat{x}\hat{z}\big]\,,
\end{align}
and with $\hat{z}\to1-\hat{z}$
\begin{align}
\label{eq:q:Cgq reg}
    \mathcal F^{\rm reg}_{gq}(\hat{x},\hat{z})=\frac{P_{g\leftarrow q}(\hat{z})}{1-\hat{x}}+C_\mathrm{F}\bigg[2(1+\hat{x}-\hat{x}\hat{z})-\frac{1+\hat{x}}{\hat{z}}+4\hat{x}(1-\hat{z})\bigg]\,.
\end{align}
The poles at $\hat{z}=1$ and $\hat{x}=1$, with residues $P_{q\leftarrow q}(\hat{x})$ and $P_{q\leftarrow q}(\hat{z})$, are regularized by the two collinear endpoints. The soft-gluon limit $z\to1$ at fixed $\hat{x}<1$, where $\boldsymbol{\mathrm{k}}_q\to\boldsymbol{\mathrm{q}}$ and $\hat{z}\to\hat{x}$, is not a collinear region, and the regulators are dropped there.

\paragraph{Initial-state endpoint $\hat{z}\to1$.} This is the initial-state collinear region $\boldsymbol{\mathrm{k}}_g\to0$ of Eq.\,\eqref{eq:collinear}. For $\eta_z\equiv1-\hat{z}\to0$ at fixed $\hat{x}<1$ one has $\delta\to0$ and $z\to\hat{x}$, and only $d_1\to(\eta_z+\hat{x}^2r_\mathrm{in})/\hat{x}$ becomes small, while $(1-z)/d_2\to1$. As in the gluon channel, the terms $zr_\mathrm{in}+zr_\mathrm{out}$ in the numerator of the first term of Eq.\,\eqref{eq:q:zW2 SIDIS} contribute at $\mathcal O(r_{\rm in,out}\ln(1/r_{\rm in,out}))$ after integration over $\eta_z$, and the last term at $\mathcal O(r_\mathrm{out})$. The remaining terms give
\begin{align}
\label{eq:q:Cqq zend endpoint}
    \mathcal F^{\rm end}_{qq}(\hat{x},\hat{z};r_\mathrm{in})=P_{q\leftarrow q}(\hat{x})\bigg[\frac{1}{\eta_z+\hat{x}^2r_\mathrm{in}}-\frac{\hat{x}^2r_\mathrm{in}}{(\eta_z+\hat{x}^2r_\mathrm{in})^2}\bigg]=P_{q\leftarrow q}(\hat{x})\,\frac{\eta_z}{(\eta_z+\hat{x}^2r_\mathrm{in})^2}\,,
\end{align}
the endpoint distribution of the gluon channel (Appendix~\ref{sec:app:SIDIS}) with $P_{q\leftarrow g}\to P_{q\leftarrow q}$. Hence the pole $P_{q\leftarrow q}(\hat{x})/(1-\hat{z})$ of Eq.\,\eqref{eq:q:Cqq reg} becomes
\begin{align}
\label{eq:q:Cqq zend}
    P_{q\leftarrow q}(\hat{x})\bigg\{\frac{1}{(1-\hat{z})_+}+\delta(1-\hat{z})\bigg[\ln\frac{Q^2}{\mu_{\mathrm{in}}^2}-2\ln\hat{x}-1\bigg]\bigg\}\,.
\end{align}

\paragraph{Final-state endpoint $\hat{x}\to1$.} This is the final-state collinear region $\boldsymbol{\mathrm{m}}\to0$ of Eq.\,\eqref{eq:collinear}. For $\eta_x\equiv1-\hat{x}\to0$ at fixed $\hat{z}<1$ one has $\delta\to0$ and $z\to\hat{z}$, and only $d_2\to\hat{z}(\eta_x+r_\mathrm{out})$ becomes small, while $(1-z)/d_1\to1$. Dropping the regulators in the numerator of the first term and the second term of Eq.\,\eqref{eq:q:zW2 SIDIS} for the same reason, the first and last terms give
\begin{align}
\label{eq:q:Cqq xend endpoint}
    \frac{1}{\hat{x}^2}\,P_{q\leftarrow q}(\hat{z})\bigg[\frac{\hat{z}}{d_2}-\frac{\hat{z}^2r_\mathrm{out}}{d_2^2}\bigg]=P_{q\leftarrow q}(\hat{z})\,\frac{\eta_x}{(\eta_x+r_\mathrm{out})^2}\,,
\end{align}
the distribution of the initial-state endpoint with $\eta_z\to\eta_x$ and $\hat{x}^2r_\mathrm{in}\to r_\mathrm{out}$. Hence the pole $P_{q\leftarrow q}(\hat{z})/(1-\hat{x})$ of Eq.\,\eqref{eq:q:Cqq reg} becomes
\begin{align}
\label{eq:q:Cqq xend}
    P_{q\leftarrow q}(\hat{z})\bigg\{\frac{1}{(1-\hat{x})_+}+\delta(1-\hat{x})\bigg[\ln\frac{Q^2}{\mu_{\mathrm{out}}^2}-1\bigg]\bigg\}\,.
\end{align}

For the tagged gluon the quark carries $1-\hat{z}$, so that $z\to1-\hat{z}$ and $d_2\to(1-\hat{z})(\eta_x+r_\mathrm{out})$, and as in Eq.\,\eqref{eq:q:Cqq xend endpoint} the pole $P_{g\leftarrow q}(\hat{z})/(1-\hat{x})$ of Eq.\,\eqref{eq:q:Cgq reg} becomes
\begin{align}
    P_{g\leftarrow q}(\hat{z})\bigg\{\frac{1}{(1-\hat{x})_+}+\delta(1-\hat{x})\bigg[\ln\frac{Q^2}{\mu_{\mathrm{out}}^2}-1\bigg]\bigg\}\,,
\end{align}
which with the remaining terms of Eq.\,\eqref{eq:q:Cgq reg} gives $\mathcal F_{gq}$, and by Eq.\,\eqref{eq:C from F} Eq.\,\eqref{eq:q:Cgq result}. This is the only singular region of the tagged gluon: its initial-state collinear region lies at $\hat{z}=0$, outside the current fragmentation region, and the virtual correction does not contribute.

\paragraph{Double endpoint.} Replacing the poles of Eq.\,\eqref{eq:q:Cqq reg} by Eqs.~\eqref{eq:q:Cqq zend} and \eqref{eq:q:Cqq xend} determines $\mathcal F_{qq}$ up to a term $K\,\delta(1-\hat{x})\delta(1-\hat{z})$, which contains the virtual correction and the soft region $1-\hat{x},1-\hat{z}=\mathcal O(r_{\rm in,out})$. Since every final state contains exactly one quark, $\int_0^1\mathrm{d}\hat{z}\,\mathcal F_{qq}(\hat{x},\hat{z};r_\mathrm{in},r_\mathrm{out})=\mathcal F_q(\hat{x};r_\mathrm{in},r_\mathrm{out})$. With $\int_0^1\mathrm{d}\hat{z}\,(1+\hat{z}^2)/(1-\hat{z})_+=-\frac32$, the integral over $\hat{z}$ reproduces $\mathcal F_q$ of Eq.\,\eqref{eq:q:Fq} at $\hat{x}<1$, and the coefficient of $\delta(1-\hat{x})$ gives
\begin{align}
    K=\frac32C_\mathrm{F}\bigg[\ln\frac{Q^2}{\mu_{\mathrm{in}}^2}+\ln\frac{Q^2}{\mu_{\mathrm{out}}^2}\bigg]-C_\mathrm{F}\bigg(4+\frac{2\pi^2}{3}\bigg)\,,
\end{align}
where the logarithms complete the terms of Eqs.~\eqref{eq:q:Cqq zend} and \eqref{eq:q:Cqq xend} that are proportional to $\ln(Q^2/\mu_{\mathrm{in}}^2)$ and $\ln(Q^2/\mu_{\mathrm{out}}^2)$ to $\mathcal P_{q\leftarrow q}(\hat{x})$ and $\mathcal P_{q\leftarrow q}(\hat{z})$. In the remaining terms $1/(1-\hat{x})$ and $1/(1-\hat{z})$ become plus distributions, and with $C_\mathrm{F}(1+\hat{x}^2)/(1-\hat{x})_+=C_\mathrm{F}[2/(1-\hat{x})_+-(1+\hat{x})]$, and likewise for $\hat{z}$, the regular part takes the form of Eq.\,\eqref{eq:q:Cqq result}. With Eq.\,\eqref{eq:C from F} this gives the quark-to-quark SIDIS coefficient function in Eq.\,\eqref{eq:q:Cqq result}.

\subsection{Fragmentation functions into off-shell partons}
\label{sec:app:FF}

Here we derive the $\overline{\textrm{MS}}$ fragmentation function from a bare quark to dressed quarks and gluons, Eq.\,\eqref{eq:q:Dqq}. The parton distribution \eqref{eq:app:PDF def} counts bare partons in a dressed incoming state. The fragmentation function counts, conversely, outgoing dressed partons in a bare state, namely the bare parton $i$ produced by the hard scattering. With the time evolution \eqref{eq: LCPT: dressed states}, the operators of the outgoing dressed partons are the bare ones dressed in the same way,
\begin{align}
\label{eq:app:aout}
    a_j^{\rm out}\equiv U_\mathrm{I}^\dagger(\infty,0)\,a_j\,U_\mathrm{I}(\infty,0)\,,\qquad|j\,X\rangle^{\rm out}_{\rm D}=a_j^{{\rm out}\dagger}|X\rangle^{\rm out}_{\rm D}\,,
\end{align}
and the distribution of the off-shell parton $j^*$ with the plus-momentum fraction $\hat{z}$ in the frame where $i$ has plus momentum $k^+$ and no transverse momentum is
\begin{align}
\label{eq:app:FF def}
    D^0_{j^*/i}(\hat{z})\equiv\frac{1}{4\pi\hat{z}}\int\frac{\mathrm{d}^{D-2}\boldsymbol{\rho}}{(2\pi)^{D-2}}\,\frac{\big\langle i(k^+)\big|a_j^{{\rm out}\dagger}(\hat{z}k^+,\boldsymbol{\rho})\,a_j^{\rm out}(\hat{z}k^+,\boldsymbol{\rho})\big|i(k^+)\big\rangle}{\big\langle i(k^+)\big|i(k^+)\big\rangle}\,,
\end{align}
with the average over the color and polarization of $i$. Inserting the complete set of outgoing dressed states, the matrix element is $\sum_X\big|{}^{\rm out}_{\ \,\rm D}\langle j\,X|i\rangle\big|^2$, the sum over the unobserved final states $X$ of the operator definition of Ref.~\cite{Collins:2011zzd}, whose Wilson lines are unity in light-cone gauge, with $\boldsymbol{\rho}$ the transverse momentum of $j$ in the frame of $i$.

For $i=q$ the bare quark is produced by the hard scattering with $k^+$ and no transverse momentum, and the gluon is emitted after the scattering. Its overlap with the outgoing dressed quark \eqref{eq:q:Fock} is the one-particle component $\sqrt{Z_q^{\rm out}}$. Its overlap with the outgoing dressed $qg$ state, whose quark carries the fraction $\zeta$ and the relative transverse momentum $\boldsymbol{\rho}$, is the single-quark component of the bra, Eq.\,\eqref{eq:q:Fock qg}, $-\Psi^{q\to qg}(\boldsymbol{\rho},\zeta;\mu_{\mathrm{out}}^2)$ at $\mathcal O(g)$, with the energy denominator of the outgoing state regulated by $\mu_{\mathrm{out}}^2$. Since the parent has no transverse momentum, $\boldsymbol{\rho}$ is the transverse momentum of the quark of the pair and $-\boldsymbol{\rho}$ that of the gluon. For the observed quark, $\zeta=\hat{z}$, and
\begin{align}
\label{eq:app:FF LCPT}
    D^0_{q^*(\mu_{\mathrm{out}})/q}(\hat{z})=Z_q^{\rm out}\,\delta(1-\hat{z})+\frac{1}{4\pi\hat{z}(1-\hat{z})}\int\frac{\mathrm{d}^{D-2}\boldsymbol{\rho}}{(2\pi)^{D-2}}\,\frac12\sum_h\sum_{i,a,h_q,\sigma}\Big|\Psi^{q\to qg}_{h_q,\sigma;h}(\boldsymbol{\rho},\hat{z};\mu_{\mathrm{out}}^2)\Big|^2\,.
\end{align}
By Eq.\,\eqref{eq:app:Psi2} with $\mu_0=\mu_{\mathrm{out}}$ and the integral of Appendix~\ref{sec:app:PDF},
\begin{align}
\label{eq:app:FF qq PD}
    D^0_{q^*(\mu_{\mathrm{out}})/q}(\hat{z})=Z_q^{\rm out}\,\delta(1-\hat{z})+\frac{\alpha_\mathrm{s}}{2\pi}\,P^{(D)}_{q\leftarrow q}(\hat{z})\,(1-\varepsilon)\,\Gamma(\varepsilon)\Big(\frac{4\pi\bar\mu^2}{\hat{z}(1-\hat{z})\mu_{\mathrm{out}}^2}\Big)^\varepsilon\,.
\end{align}
By Eq.\,\eqref{eq:q:Z} at $\boldsymbol{\mathrm{k}}_q=0$, $Z_q^{\rm out}-1$ is minus the integral over $\hat{z}$ of the second term: the bare quark is found either as a single dressed quark or as the quark of a dressed $qg$ pair, which conserves the quark number. With Eq.\,\eqref{eq:app:Peps},
\begin{align}
    D^0_{q^*(\mu_{\mathrm{out}})/q}(\hat{z})=\delta(1-\hat{z})+\frac{\alpha_\mathrm{s}}{2\pi}\bigg\{P_{q\leftarrow q}(\hat{z})\bigg[\frac1\varepsilon+\ln\frac{\mu^2}{\mu_{\mathrm{out}}^2\hat{z}(1-\hat{z})}-1\bigg]-C_\mathrm{F}(1-\hat{z})\bigg\}_++\mathcal O(\varepsilon)\,.
\end{align}
For the observed gluon, $\zeta=1-\hat{z}$, and there is no one-particle component,
\begin{align}
\label{eq:app:FF gq}
    D^0_{g^*(\mu_{\mathrm{out}})/q}(\hat{z})&=\frac{1}{4\pi\hat{z}(1-\hat{z})}\int\frac{\mathrm{d}^{D-2}\boldsymbol{\rho}}{(2\pi)^{D-2}}\,\frac12\sum_h\sum_{i,a,h_q,\sigma}\Big|\Psi^{q\to qg}_{h_q,\sigma;h}(\boldsymbol{\rho},1-\hat{z};\mu_{\mathrm{out}}^2)\Big|^2\nonumber\\
    &=\frac{\alpha_\mathrm{s}}{2\pi}\,P^{(D)}_{g\leftarrow q}(\hat{z})\,(1-\varepsilon)\,\Gamma(\varepsilon)\Big(\frac{4\pi\bar\mu^2}{\hat{z}(1-\hat{z})\mu_{\mathrm{out}}^2}\Big)^\varepsilon\nonumber\\
    &=\frac{\alpha_\mathrm{s}}{2\pi}\bigg\{P_{g\leftarrow q}(\hat{z})\bigg[\frac1\varepsilon+\ln\frac{\mu^2}{\mu_{\mathrm{out}}^2\hat{z}(1-\hat{z})}-1\bigg]-C_\mathrm{F}\,\hat{z}\bigg\}+\mathcal O(\varepsilon)\,.
\end{align}
Subtracting $\mathcal P_{q\leftarrow q}(\hat{z})\big[1/\varepsilon+\ln(\mu^2/\mu_D^2)\big]$ and $P_{g\leftarrow q}(\hat{z})\big[1/\varepsilon+\ln(\mu^2/\mu_D^2)\big]$ in the $\overline{\mathrm{MS}}$ scheme gives the fragmentation functions in Eq.\,\eqref{eq:q:Dqq}. The fragmentation function into the quark has the form of the distribution $f_{q/q^*}$ with $x\to\hat{z}$ and $\mu_{\mathrm{in}}\to\mu_{\mathrm{out}}$, since the splitting and merging wave functions differ only by their sign and the conjugation, but it counts the emission after the scattering, in the outgoing states regulated by $\mu_{\mathrm{out}}$.

\bibliographystyle{JHEP-2modlong}
\bibliography{biblio}

\end{document}